\documentclass[fleqn,usenatbib]{mnras}

\usepackage{newtxtext,newtxmath}

\usepackage[T1]{fontenc}

\DeclareRobustCommand{\VAN}[3]{#2}
\let\VANthebibliography\thebibliography
\def\thebibliography{\DeclareRobustCommand{\VAN}[3]{##3}\VANthebibliography}

\usepackage{graphicx}	
\usepackage{amsmath}	
\usepackage{caption}
\usepackage{subcaption}
\usepackage[nopatch]{microtype}
\usepackage{booktabs}
\usepackage{threeparttable}
\usepackage{hyperref}
\usepackage{verbatim}

\title[The Illusion of Morphology]{The Illusion of Morphology in Tidal Structures: Changes to Stellar Shells and Streams in Non-Spherical Haloes}

\author[Gireesh et al.]{
Smrithi Gireesh Babu,$^{1}$\thanks{SGB: smrithi.gireeshbabu@sydney.edu.au, VE: naduran.ekanayakagedon@sydney.edu.au}
Viraj Ekanayaka,$^{1}$ \thanks{Joint first author}
William H. Oliver$^{2,3}$
and Geraint F. Lewis$^{1}$
\\
$^{1}$Sydney Institute for Astronomy, School of Physics A28, The University of Sydney, NSW 2006, Australia\\
$^{2}$Interdisziplin\"{a}res Zentrum f\"{u}r Wissenschaftliches Rechnen, Universit\"{a}t Heidelberg, Im Neuenheimer Feld 205, D-69120 Heidelberg, Germany\\
$^{3}$Zentrum f\"{u}r Astronomie, Institut f\"{u}r Theoretische Astrophysik, Universit\"{a}t Heidelberg, Albert-Ueberle-Stra{\ss}e 2, D-69120 Heidelberg, Germany
}

\date{Accepted XXX. Received YYY; in original form ZZZ}

\pubyear{\the\year{}}

\begin{document}
\label{firstpage}
\pagerange{\pageref{firstpage}--\pageref{lastpage}}
\maketitle

\begin{abstract}
We identify shell-like tidal structures in flattened haloes that appear stream-like when viewed under different projections. This dependence on projection demonstrates how changes in the host halo can directly impact the formation and classification of tidal debris, highlighting the challenges of relying solely on visual inspections. To address this, we employ our clustering-based classification framework to systematically categorise the tidally disrupted satellites into stream-like and shell-like structures. Our host consists of a static, three component MW model with flattening introduced along the $z$-axis NFW dark halo. We consider three halo shape scenarios: a spherical halo $q=1$ , an extremely oblate halo with $q=0.5$ , and a prolate halo where $q=1.5$. We evolve three types subhalos: a highly radial, massive subhalo favouring shell formation, an eccentric orbit leading to stream formation and an intermediate orbit case. We first classify the tidal structures visually using face-on and edge-on density projections of the 3D position distribution. This visual inspection reveals shell-like and stream-like formations across all the face-on projections of different halo shapes, while the edge-on projection leads to contrary classifications in some cases. To resolve these ambiguities, we apply the classification method developed in our earlier work analysing the structures in ordered density, radial and energy angle space. We further investigate the spatial dispersion of stream-like structures and the rate at which core density reduces as the flattening parameter varies. Our results demonstrate that the variations in halo shape can affect the formation and classification of tidal debris, as well as the spatial dispersion and core density evolution of streams. This offers insights on how not only the initial condition of the subhalo, but also the structural properties of the host halo, plays a crucial role in determining the morphology of tidal features. These findings offer new insights into the role of dark matter halo geometry in shaping the tidal structure formation and its contribution to hierarchical galaxy formation and evolution.
\end{abstract}




\section{Introduction}
The $\Lambda$ Cold Dark Matter ($\Lambda$CDM) model stands as the prevailing cosmological framework, explaining the formation and evolution of structure in the Universe through hierarchical assembly. In this paradigm, structures grow via the accretion and merging of smaller substructures \citep{Zeldovich_1978}, with Cold Dark Matter (CDM) acting as the dominant component driving gravitational influence on galactic \citep{Kaufman_white_1999} and cosmic scales. Dark matter haloes provide the grounds upon which baryonic matter accumulates, cools, and condenses to form galaxies \citep{Blumenthal_1984, Frenk_1988}. The predictive successes of $\Lambda$CDM have been substantiated by observations of large-scale structure in redshift surveys such as SDSS \citep{SDSS_2000} and cosmic microwave background measurements from Planck \citep{Planck_2020}.

Galaxies themselves are typically embedded within massive dark matter haloes that far outweigh their baryonic components \citep{Simon_geha_2007}. These haloes not only govern the gravitational dynamics of galaxy formation but also facilitate the accretion of smaller systems, such as satellite galaxies and globular clusters \citep{Peebles_1981, White_Rees_1978}. The Local Group provides a clear example of hierarchical structure formation with a few dominant galaxies, including the Milky Way, surrounded by numerous smaller satellites. The existence of these satellite galaxies is a natural outcome of $\Lambda$CDM cosmology which predicts the survival of subhalos within the larger host halo environment.

As satellite galaxies orbit within the potential well of their host, they are subjected to tidal forces that progressively strip material from their outskirts. This tidal stripping gives rise to low surface brightness (LSB) structures, composed primarily of older stellar populations, dispersed into the stellar halo of the host galaxy. The morphology of these tidal remnants is strongly influenced by the initial orbital parameters and mass of the satellite. Satellites on highly eccentric orbits tend to form narrow, elongated streams and tails \citep{Newberg_2016}, whereas more massive satellites on radial orbits produce shell-like features through repeated pericentric passages \citep{Hernquist_Quinn_1988}.

A wealth of observational evidence supports the presence of such tidal structures in the Milky Way’s halo. Prominent examples include the Sagittarius Stream \citep{Ibata_1994, Ibata_2001}, Palomar 5 \citep{Odenkirchen_2003}, and more recently discovered streams such as ATLAS-Aliqa Uma, Chenab, and GD-1 \citep{Ji_s5_streams_2020}. These discoveries have been made possible through wide-field surveys, notably the Gaia mission \citep{Gaia_2016}, Sloan Digital Sky Survey (SDSS; \citep{SDSS_2000}), the ATLAS-3D project \citep{Atlas_3d_2001}, and the Dark Energy Survey \citep{DES_2005}. Targeted spectroscopic programs like the S5 survey \citep{s5_2019} and APOGEE \citep{APOGEE_2017} have further enriched our understanding by providing precise kinematic and chemical abundance measurements of these structures.

Complementing observational efforts, numerical N-body simulations have been instrumental in elucidating the dynamical evolution of tidal debris and the role of dark matter haloes in shaping their morphology. Simulations have convincingly demonstrated how galaxies assemble within the potential wells of dark matter haloes, reinforcing the hierarchical framework predicted by $\Lambda$CDM \citep{Frenk_1988, springel_2005, Wang_2011}. Moreover, the ongoing detection of tidal streams and ultra-faint dwarf galaxies \citep{Kravtsov_2004} offers a promising avenue for addressing the "missing satellites problem" \citep{Missing_satellites}, which refers to the apparent discrepancy between the observed number of satellite galaxies and the larger population predicted by cosmological simulations.

Tidal structures serve as powerful probes of the underlying dark matter distribution. Their spatial and kinematic properties encode information about the gravitational potential of the host galaxy, including the shape, orientation, and depth of its dark matter halo. For instance, the triaxiality of the Milky Way's halo has been inferred through detailed modelling of the Sagittarius Stream’s phase-space distribution \citep{Law_Majewski_2010}. Thus, the study of tidal debris not only provides insights into the accretion history of galaxies but also offers a unique pathway to constraining the elusive properties of dark matter haloes.

Extending this area, we present our findings in classification of tidal structures that are evolved in a range orbits for different halo shapes. We also study how variations in the host halo’s shape influence the resulting stream properties. Our study demonstrates that, beyond the initial conditions of the infalling subhalo, the geometry of the host halo plays a significant role in shaping the formation and evolution of tidal structures and their observable characteristics.
This paper is structured as follows: Section~\ref{Section: Background} provides an overview of dark matter haloes and tidal structures, along with N-body simulations, to infer galactic potentials. Section~\ref{Section: Methods} describes the numerical setup and simulation methodologies, with the results presented in Section~\ref{Section: Results}, followed by a discussion and concluding remarks in Section~\ref{Section: Discussion and Conclusion}.

\section{Background}
\label{Section: Background}
\subsection{Dark matter Haloes}
Dark matter haloes are the fundamental non linear units of cosmic structure \citep{Frenk_white_2012}. Their hierarchical formation occurs through the collapse of a strongly bound stable core with more material being added on the lesser bound orbits. Halo growth is driven largely by mergers, with only major mergers leading to full mixing of the old and new components. Numerical simulations show that most of the halo mass originates from numerous minor mergers rather than major ones \citep{springel_2008, Wang_2011}. Early simulation work also confirmed the predicted abundance of dark matter haloes and their correspondence with observed galaxies \citep{Frenk_1988}.

The density distribution of dark matter haloes is broadly spherical and can be described by the Navarro–Frenk–White (NFW) profile \citep{Navarro_NFW_1997}. This profile features a shallower slope in the inner regions and a steeper decline in the outskirts, and has been extensively tested and refined in subsequent studies \citep{Power_2003, Frenk_2004}.
A significant fraction of halo mass resides in subhalos, which populate the outer regions of the main halo. These gravitationally bound subhalos undergo tidal disruption at different stages depending on their mass and orbital parameters. Such processes give rise to low-surface-brightness features such as streams and shells. In general, the dark-matter-dominated outskirts of subhalos are stripped first, while the more tightly bound stellar cores survive longer but eventually follow similar trajectories. This stripping process deposits stars into extended regions of the host halo, forming diffuse stellar haloes. The resulting tidal substructures are most commonly detected at distances of 20–50 kpc from the galactic centre \citep{Reino_2021}. 
The structure of dark matter haloes is often quantified using the concentration parameter, defined as the ratio of $R_{200}/r_s$, where $R_{200}$ corresponds to the radius within which the mean density is 200 times the critical density of the Universe and $r_s$ is the scale radius \citep{NFW_1995}. While virial concentration uses $R_{\text{virial}}$ instead of $R_{200}$\citep{Bullock_2001}. The concentration parameter helps us to better constrain the amount of dark matter within the region; although it does not include the total mass of the halo completely, it hints at a shape of the dark matter halo.

\subsection{Tidal structures and Galactic potential}
Tidal structures have become powerful probes to infer the host galactic potential and formation history. Streams have been used to infer the host potential by fitting their orbit \citep{Newberg_2016}, particle spray modelling \citep{kupper_2015}, N-body simulations \citep{Law_Majewski_2010}, action angle method \citep{Bovy_2016} and generative models in the recent years\citep{Nibauer_2025}.

The shape of host halos, particularly the Milky Way's, has been a pertinent point of consideration for tidal debris morphology. The general consensus leans away from a perfectly spherical model with studies such as \citep{Vera_2011}, finding the MW halo to be prolate in the inner regions, transitioning to a more triaxial/oblate configuration in the outskirts. In the case of \citep{Bariego_2024, Bariego_2022}, independent analysis of the SPARC \citep{SPARC} rotational curves showed that slightly prolate configurations provide significantly improved fits compared to spherically symmetric models, underscoring that halo asymmetry is observationally motivated and should be addressed in debris morphology studies.

The detection of numerous tidal streams around the MW, like the Saggitarius \citep{Ibata_1994}, Palomar5 \citep{kupper_2015} and GD-1 \citep{Koposov_2010} have all contributed to constraining the halo. The Sagittarius stream was among the first to be used for such studies: early work indicated a nearly spherical halo \citep{Ibata_2001b}, later revised to a more prolate structure based on radial velocity data from the leading arm \citep{Helmi_2004}, and subsequently to oblate or triaxial configurations \citep{Law_Majewski_2010}. With Gaia DR2 and DR3, stream modelling has significantly advanced. For example, Palomar 5 has been used to infer a nearly spherical halo with $q \approx 0.95$ \citep{kupper_2015}, GD-1 suggests a tilted halo with $q \approx 0.81$ \citep{Nibauer_2025}, and combined analyses of Pal~5 and GD-1 in action--angle coordinates point to a global flattening of $q \approx 0.95$ \citep{Bovy_2016, Bonaca_2018, Koposov_2023}. More recent studies highlight that triaxial or resonant halo geometries can alter stream morphologies, producing diffuse “fans” or bifurcations, and driving dynamical diffusion through separatrix divergence \citep{Pearson2015, PriceWhelan2016, Yavetz2020}. Time-dependent effects, such as perturbations from the Large Magellanic Cloud, also deform streams and complicate static potential inferences \citep{Brooks2024, BuistHelmi2015}. \cite{Johnston_2005} and \cite{Fellhauer_2006} use streams to suggest that halos are slightly oblate. In contrast, with the application of kinematic data \citep{Helmi_2004} argue for a slight prolate profile. \citep{Bowden_2016} independently supports the prolate stance using SDSS \citep{SDSS_2000} data. However, the exact shapes continues to remain a point of contention with various studies such as \citep{Posti_2019} and \citep{Bovy_2016} leaning towards prolate and oblate configurations respectively. The use of stellar streams projection and their sensitivity to local gravitational acceleration was utilised by \citep{Nibauer_2023} to constrain the dark matter halo of NGC5907. 
Similarly, streams with multiple wraps serve as a strong tool for extracting the underlying potential they were disrupted in. Making use of this, \citep{Walder_2025} derived the radial density and the scale radius of the host halo using positional data alone. Additional radial velocity measurements provided a significant tightening of constraints on the enclosed mass. On an extending scale, \citep{Chemaly_2026} developed a hierarchical Bayesian framework that combines posteriors on halo flattening from many individual projected stream tracks, demonstrating that stream morphology aggregated across galaxy populations can constrain the distribution of halo oblateness without kinematic data.

Shells, in contrast to streams, are typically associated with radial or near-radial mergers and are more commonly detected in external galaxies \citep{1991_Thomson}. Early simulations demonstrated that line-of-sight velocities across shells could trace the radial gradient of the potential, effectively mapping the host halo under the assumption of radial infall \citep{Merrifield_1998}. Observations of shells in M31 revealed their kinematics to be strongly dependent on the halo potential \citep{Fardal_2007}, and further work has reinforced their use as probes of halo structure in prominent shell systems like NGC474, NGC3923 \citep{Bilek_2022, Fensch_2020}. Within the Milky Way, while classical shell systems have not been definitively identified, shell-like debris has been linked to past merger events. The Gaia-Sausage/Enceladus (GSE) merger is thought to have significantly contributed to the thick disk and stellar halo \citep{Helmi_2018, Belokurov_2018}, while the Virgo Radial Merger has been interpreted as a large shell-like structure arising from a major radial accretion event \citep{Donlon_2020, Donlon_2022}.

\section{Simulations}
\label{Section: Methods}
\subsection{Host Halo potential}
GADGET-4 \citep{2021_springel} is employed to model the tidal evolution and disruption of subhalos within an external Milky Way potential. Our simulations follow the static three-component framework described in \citet{paper_1}. The bulge and disk are represented by Hernquist and Miyamoto–Nagai profiles respectively, while the NFW dark matter halo is modified to include a flattening along its $z$-axis. This is implemented by introducing a flattening parameter, $q$, into the radial coordinate. The parameter regulates the halo curvature, that is, $q = 1$ recovers a spherical NFW halo, $q < 1$ produces an increasingly oblate configuration (with q = 0.5 marking an extreme case), and $q > 1$ yields a prolate structure. Minor updates to several model parameters have also been applied and are summarised in Table \ref{IC_table1}. The three components are characterised as follows. 

The bulge is defined as a Hernquist potential \citep{Hernquist_1990},
\begin{equation}
    \Phi_{\text{bulge}} = -\frac{GM_{\text{bulge}}}{r_{\text{bulge}} +r}.
\end{equation}
For our purpose, we set radii scale $r_{\text{bulge}}=0.7$ kpc and bulge mass $M_{\text{bulge}}=0.7\times10^{10}$ ${\rm M_\odot}$. 
The disk is comprised of the Miyamoto-Nagai potential \citep{Miyamoto_Nagai_1975},
\begin{equation}
    \Phi_{\text{disk}}(R,z)=-\frac{GM_{\text{disk}}}{\left(R^2+\left(r_\text{disk}+\sqrt{\left(z^2+b^2\right)}\right)^2\right)^{1/2}},
\end{equation}
where $b$ is the scale height which we set to a fifth of the disk radius. Radii scale $r_{\text{disk}}=3.5$ kpc and the disk mass is set at $M_{\text{disk}}=0.7\times10^{10}$ ${\rm M_\odot}$. 
Finally, the encompassing DM halo is represented as the typical Navarro-Frenk-White (NFW) potential, 
\citep{Navarro_NFW_1997}, defined as:
\begin{equation}
\Phi_{\text{halo}} = -\frac{GM_{\text{halo}}}{r} \ln\left(\frac{r}{r_{\text{halo}}}+1\right).
\end{equation}
with r being,
\begin{equation}
    \sqrt{x^2 + y^2 + (z/q)^2}
\end{equation}
The Halo parameters are influenced by the virial radius and mass and the concentration. The virial radius is given by 
\begin{equation}
    r_{\text{vir}} = 258\left(\frac{\Delta_{\text{vir}}\Omega_M}{102}\right)^{-1/3}  \left(\frac{M_{\text{vir}}}{10^{12}{\rm M_\odot}}\right)^{1/3} \text{ kpc}
\end{equation}
with $\Delta_{\text{vir}}= 340$ and $\Omega_M = 0.3$ and with the $M_{\text{vir}} = 1.3\pm 0.3 \times 10^{12} {\rm M_\odot}$ from \citet{Bland-Hawthorn_2016}, we fixed our $r_{{\rm vir}} \approx 276$ kpc and the radius of the halo is the ratio of the virial radius and the concentration parameter \citep{Bullock_2001}.

\subsection{Initial Conditions}
We continue using the same custom developed initial code that generates our initial condition and parameter files. This follows creating progenitors with Plummer sphere potentials as per \cite{Aarseth_1974}. We use the following initial conditions and vary the flattening parameter $q$ from 0.5 (extreme oblate) to 1.5 (extreme prolate). The simulations have three subhalos characterised by the following properties:

\begin{table}
\centering
\caption{Summary of the initial conditions for evolved subhalos.}
\label{IC_table1}
\begin{tabular}{cc}

\hline \textbf{Mass} & \textbf{Initial}\\
 ${\rm M_\odot}$ & \textbf{Orbit} \\ \cline{1-2}
\midrule
$10^7$ &  Radial \\ 
\midrule
$10^5$ &  Elliptical \\
\midrule
$10^6$ & Intermediate$^{a}$ \\ 
\bottomrule
\end{tabular}
\begin{tablenotes}
\item[a]$^a$ initial orbit condition is between a radial and eccentric orbit
\end{tablenotes}
\end{table}

All subhalos are evolved for approximately $8.6$ Gyrs to study the effect of the halo flattening on the tidal evolution and structure formation. While we acknowledge that using a flattened NFW potential does not perfectly capture the complexity of realistic triaxial haloes, the implications of this model choice is discussed further in Section~\ref{Section: Discussion and Conclusion}.

\section{Results}
\label{Section: Results}
\subsection{Structure and Morphology}
\label{Subsection: Structure and Morphology}

\begin{figure*}
\centering
    \setlength{\tabcolsep}{1pt} 
    \renewcommand{\arraystretch}{0} 
\includegraphics[width=.9\textwidth, height=.6\textwidth]{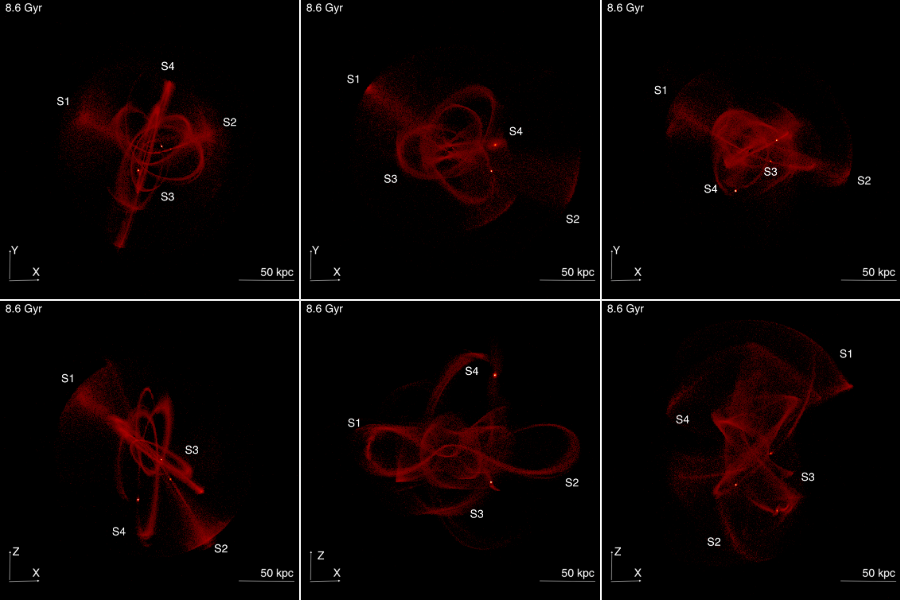} 
       
    \caption{The position density distribution of all three progenitor conditions evolved to $8.6$ Gyrs. Row 1 shows the face-on projection while Row 2 is the edge-on perspective. From left to right; Column 1 depicts the spherical NFW halo with \(q=1\), Column 2 contains the oblate NFW halo with \(q=0.5\), and Column 3 contains the prolate halo with \(q=1.5\). All plots are shown to the same scale, spanning 50 kpc.}
   
    \label{fig_1:position_distribution_for_all_q_values}
\end{figure*}

We present the progenitors listed in Table~\ref{IC_table1} to illustrate how their structural evolution varies as a function of the halo flattening parameter $q$. Figure~\ref{fig_1:position_distribution_for_all_q_values} shows the face-on projections in the first row and the edge-on projections in the second row for the three values of $q$: from left to right, we have the oblate ($q=0.5$), the spherical ($q=1$), and the strongly prolate case ($q=1.5$). We refer to the tidal structures formed from the radial progenitor as structures S1 and S2 (with S5 in the oblate). Structure S3 emerges from the progenitor with eccentric orbit and S4 is produced by the progenitor with the intermediate orbit. All the tidal structures are labelled in Figure \ref{fig_1:position_distribution_for_all_q_values}.

\subsubsection{Spherical Case}
We first examine the spherical scenario ($q=1$). Here, the two progenitor cores with eccentric and intermediate initial conditions follow stream producing trajectories. Each progenitor contributes to the formation of streams S3 and S4 labelled in Fig \ref{fig_1:position_distribution_for_all_q_values}. Although these streams appear to lie on perpendicular orbital planes in both face-on and edge-on projections, the full 3D view shows that each progenitor generates a distinct type of stream orbit.

The progenitor located at the top of the distribution, which had an intermediate initial condition, contributes to a ring-like orbit which in 3D remains well aligned along the $y$-axis. The extended orbital path for this structure could be due to its initial condition being intermediate between radial and eccentric orbits. In contrast, the subhalo on the eccentric orbit produces stream S3, which is responsible for the formation of the rosette-like orbital patterns. The orbital paths of these two streams differ in distance from the host centre, their orbital shapes and the masses of their progenitors. We also identify the progenitor with the initially radial orbit, forms plume-like features in the face-on view. These later appear as shells in the edge-on view and diffuse more slowly. 

It is important to acknowledge that the appearance of tidal structures is strongly time dependent. As these systems evolve, the visibility, coherence, and morphology change continuously. While we choose to illustrate the structures at a specific timestep of $8.6$ Gyrs, it should be emphasised that other evolutionary stages would be just as revealing. For example, approximately $4$ Gyr later, the same shell forming progenitor forms further shells that are easily identifiable in both projections. The progenitor later transitions into a semi-radial orbit, it eventually forms larger shells. This produces a Type~I shell system where the shells occur on opposite sides and as a result, undergoes the largest mass loss among the progenitors.

\subsubsection{Oblate Case}

In the oblate case ($q=0.5$), the eccentric orbit produces the S3 stream in a different orbital plane compared to the spherical scenario. The progenitor with intermediate initial condition forms a less distinct stream in the face-on view. Both S3 and S4 appear significantly dispersed, making their structures harder to distinguish. 

The remaining bound particles of the subhalo on the intermediate orbit is more disrupted than in the $q=1$ case. This results in the formation of three shells in the face-on projection labelled S1, S2, and S5. The third shell, S5, is not seen in the spherical case. The presence of an additional shell shifts the overall shell system's morphology from the Type~I (spherical) to a Type~II shell system, where the shells are distributed in all directions. Shell formation begins earlier than in the spherical case, the first shells appear at $\sim 4.06$~Gyrs and additional shells forming roughly $0.2$~Gyrs later. These shells do not form via typical pericentric motion of the progenitor. Instead, they emerge in close succession and appear to brighten over time. The shell formation rate is significantly higher than in the spherical case, and shells S1 and S2 are noticeably wider. Streams on the other hand, have longer orbital paths and diffuse alongside already-dispersed shell particles. This produces a more phase-mixed appearance. Overall, the particle distribution is more dispersed in comparison to the spherical case. Shells S1, S2, S5 remain clearly visible as they are well outside of the phase-mixed population. Streams S3 and S4 are only partly visible in the face-on projection.

The edge-on projection paints a very different picture: Shells S1, S2, and S5 identified in the face-on view, are part of a single wide stream with a wave-like orbit. This stream contains denser regions at its peaks and troughs. The three shell-like features therefore correspond to the peaks of this extended stream. Despite its initially radial orbit, and the progenitor responsible for it, the trajectory of this shell-stream structure changes in the strongly oblate halo. This leads to the formation of a stream with an oscillatory structure and a wider particle distribution. This contributes to the shell-like resemblance in different viewing angles. Stream S4 appears as a massive ring-like structure in the edge-on projection, even though it was barely visible in the face-on view. Stream S3 appears extremely dispersed in the edge-on projection and is only visible at the edges of the structure, hinting its stream-like nature.

\subsubsection{Prolate Case}

In the prolate scenario ($q=1.5$), the face-on projection shows that the cores of all three subhalos remains identifiable. However, the particle distribution is substantially more dispersed, making morphological distinctions more difficult. The stream-forming progenitors produce noticeably wider streams which, in some projections, resemble small shells. These broadened streams, combined with the high degree of dispersion, produce shell-like features that mix rapidly and lead to a much more phase-mixed appearance than the previous case.
Shell formation in S1 and S2 follows a predominantly Type~I morphology. The first shell appears at $\sim 6$~Gyrs and subsequent shells emerging $\sim 3.64$~Gyrs later. Although the shell formation rate is comparable to the spherical case, shells appear earlier. They also diffuse more slowly than in both the spherical and oblate scenarios. Streams S3 and S4 are visible on different orbital planes. In certain viewing angles, a bulge-like structure appears between S1 and S3. This overdensity seems to originate from S4, possibly due to interactions among the progenitors.

The edge-on projection again reveals a different morphology. Shells S1 and S2, produced by the radial orbit progenitor, are barely visible. In contrast, streams S3 and S4 are clearly seen. These streams show folds that increase the density contrast and give them a shell-like appearance. All three progenitors appear substantially more diffused compared to the other halo configurations.

Despite the change in $q$ leading to significant variations in the eccentric subhalo’s orbital path, it ultimately forms a stream-like structure S3 in all three cases. We discuss the impact of these changing orbital paths in Section~\ref{Subsection: Orbital Evolution and Dispersion}. The intermediate orbit subhalo also forms a stream, S4, in all three scenarios. Although S4 exhibits characteristics of both streams and shells, it results in a stream with an extended orbital path.

Finally, the radial progenitor forms shells S1 and S2 in both the spherical and prolate cases. In the oblate case, however, it transforms into a massive and wide stream. The timing of the first shell formation, the total number of shells, their radii, and their survival times differ between halo scenarios. As a result, many structures appear transitional or ambiguous depending on the projection. This makes visual classification inherently challenging.

\subsection{6D phase space clustering with AstroLink}
\label{Subsection: AstroLink}

The morphological classification of tidal structures based on visual inspection is inherently ambiguous. The appearance of shells and streams can vary significantly depending on the projection angle, with the same structure appearing shell-like in one view and stream-like in another. This projection dependence complicates a purely morphological classification based on face-on and edge-on views alone. To help resolve this ambiguity, we turn to the full six-dimensional phase-space information of the particles and examine whether this provides a more informative classification of structures. 

\begin{figure}
\centering
\includegraphics[width=\linewidth]{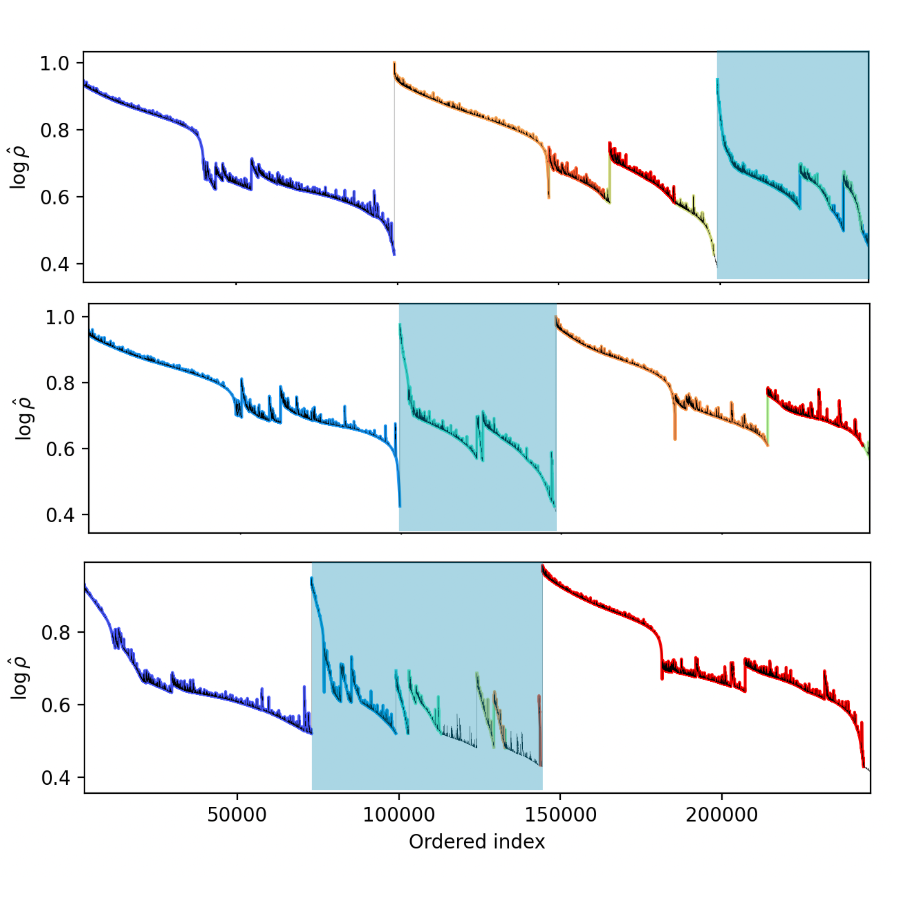}
\caption{From top to bottom, the three panels correspond to $q=0.5$, $q=1$ and $q=1.5$. AstroLink ordered density plots with the identified clusters coloured and the highlighted cluster representing the shell-like structure in all the three q scenarios. Shells S1 and S2 (and S5 in the case of q = 0.5) is highlighted in turquoise blue cluster, Stream S3 is reddish orange cluster and Stream S4 is the blue cluster.From Top to bottom q values for q=0.5, q=1, q=1.5}
\label{fig:od_plot}
\end{figure}

\begin{figure}
\centering
\includegraphics[width=\linewidth]{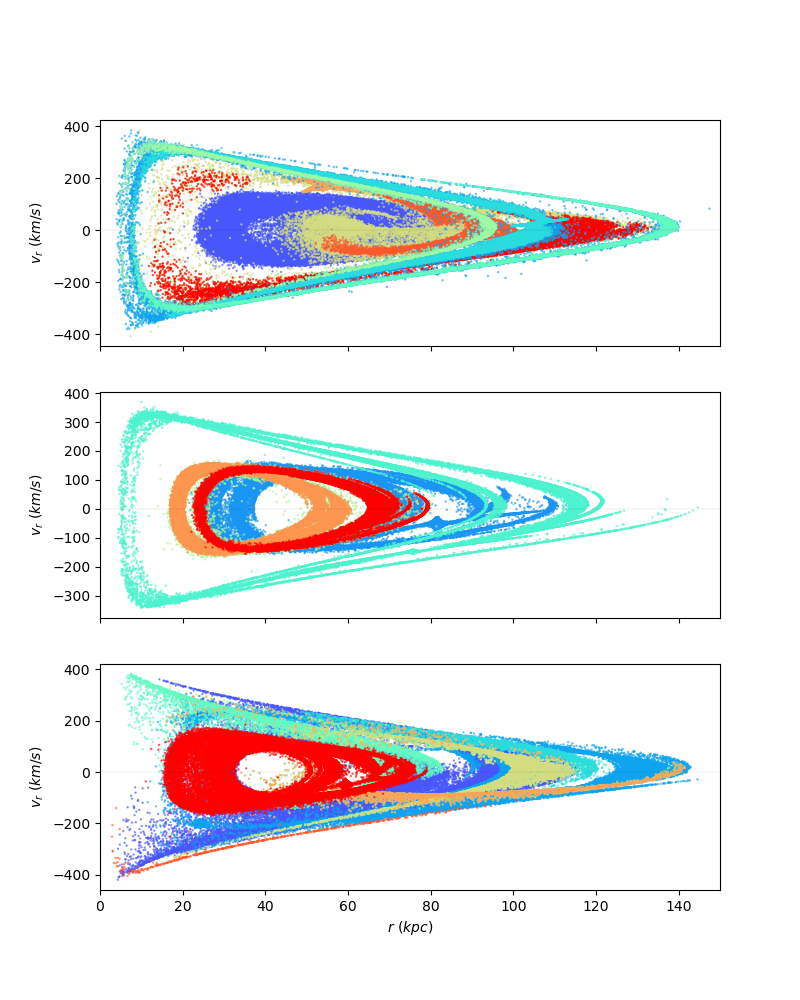}
\caption{From top to bottom, the panels show the radial-velocity–distance distributions of all the subhalos evolved to $8.6 Gyrs$ for the extreme axis-ratio values $q = 0.5$, $1$, and $1.5$. Cluster colours are consistent with those used in the ordered-density plots figure \ref{fig:od_plot}. Shell-like signatures are visible in all three cases, with the oblate and prolate potentials exhibiting more phase-mixed distributions.}
\label{fig:rad_plot}
\end{figure}

To differentiate shell-like from stream-like morphologies, we apply the clustering algorithm AstroLink \citep{2024_Oliver}. With it's use, we are able to produce hierarchical density ordered clusters in six dimensional phase space. The resulting clusters are then used to generate a set of diagnostic plots to examine how the structures appear in the algorithm's native ordered density space (figure \ref{fig:od_plot}), as well as in radial (figure \ref{fig:rad_plot}) and energy–angle space (figure \ref{fig:energy angle plot}).
These diagnostics provide a direct way to compare the visual classification of the face-on and edge-on projections described in Section \ref{Subsection: Structure and Morphology} with the classifications inferred from the 6D clustering. We adopt our classification framework from \citep{paper_1}, which highlights the signature profile markers produced by shells and streams in their ordered density distribution.  Typically streams map a three-curve cluster representing the presence of the leading and tailing tidal tails and the remaining core remnant. While shells are generally more homogenised with little to no subclusters and hence appear as a single smooth cluster curve. However, as identified in figure \ref{fig:od_plot}, atypical structures may present more fractured cluster curves that represent greater phase mixing and degeneracy. We also use the clustered results of position and velocity information from AstroLink to study the structures in radial and energy angle space to search for their characteristic signatures that further verify our classification.

\begin{figure}
\centering
\includegraphics[width=\linewidth]{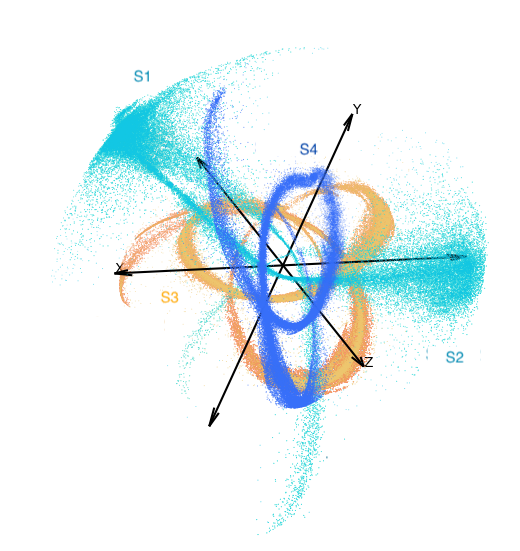}
\caption{3D position distribution of the q = 1 scenario where this view reveals the 2 shells, 2 streams and faint tail like structures that seems to be in the same cluster as the Shells.All structures are labelled in their cluster colour.}
\label{fig:pos_distribution_q1}
\end{figure}

Figure~\ref{fig:pos_distribution_q1} shows a clear example of the cluster-coloured position distribution for the spherical case with $q = 1$. In this configuration, the two plume-like structures S1 and S2, identified earlier in Section~\ref{Subsection: Structure and Morphology}, appear more clearly as shells in 3D position space. Shell S1 is oriented predominantly along the $z$-axis, while S2 lies mainly along the $x$-axis. We also recover the two streams, S3 and S4, associated with the eccentric and intermediate progenitors.

The shell-forming progenitor additionally produces tail-like features that lie on either side of the shells, extending below them. These tail structures were not evident in Figure \ref{fig_1:position_distribution_for_all_q_values} and were therefore missed by visual classification. This highlights the limitations of 2D projection view when compared to the clustered 6D phase space distribution. Shell S1 exhibits a broader particle spread than S2, suggesting that S2 is the more recently formed shell, while S1 has already begun to diffuse. 

The visibility of progenitor cores in Figure~\ref{fig_1:position_distribution_for_all_q_values} provides an advantage over the 3D position plots. In density projections, the cores appear as strong overdensities because most particles remain bound to the subhalo. In contrast, the cores are more difficult to visually identify in 3D position space. This limitation is overcome by the ordered-density plots produced by AstroLink, where the cores are easier to locate. Each density peak corresponds to a physically distinct cluster, and subsequent peaks of the same colour represent subclusters belonging to that main structure.

The ordered-density plots in figure~\ref{fig:od_plot} are arranged in rows corresponding to the oblate, spherical, and prolate halo cases. Starting with the first row ($q = 0.5$), the dark blue cluster corresponds to S4. Its ordered-density profile is characteristic of a stream, with subclusters representing the core and tails. The reddish-orange cluster corresponds to S3, which exhibits the characteristic stream-like structure, with distinct subclusters associated with the head, tail, and surviving core.

Following the sharp density drop-off, the three shell-like features that are most prominent in the density projections appear as shell-like substructures in ordered-density space. Here, shell S1 corresponds to the first peak with the highest ordered density, while S2 and S5 correspond to the subsequent peaks. This demonstrates that, despite the edge-on view reclassifying all three shells as part of the wide stream, their ordered-density distribution retain shell-like signatures.

This behaviour arises from the combination of a highly radial orbit and an extreme oblate halo shape. The flattened host potential reshapes the debris flow, causing the massive progenitor to form a wide stream with a long, wave-like orbital path. Overdensities develop at the turning points of this orbit and appear as shell-like features in some projections. As a result, S1, S2, and S5 occupy an intermediate classification regime between classical shells and streams. While shells do form in this configuration, they dissolve more rapidly. When the shell material reaches apocentre along the $z$–axis, the flattening deflects the debris into the plane of symmetry of the halo. Making the remaining high-density features appear as thick, stream-like structures. This provides direct evidence of host potential strongly reshaping massive subhalo on radial orbits, producing significant changes in the orbit causing the subhalo to form wide streams with thick tails that retain shell-like signatures depending on the viewing angle. 

Importantly, these structures retain shell-like signatures in both radial and energy–angle space, as discussed in Sections~\ref{Subsection:Radial phase space} and~\ref{Subsection: Energy Angle space}. This suggests that while their morphology in position space becomes more stream-like due to flattening and subsequent orbital evolution, their underlying kinematics preserve the imprint of shell formation. Radial and energy–angle diagnostics therefore support our classification with  than position-space morphology alone.
In the spherical case, the blue cluster S4, shows a less clearly defined core and tail separation in the ordered-density space. The clusters corresponding to S1 and S2 highlighted, exhibit the smooth, single-curve profiles expected for shells. Stream S3, shown in reddish-orange, displays the characteristic multi-curve structure of a stream. Compared to the oblate case, the progenitor core of S3 is denser, the head of the stream is less dense, and the tail has a similar density. This contrast likely reflects the increased precession in the oblate halo, where successive orbits lie in different planes and enhance disruption.

In the prolate scenario, the structures retain the same colour. S4 in the ordered-density plot is much harder to decompose into distinct subclusters, although its overall structure remains stream-like. Shells S1 and S2 are poorly recovered by AstroLink's clustering because of their advanced phase mixing.They also seem to contain more particles than other cases, likely due to the contamination from the less dense stream S4. Stream S3 behaves similarly to the spherical case, potentially with more particles in its subclusters, particularly in its tail relative to its core. We interpret this as a consequence of stronger phase mixing of the debris: as the structures diffuse further and become increasingly bound to the host halo, it becomes more difficult for AstroLink to disentangle them into distinct clusters.

\subsection{Radial Phase Space}
\label{Subsection:Radial phase space}
The radial and energy-angle plots are used not only to aid the classification of shells and streams, but also to probe the influence of the host halo. The series of radial phase space contains show most coherent structures in the spherical halo(middle row) in comparison to the oblate and prolate cases, which appear more widely phase mixed.
In the oblate potential, S1, S2 and S5 take on shell-like patterns, indicating that they are more like a shell than a stream at 8.6 Gyr. The slight green colour is due to the mix of particles from S1, S2, S5 and S3 being in close proximity. This overlap suggests that parts of these structures are transitioning toward stream-like behaviour. S4, on the other hand, is clearly a stream. S3, which is the red cluster, shows some shell-like radial elongations in some regions, while preserving stream signature between $50-60$ kpc. The mixed behaviour likely results from differential stretching driven by the halo distribution and initial conditions. 

In the spherical potential, the shells produce well-defined radial peaks, while streams S3 and S4 exhibit the characteristic guitar-pick morphology. The faint core of S4 is visible near $r \sim 80$ kpc. All tidal structures are clearly recovered in radial phase space, reinforcing the reduced ambiguity of the spherical case.
In the prolate halo, clusters appear more radially extended than in the other configurations. Streams develop stronger radial elongations and transition toward more egg-like shapes. Shells S1, S2 and parts of the stream S4 retain shell-like signatures, but their conical peaks are less sharp and not closed. The reduced range of radial velocities for particles closer to the host suggests slower shell evolution in the prolate halo compared to the oblate and spherical cases.

\subsection{Energy Angle Space}
\label{Subsection: Energy Angle space}
In energy-angle space, the imprint of halo shape becomes even more pronounced. The oblate halo (row 1) produces a more well-like distribution, the spherical configuration (row 2) shows relatively straight, layered features, and the prolate halo (row 3) exhibits a more irregular, ``bumpy'' pattern. The structures are most clearly separated in the spherical halo, while they become more diffuse and blended in the oblate and prolate cases.
\begin{figure}
\centering
\includegraphics[width=\linewidth]{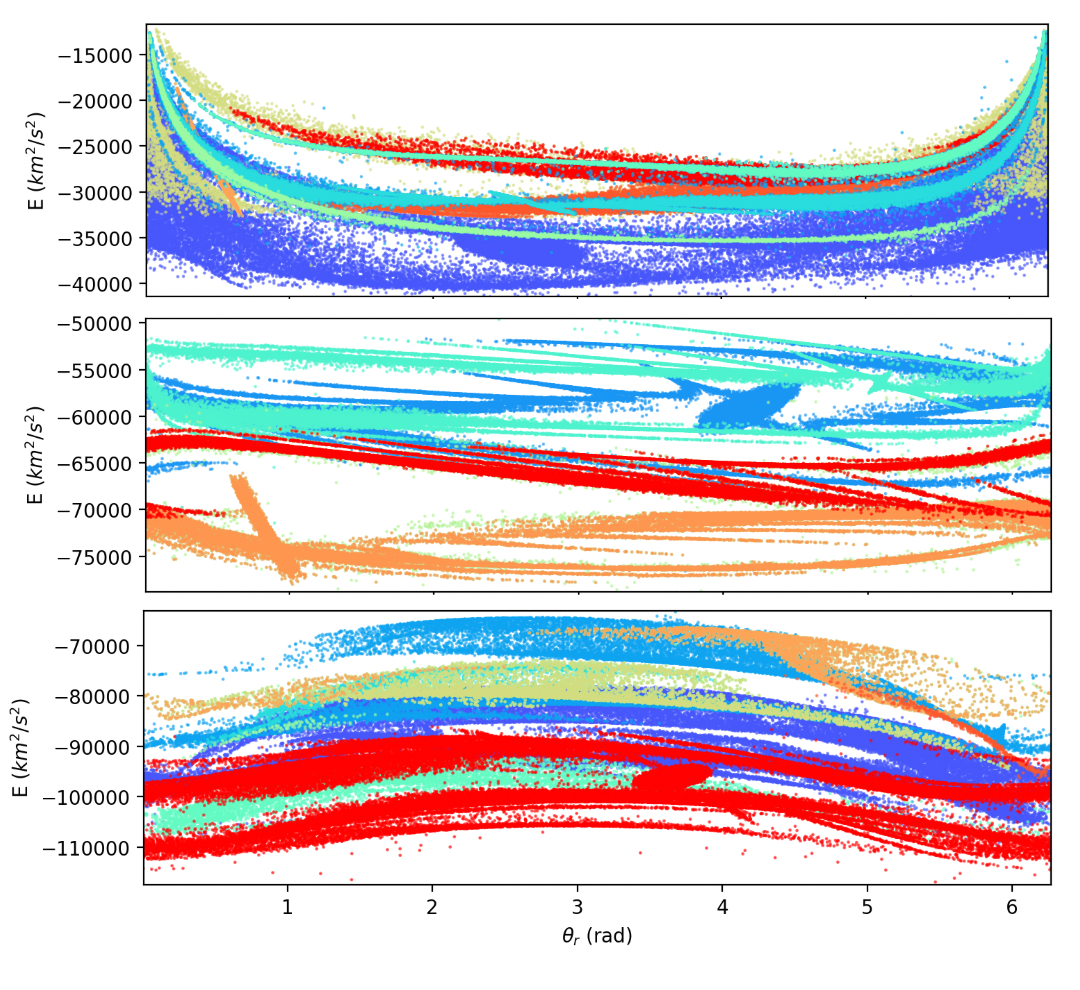}
\caption{From top to bottom, the panels show the AstroLink cluster-coloured energy–angle distributions of the structures evolved to 8.6 Gyr for q = 0.5, 1, and 1.5. In each case, the overall energy distribution mildly reflects the underlying shape of the halo.}
\label{fig:energy angle plot}
\end{figure}
in the oblate case, Stream S4 spans a broader energy range than the other structures. Shells S1, S2 and S5 split into three quasi-parallel branches,  reminiscent of multiple wraps of shell-like features. This contributes to the added shell-stream transition behaviour described in the earlier sections \ref{Subsection: Structure and Morphology}, \ref{Subsection: AstroLink}. This is consistent with their initially more radial orbit and their larger vertical excursions, which cause them to experience the halo flattening more strongly. We see S3's energy-angle distribution lying behind the shells, with two thick features in the \(-20,000\) to \(-30,000\ km^2s^{-2}\) energy range and it resembles stream pattern.

Mapping these structures onto the spherical potential, S3 spans in energies from \(-75,000\) to \(-62,500\ km^2s^{-2}\). A series of small, nearly straight lines trace the debris along the stream, while a thicker, more prominent branch at angles \(\sim 1\)-- \(2\,\mathrm{rad}\) corresponds to the stream core and its immediate surroundings. Because the angular separation between these components is small, they appear somewhat compressed in the plot. S4 shows a significantly larger core than S3. There are also nearly linear features at energies around \(-60,000\) and \(-53,000\ km^2s^{-2}\), associated with shell-like structures S1 and S2. In particular, the feature at \(-53000 \ km^2s^{-2}\) with a concentration near \(\sim 5\,\mathrm{rad}\) is consistent with the core of a shell whose debris remains relatively coherent. 

In the prolate halo, the stream S3 appears along two almost straight features, with its core lying between \(3\)–\(4\,\mathrm{rad}\). S4 spans an energy range of \(-90000\) to \(-70000 \ km^2s^{-2}\), showing a large variation in energy. The shell substructures of S1 and S2 appear more in the background of the streams. When viewed without the streams, the shells appear phase mixed yet still retaining the straight-line behaviour they are known for. S3 appears more dispersed than its spherical and oblate cases, with a wider energy distribution. The S4 in the background has a somewhat narrower energy range compared to the oblate case, but its core appears vertically stretched.

The overall energy range in figure \ref{fig:energy angle plot} varies greatly with halo shape: the oblate halo spans \(-40000\) to \(-15000 \ km^2s^{-2}\), the spherical spans \(-75000\) to \(-50000 \ km^2s^{-2}\), and the prolate configuration spans \(-110000\) to \(-70000 \ km^2s^{-2}\). This reflects how the total energy distribution changes with the flattening parameter \(q\), and this in turn is imprinted onto the structures that form, as one would expect.

\subsection{Isolated Example of the Intermediate Progenitor}
\label{subsection: Subhalo B example}

\begin{figure*}
\centering
    \setlength{\tabcolsep}{1pt} 
    \renewcommand{\arraystretch}{0} 
\includegraphics[width=1\textwidth]{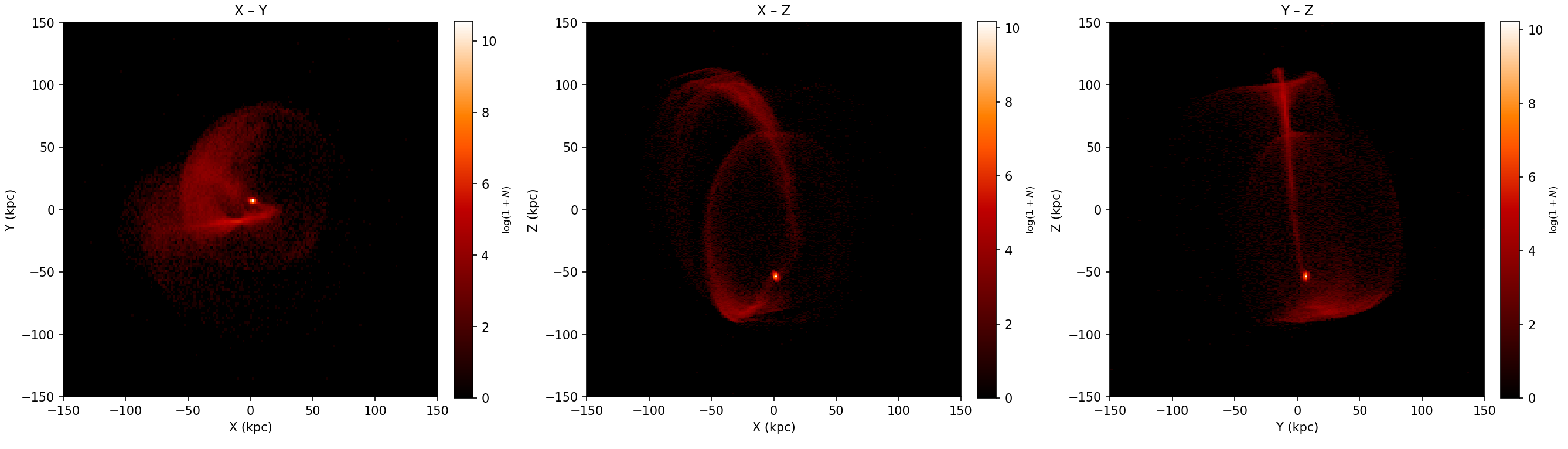} 

\caption{The isolated position density distribution of intermediate orbit progenitor S4. From left to right are the $X-Y$, $X-Z$ and $Y-Z$ face projections. S4 appears Shell like in $X-Y$ while stream-like and diffused in $X-Z$ and $Y-Z$.}
\label{fig_6:Subhalo_B_subsection}
\end{figure*}

\begin{figure*}
\centering
    \setlength{\tabcolsep}{1pt} 
    \renewcommand{\arraystretch}{0} 
\includegraphics[width=0.9\textwidth, height=.35\textwidth]{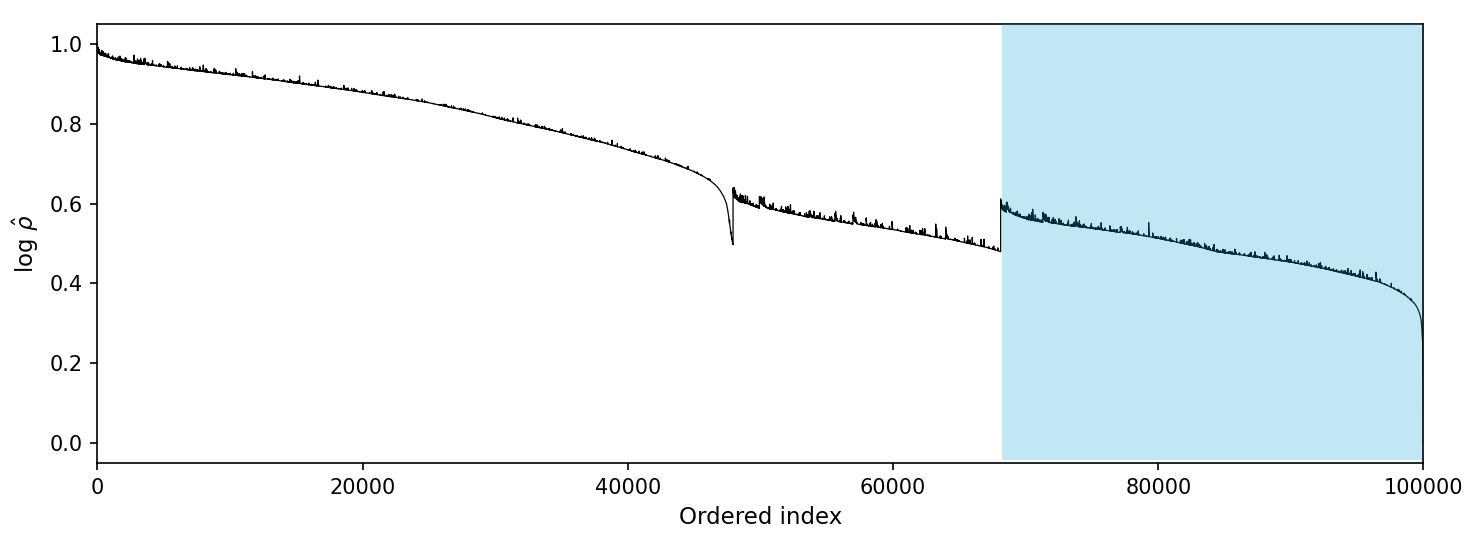}

\caption{The ordered density of the intermediate progenitor S4. We see a characteristic stream-like structure. The density curve traces the drop of stream in the first and last sub-cluster but the middle sub cluster highlighted in blue displays a shell-like distribution feature.}
\label{fig_7:Ordered_density_subhalo_B}
\end{figure*}

To better visualise the classification methodology, figure \ref{fig_6:Subhalo_B_subsection} presents an isolated depiction of S4 (as per figure \ref{fig_1:position_distribution_for_all_q_values}) in three separate projections for the extremely oblate case, with $q=0.5$. The structure experiences significant changes as the $X-Y$ projection portrays a shell like pattern contradicting the perspectives of the remaining projections. $X-Z$ portrays a structure which is visually stream-like while $Y-Z$ introduces ambiguity, displaying an intermediate structure with shell like arcs closer to the core. This is however accompanied by a thin stream-like band extending away from the host centre. We are able to provide a degree of clarification to the obscurity at hand with the employment of the ordered density plot, provided in figure \ref{fig_7:Ordered_density_subhalo_B}. There are three density subclusters which correspond to each distinct substructure of the shape. The first curve in the ordered index ranging up to 50,000 has the characteristic of a stream, the middle subcluster is the remaining core of the progenitor and the last subcluster corresponds to that of a stream tail. As a result the complete ordered density plot can be described as the head, core and tail of a stream.

It should be noted that the tail subcluster retains characteristics of a shell in the density drop off rate. The substructure when relying on visual projections of a stream especially in $X-Y$ proves to be inconclusive, the ordered density plot Fig \ref{fig_7:Ordered_density_subhalo_B}, along with radial figure \ref{fig:rad_plot} and energy angle figure \ref{fig:energy angle plot} confirms it to be a stream. The shell like distribution in certain projections can be attributed to the flattening of the halo in this scenario. This extreme oblate halo along with the intermediate orbit of the progenitor stretches the stream along the flattened axis, inducing a 'fanned out' phenomena resembling features of a shell. As a consequence, the ordered density plot has parts of the tail containing shell like density. However, the overall nature of the structure can be asserted as a stream on the basis of the three subcluster ordered density profile, characteristic of stream like structures. More projections of all of the progenitors are provided in Appendix \ref{Density Projections of Subhalos in multiple projections} along with their isolated ordered density distributions.

\subsection{Orbital Evolution and Dispersion}
\label{Subsection: Orbital Evolution and Dispersion}
The morphological consequences of flattening are primarily driven by the deviations in orbital paths for different values of $q$. Perhaps most crucially, the flattening of the halo promotes higher eccentric precession, exposing infalling satellites to greater tidal forces.

\begin{figure*}[ht!]
\centering
\hspace{-12mm}
\begin{minipage}[c]{0.95\textwidth}
\vspace{-5mm}
\centering
\setlength{\tabcolsep}{-1pt}
\renewcommand{\arraystretch}{0}
\begin{tabular}{ccccc}

    \raisebox{-2pt}[\height][0pt]{\includegraphics[width=.202\textwidth, height=.202\textwidth]{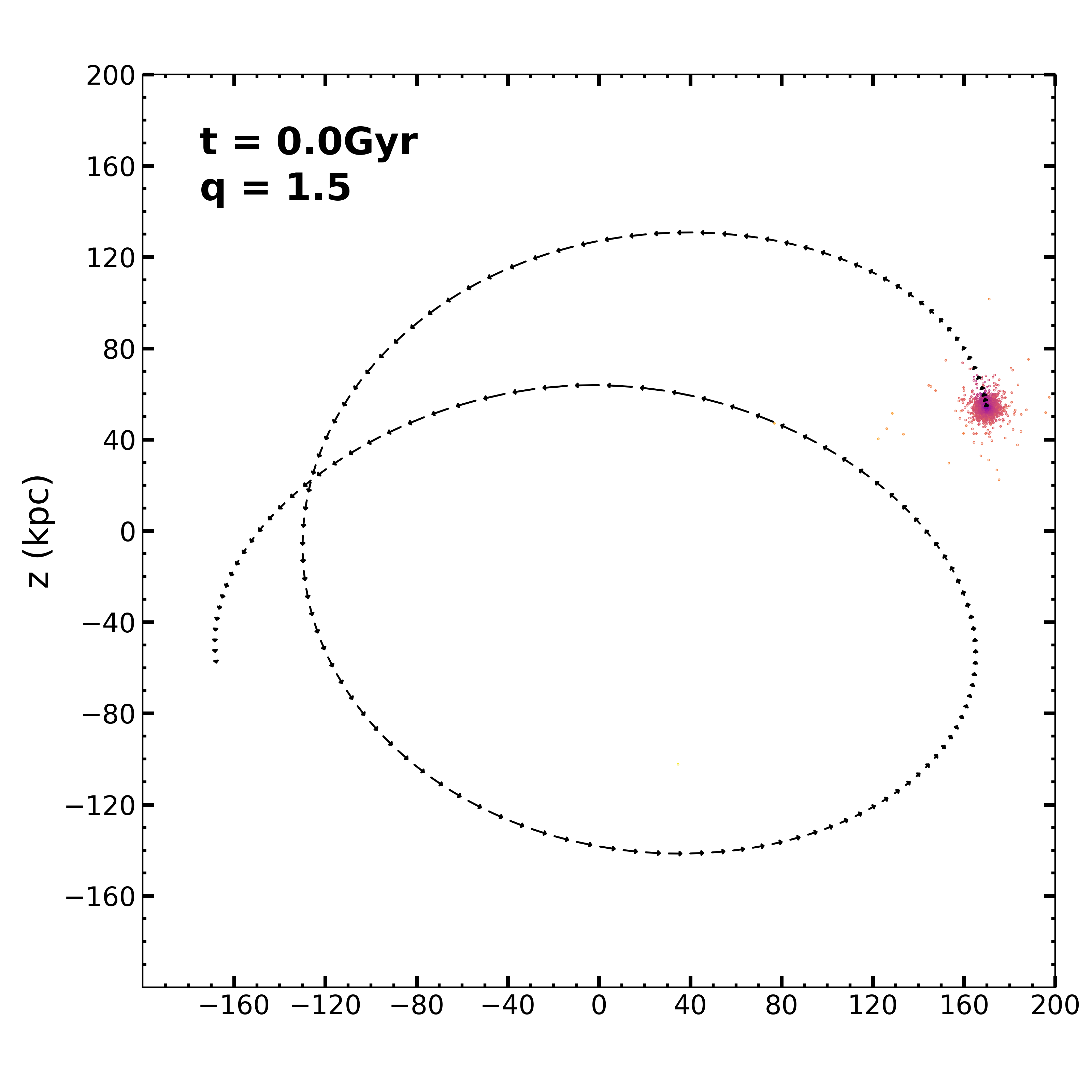}} &
    \includegraphics[width=.1940\textwidth, height=.1940\textwidth]{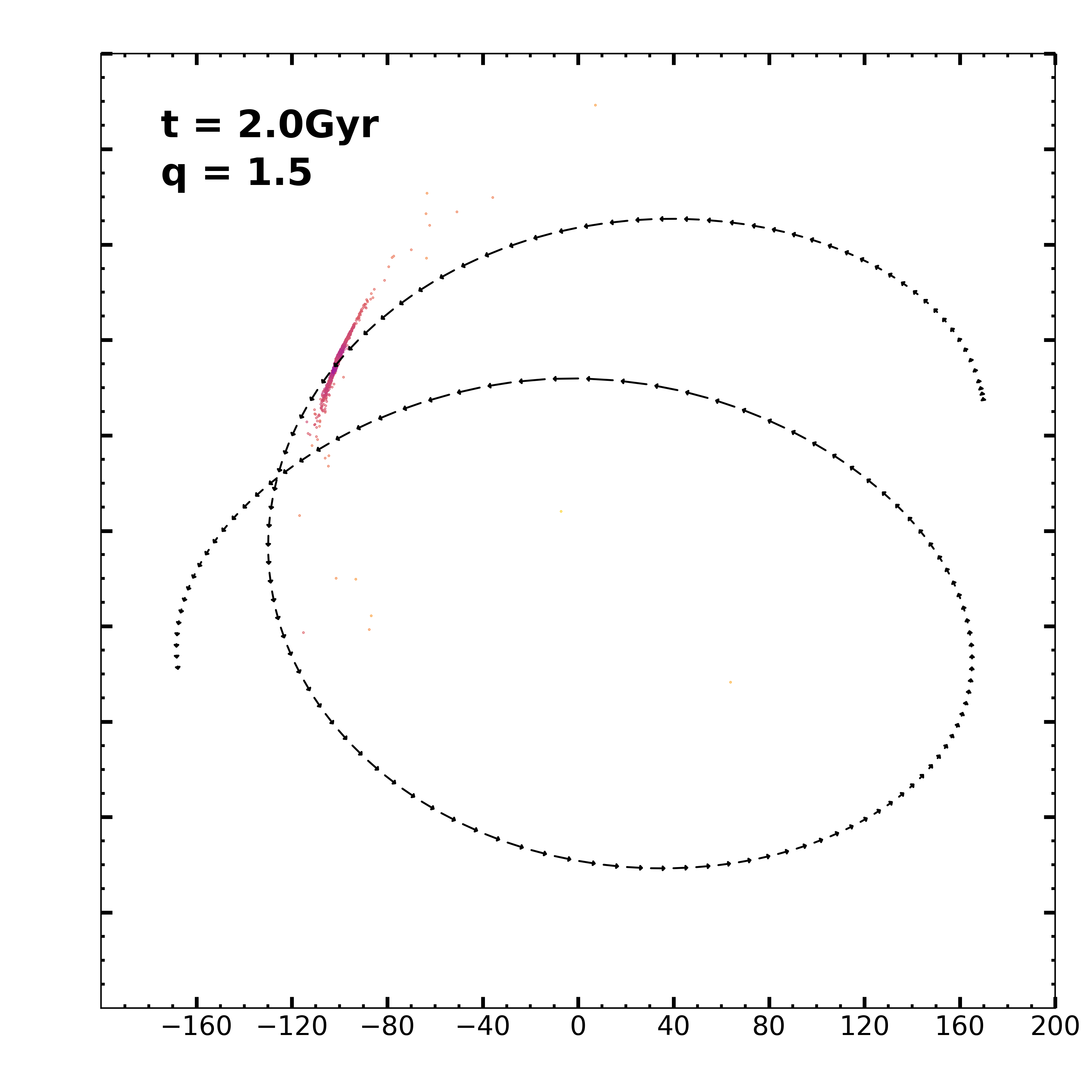} &
    \includegraphics[width=.1940\textwidth, height=.1940\textwidth]{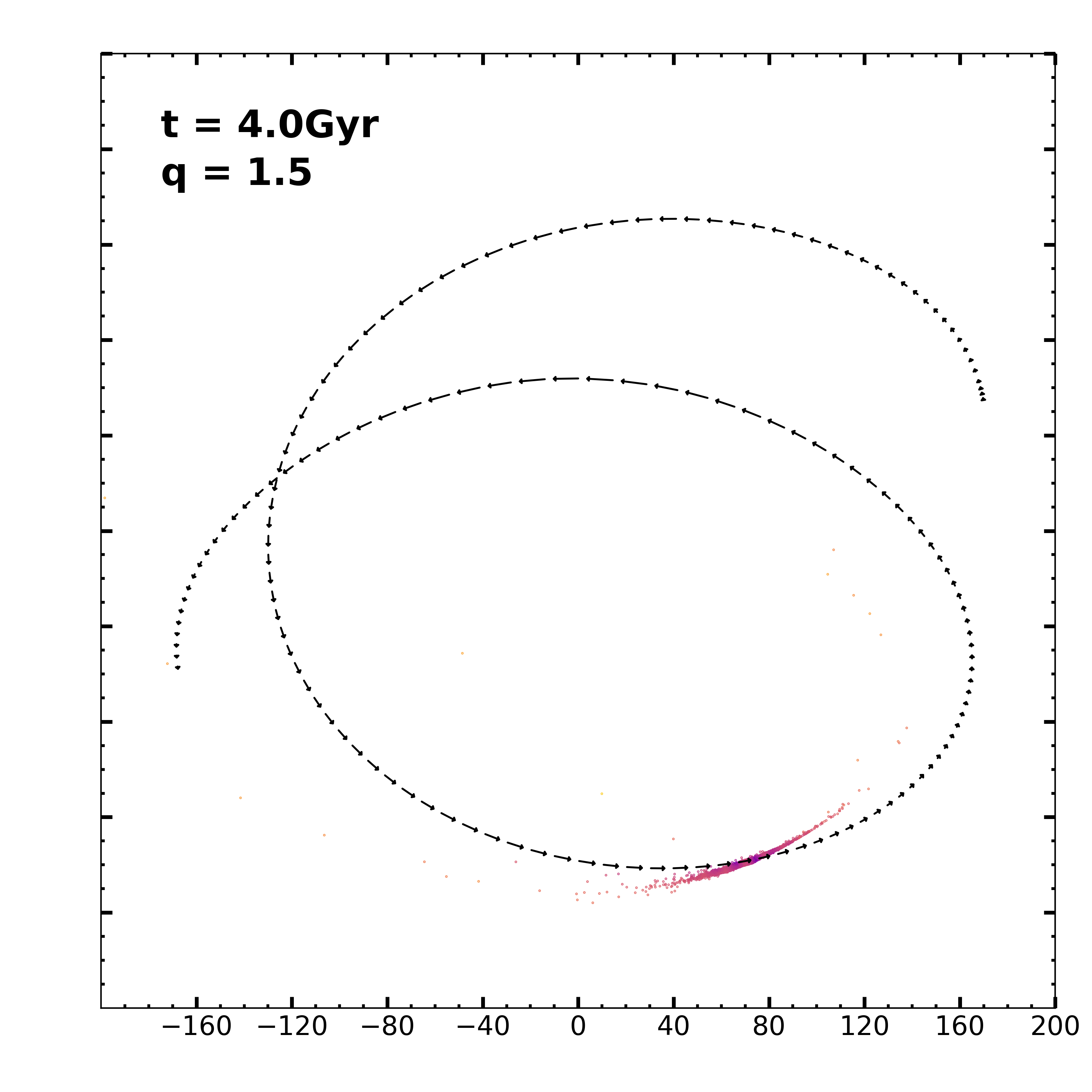} &
    \includegraphics[width=.1940\textwidth, height=.1940\textwidth]{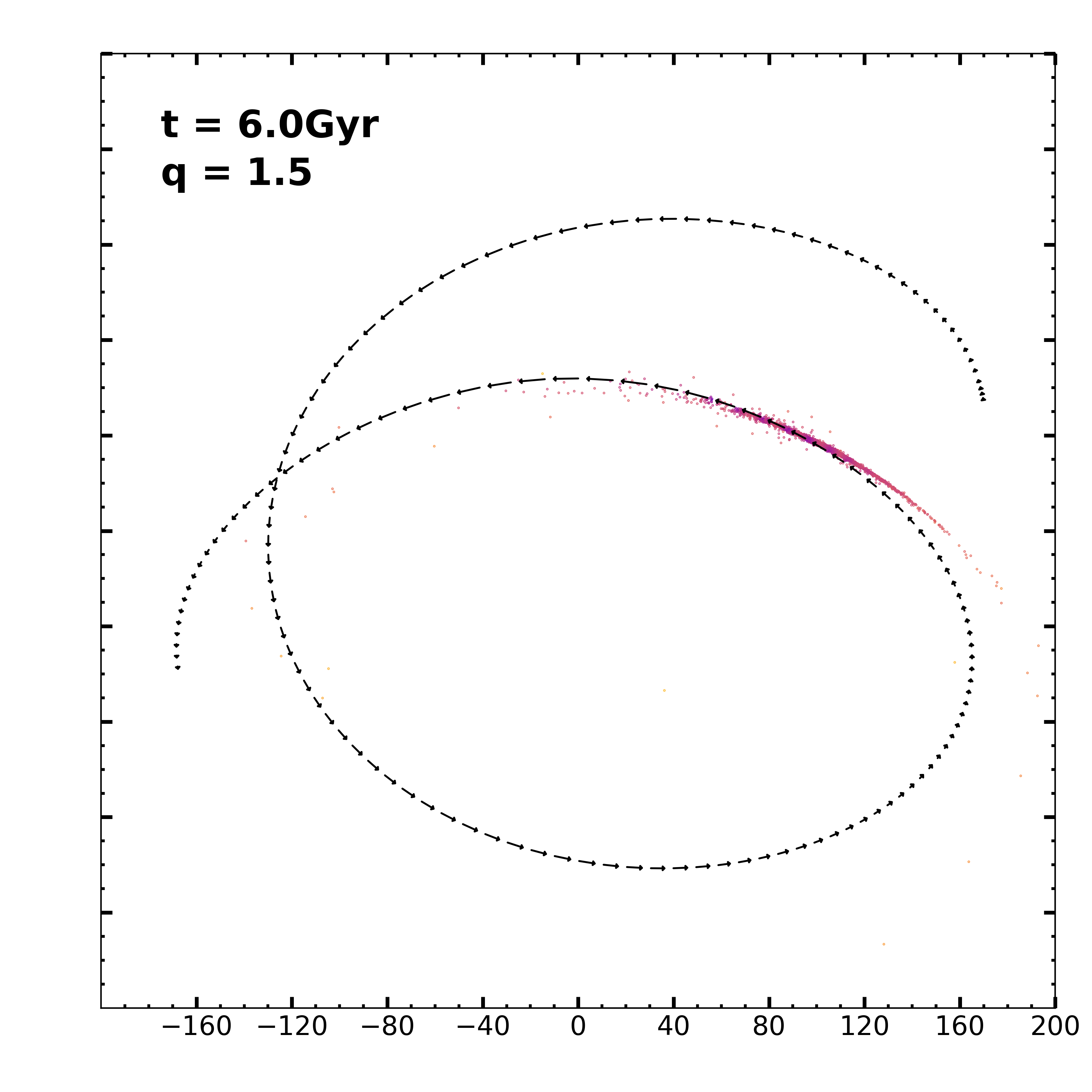} &
    \includegraphics[width=.1940\textwidth, height=.1940\textwidth]{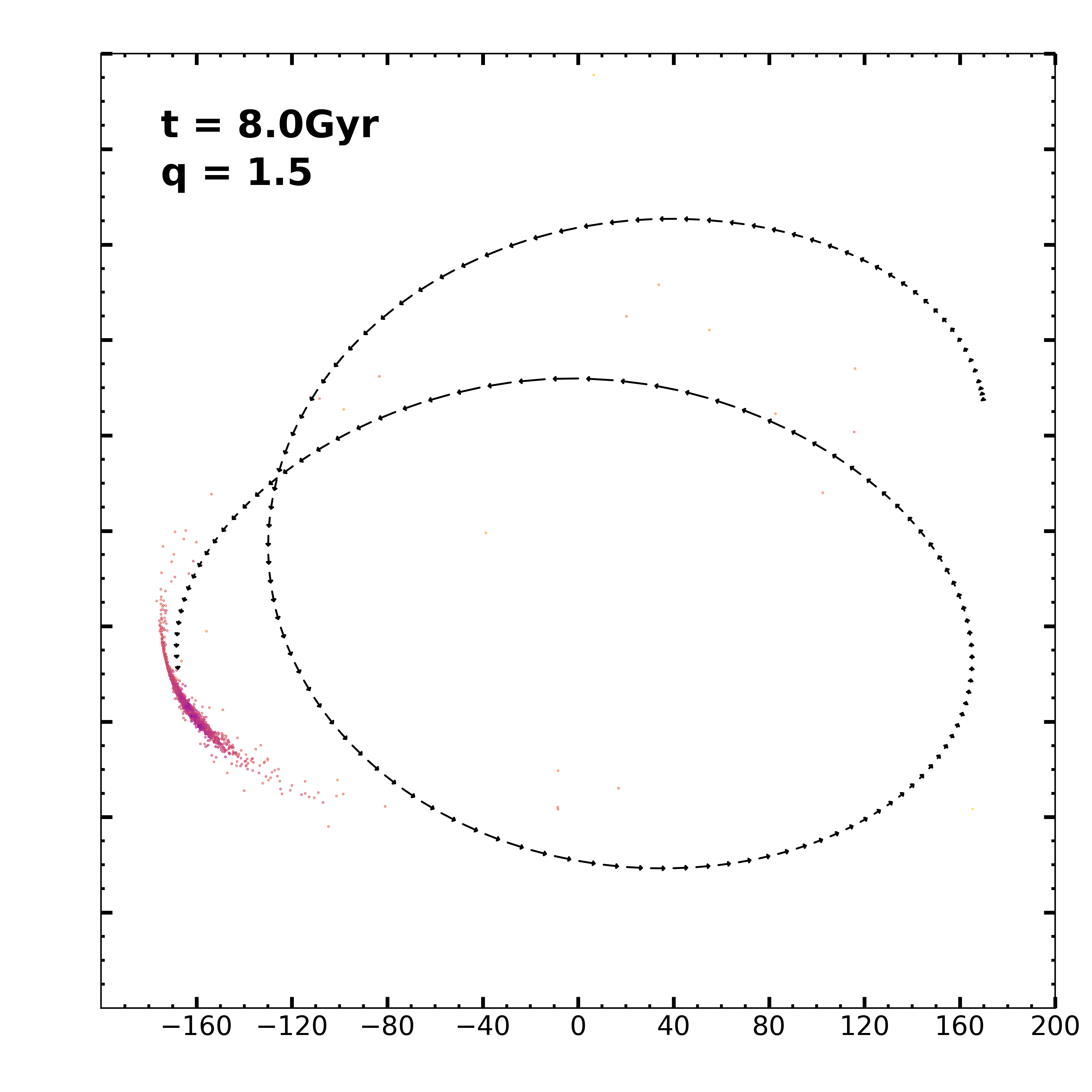} 
    \\[-5pt]

    \raisebox{-2pt}[\height][0pt]{\includegraphics[width=.2032\textwidth, height=.2032\textwidth]{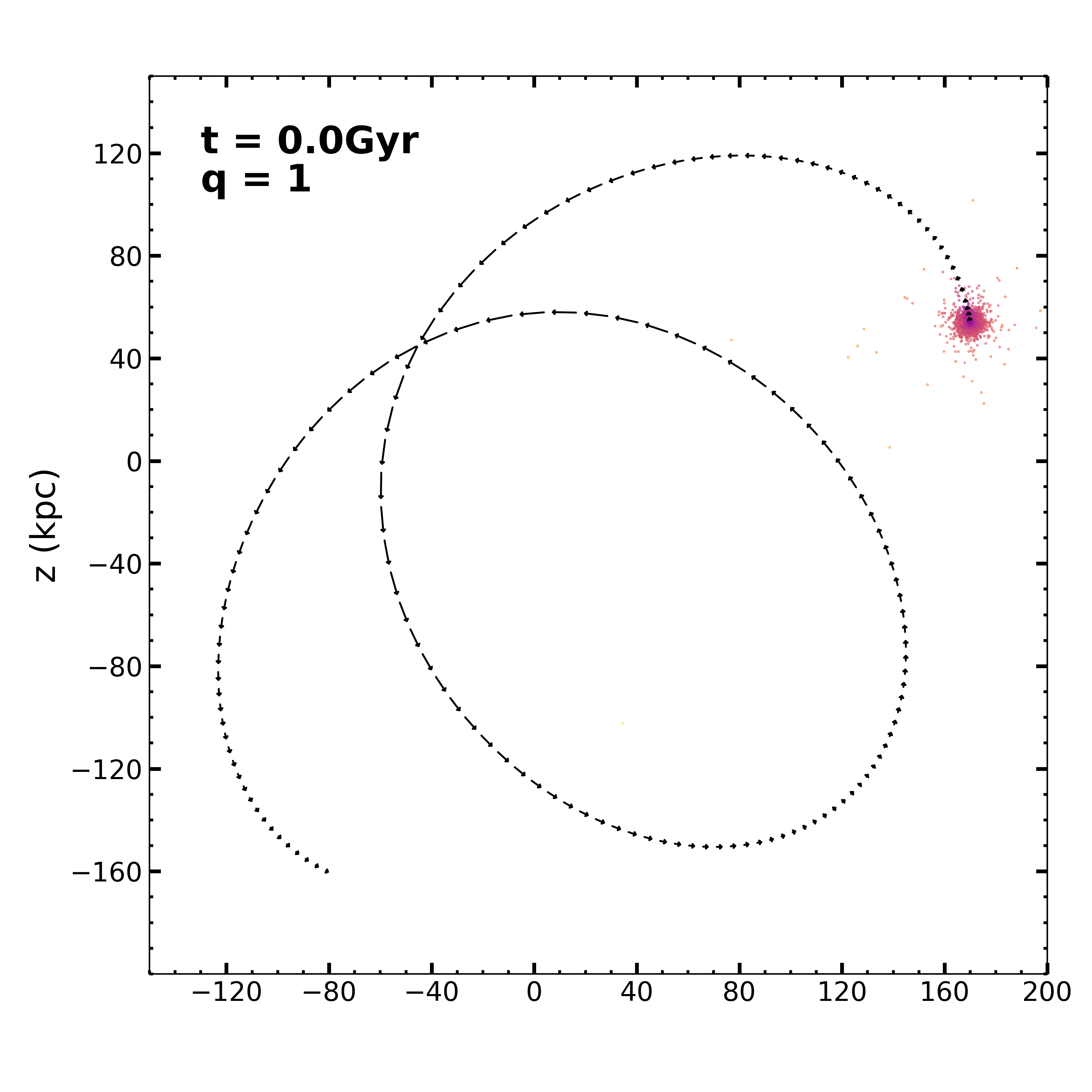}} &
    \raisebox{0pt}[\height][0pt]{\includegraphics[width=.196\textwidth, height=.196\textwidth]{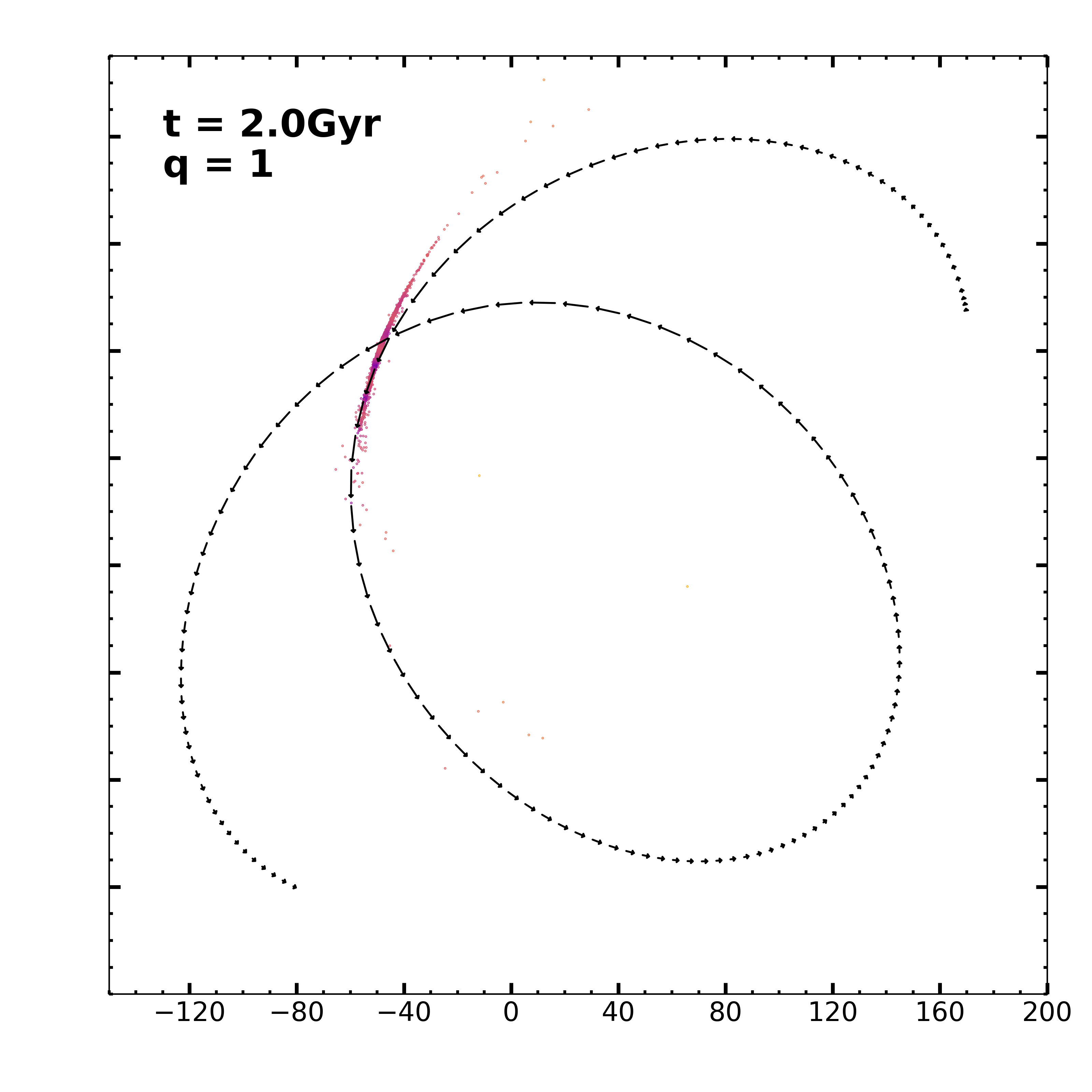}} &
    \raisebox{0pt}[\height][0pt]{\includegraphics[width=.196\textwidth, height=.196\textwidth]{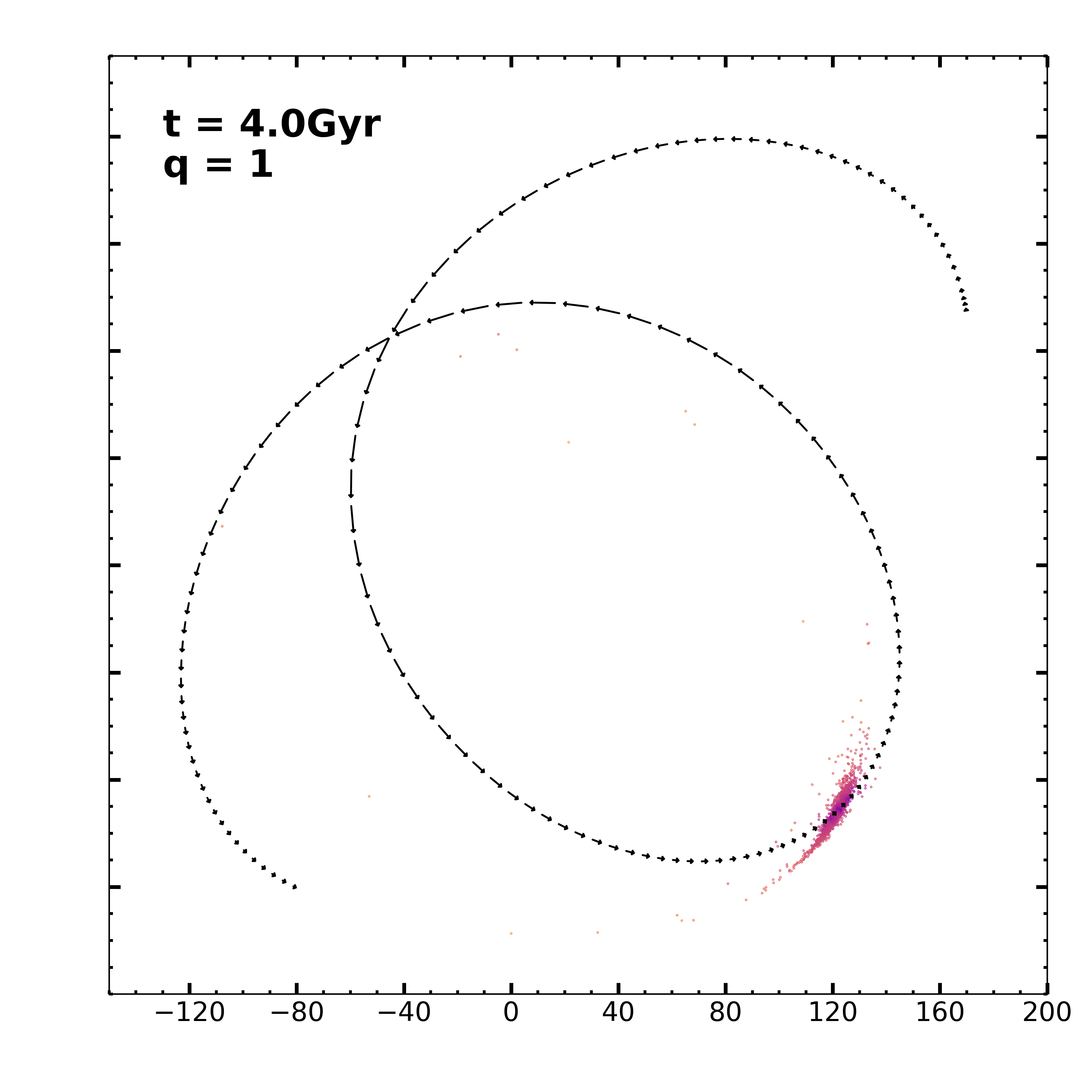}} &
    \raisebox{0pt}[\height][0pt]{\includegraphics[width=.196\textwidth, height=.196\textwidth]{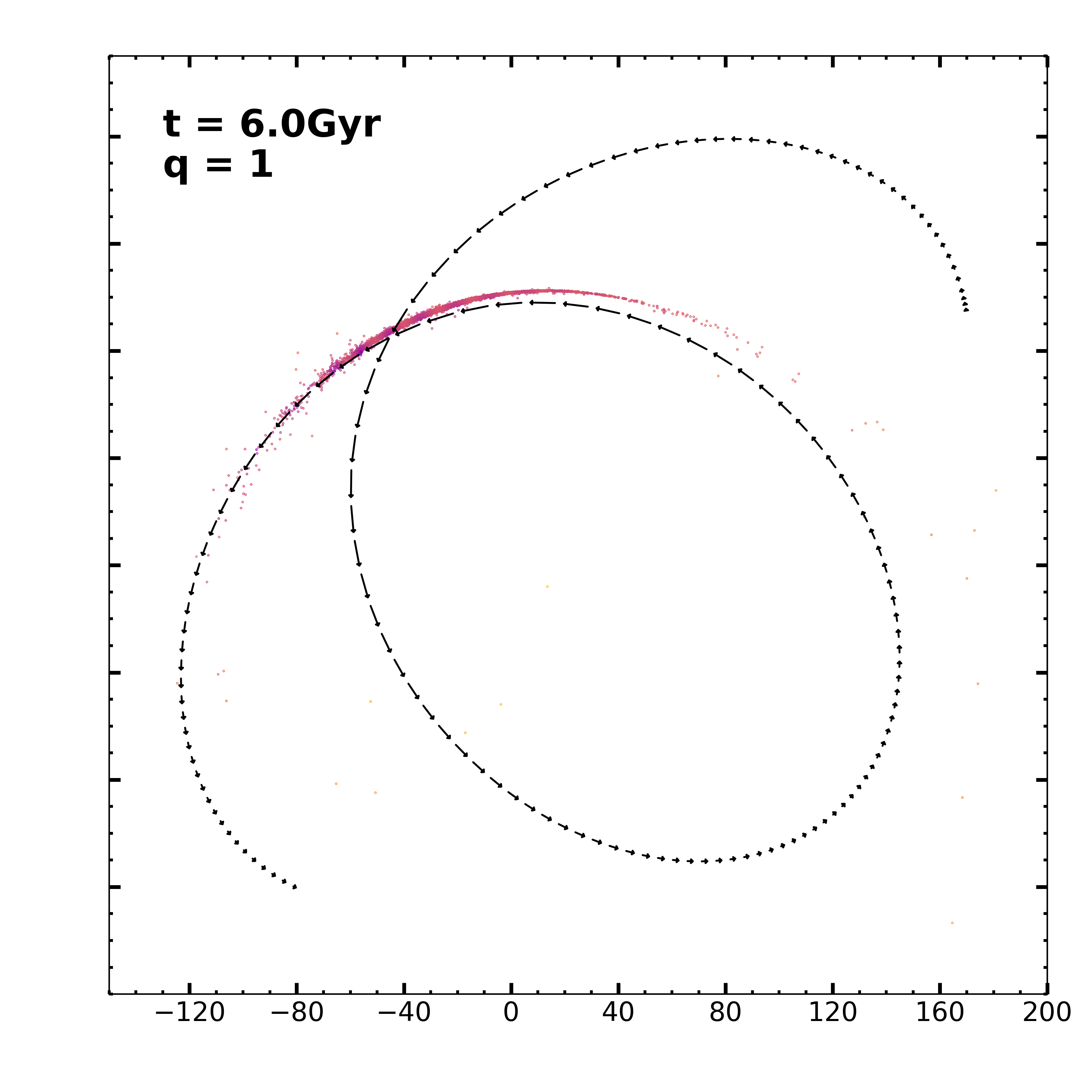}} &
    \raisebox{0pt}[\height][0pt]{\includegraphics[width=.196\textwidth, height=.196\textwidth]{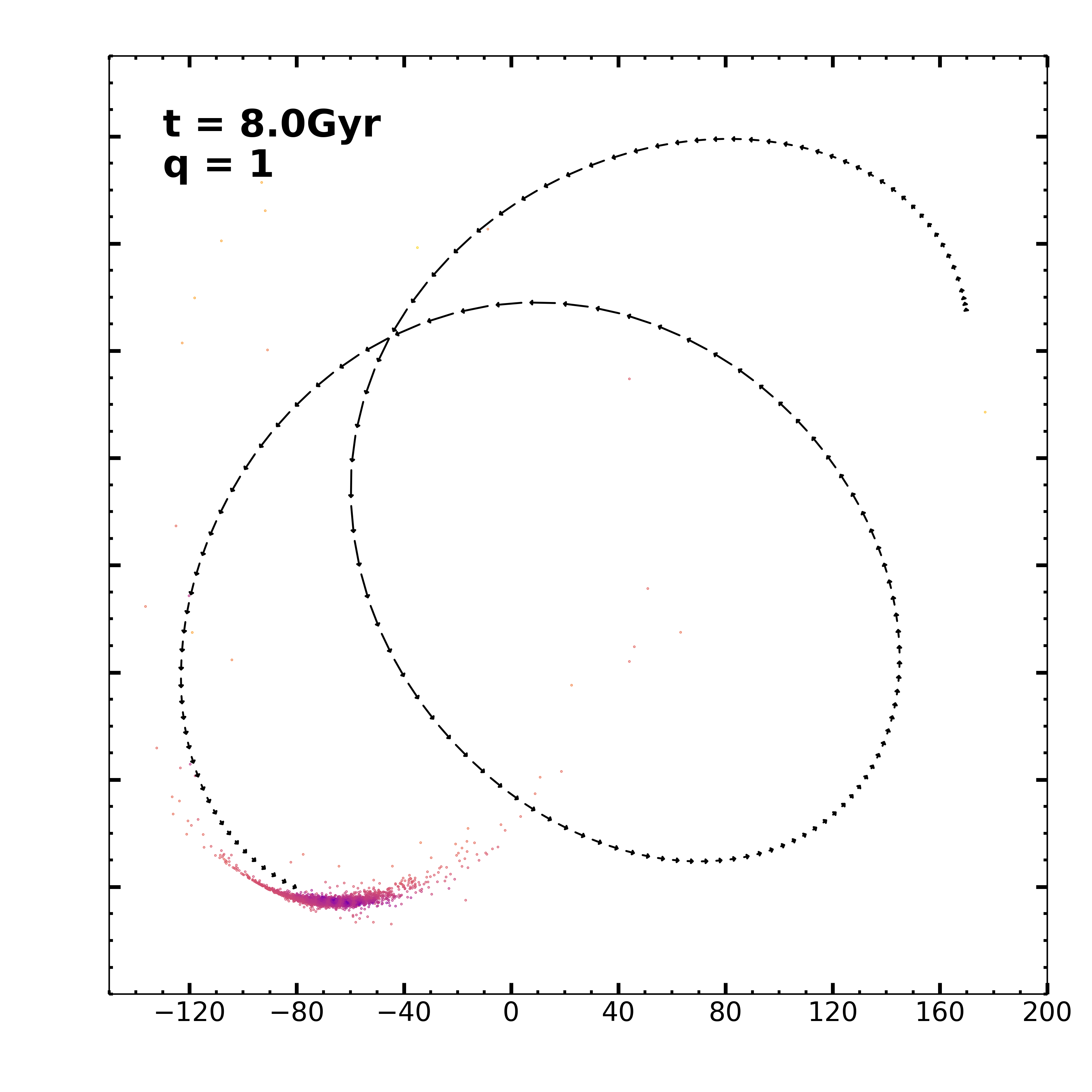}} \\[-5pt]
    
    \raisebox{-2pt}[\height][0pt]{\includegraphics[width=.2032\textwidth, height=.2032\textwidth]{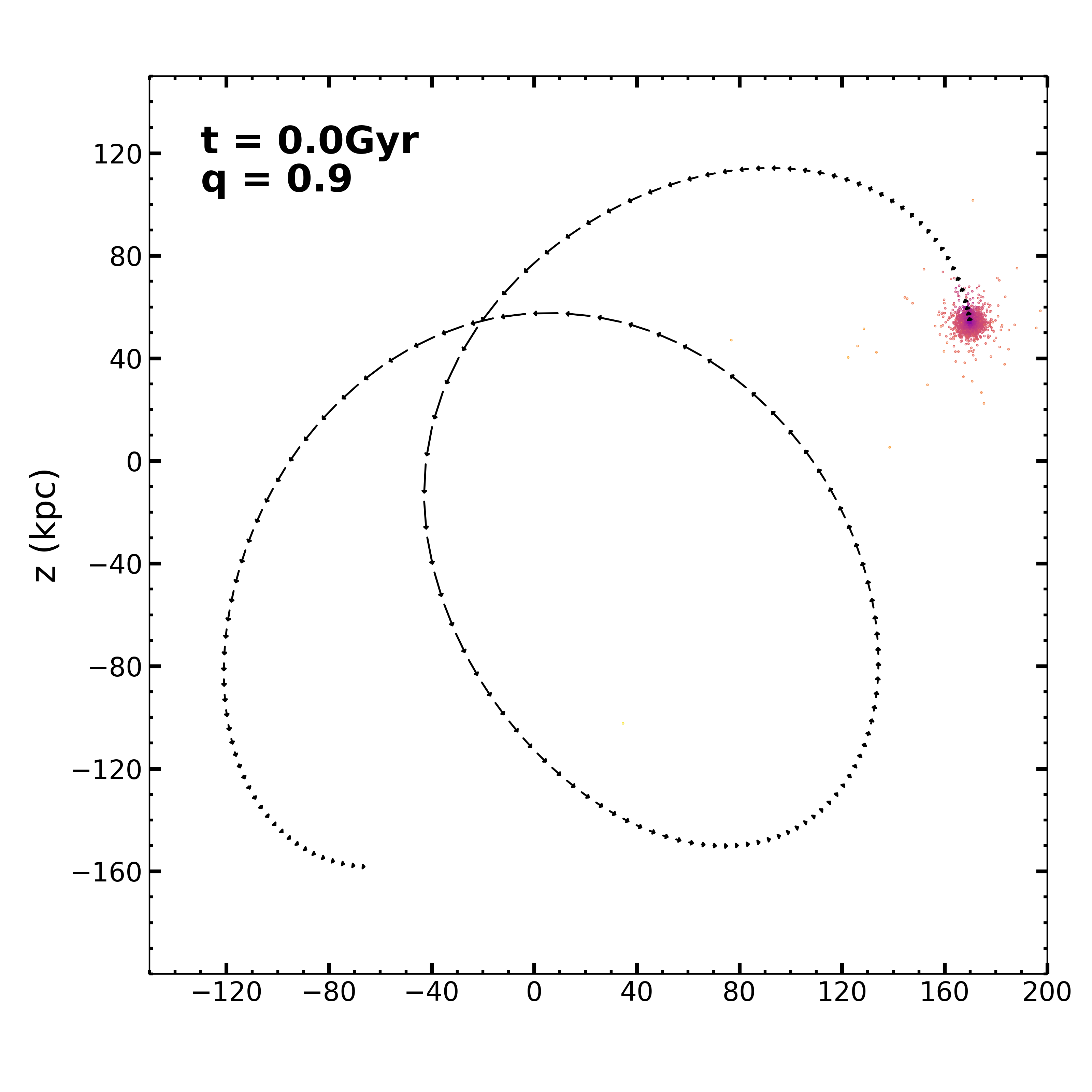}} &
    \includegraphics[width=.196\textwidth, height=.196\textwidth]{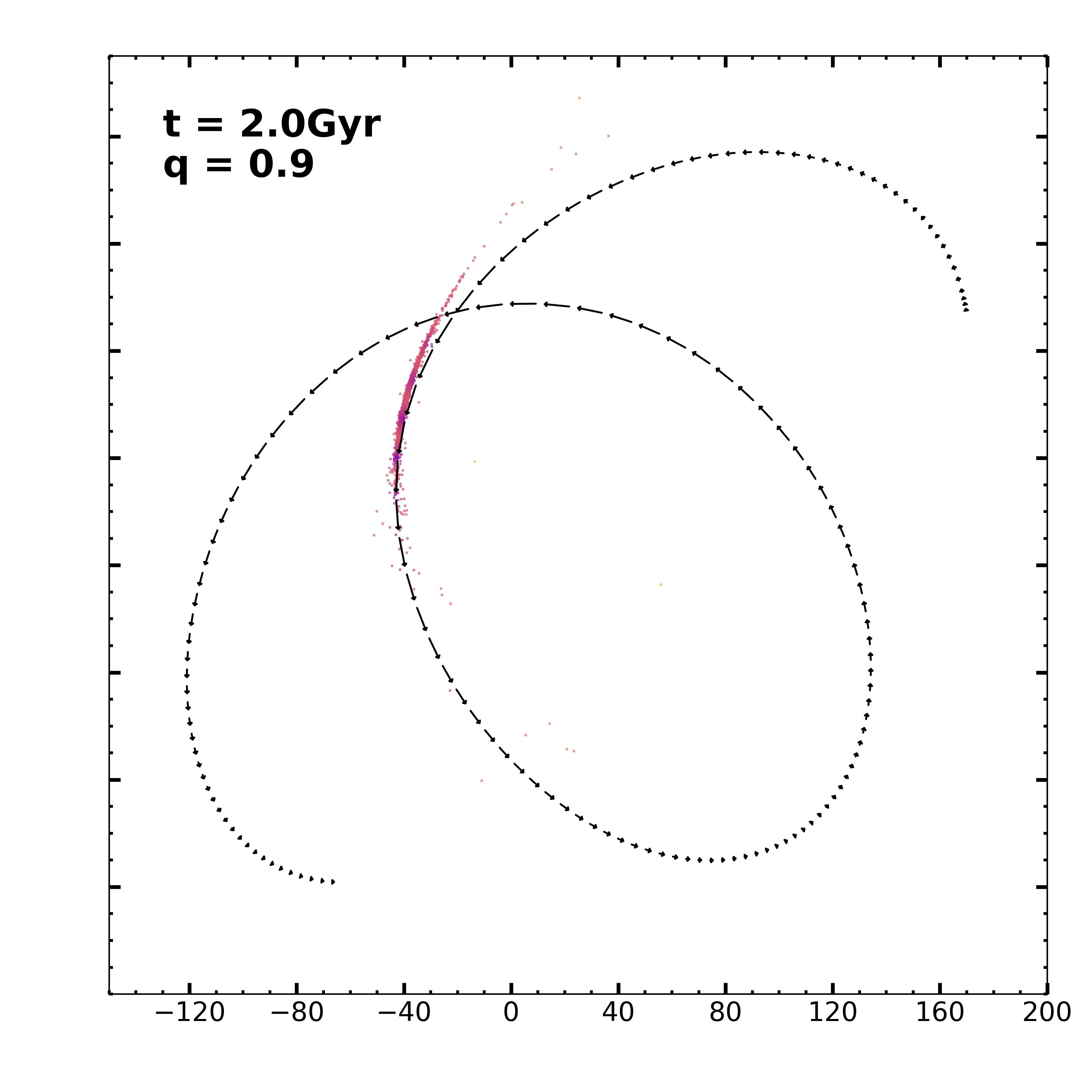} &
    \includegraphics[width=.196\textwidth, height=.196\textwidth]{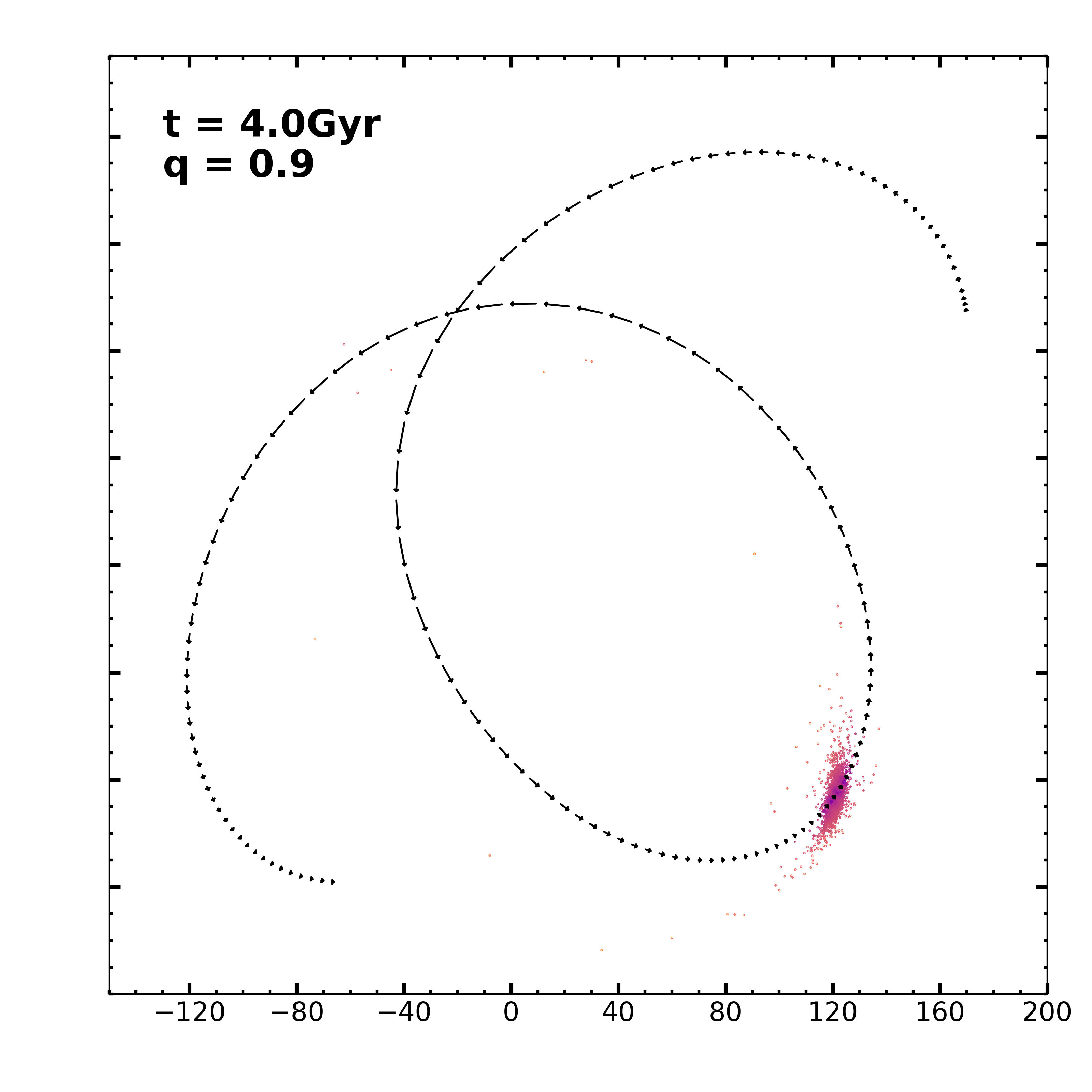} &
    \includegraphics[width=.196\textwidth, height=.196\textwidth]{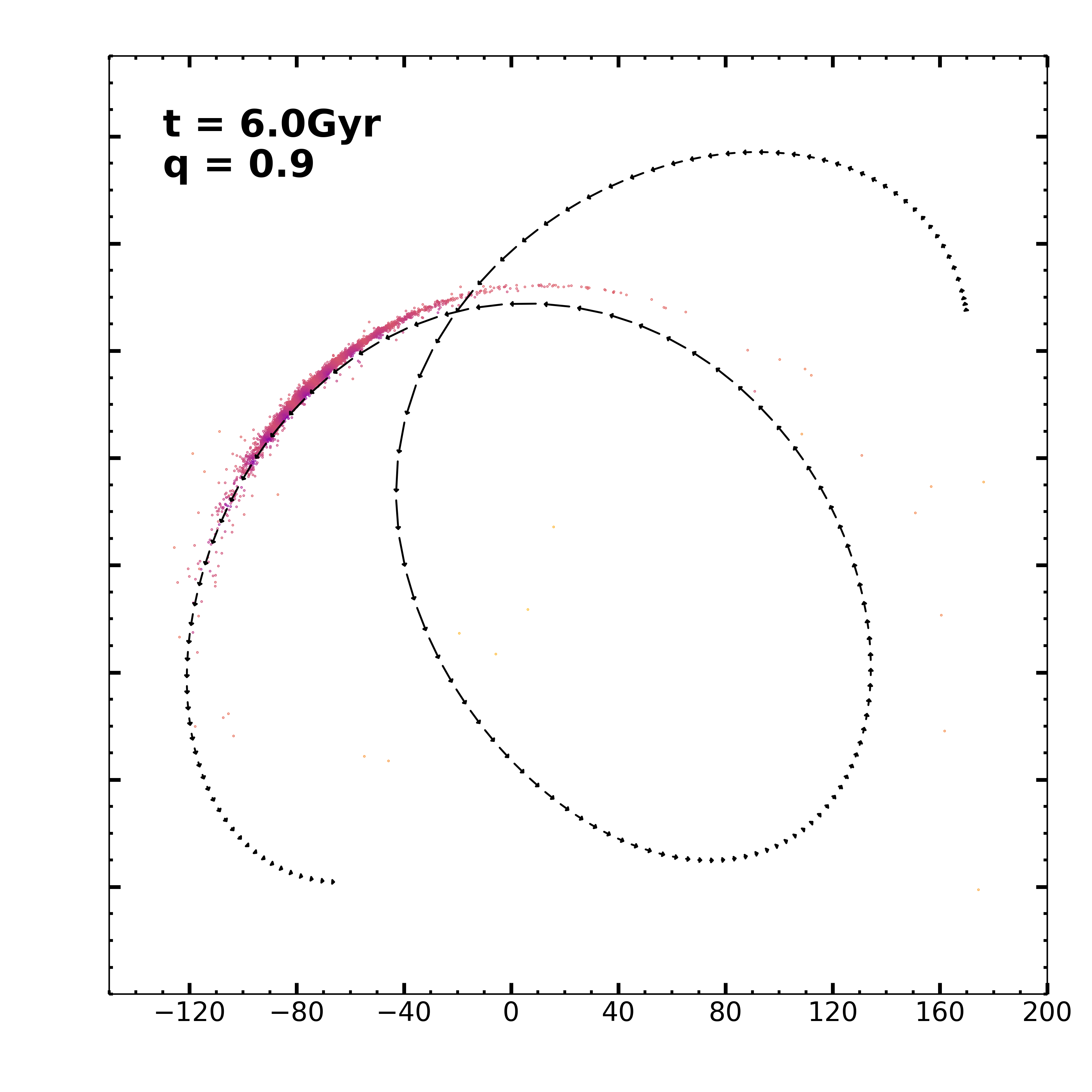} &
    \includegraphics[width=.196\textwidth, height=.196\textwidth]{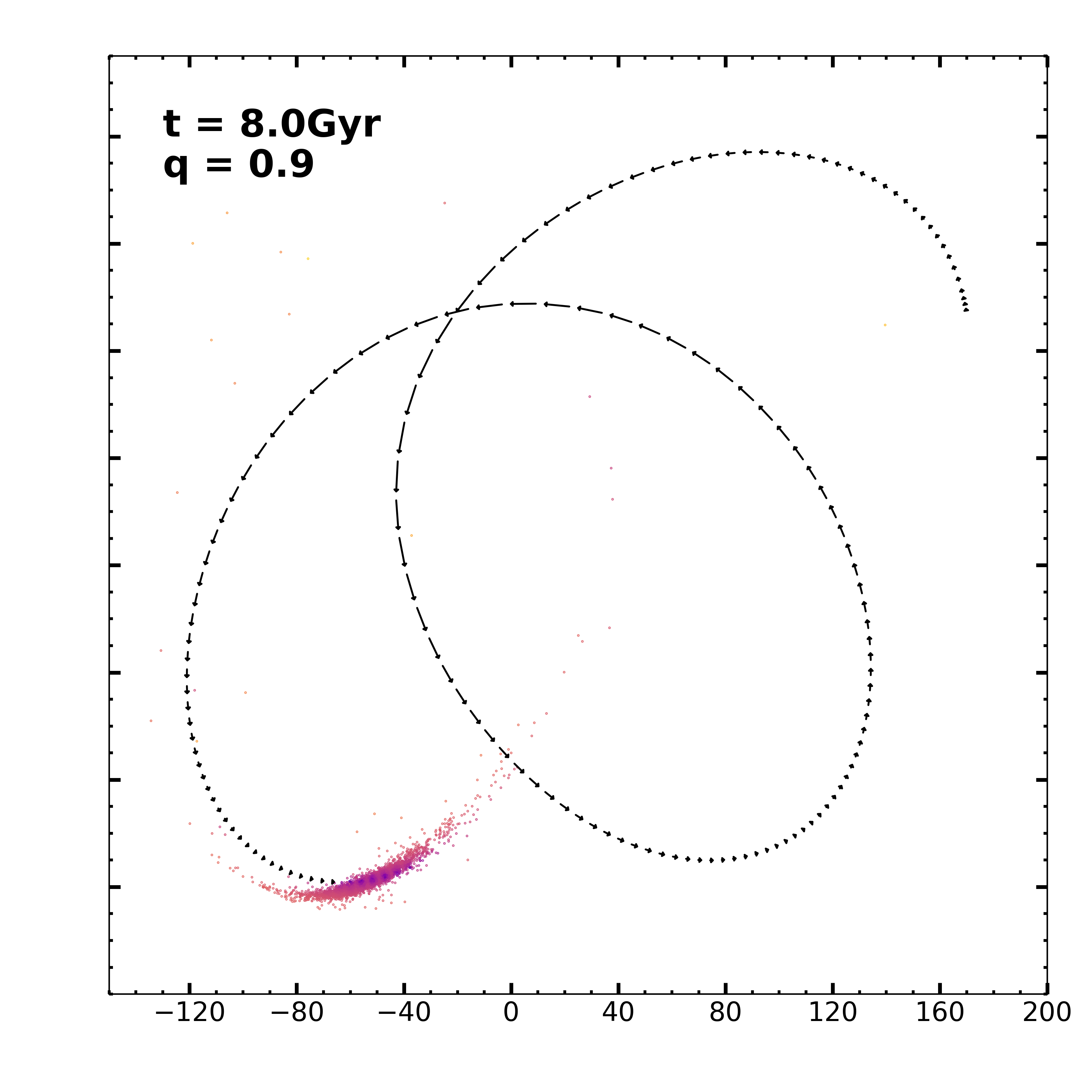} \\[-5pt]
    
    \raisebox{-2pt}[\height][0pt]{\includegraphics[width=.2032\textwidth, height=.2032\textwidth]{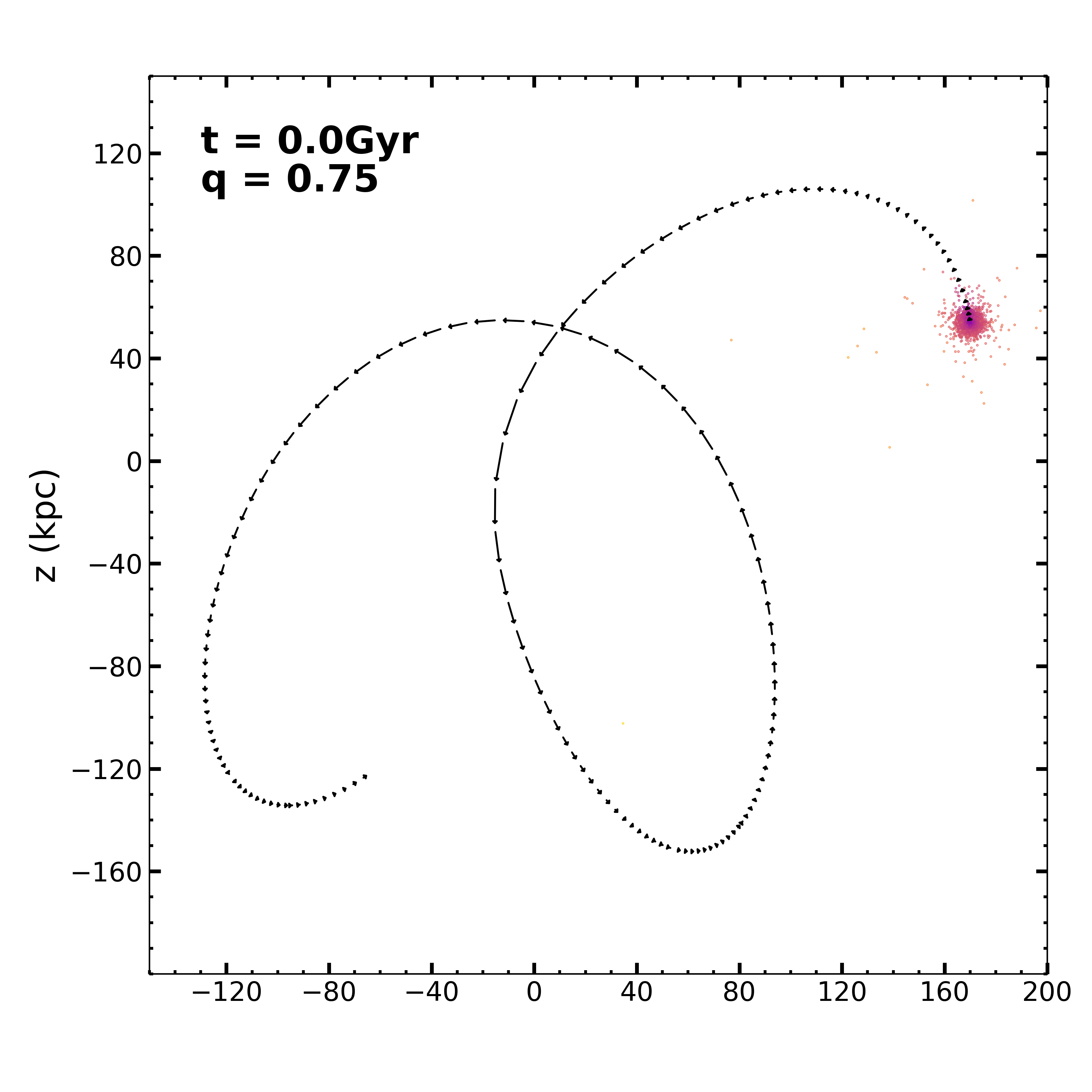}} &
    \includegraphics[width=.196\textwidth, height=.196\textwidth]{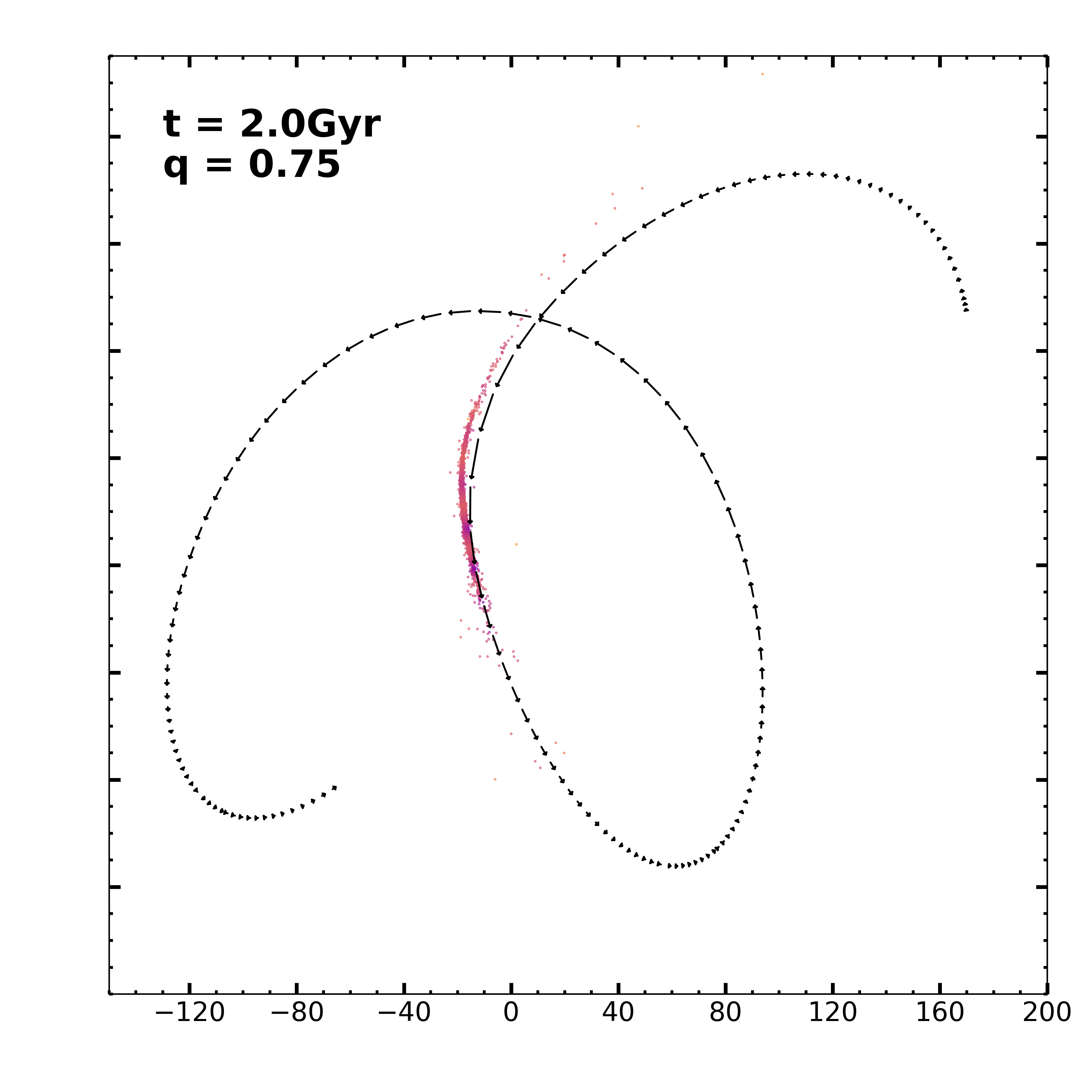} &
    \includegraphics[width=.196\textwidth, height=.196\textwidth]{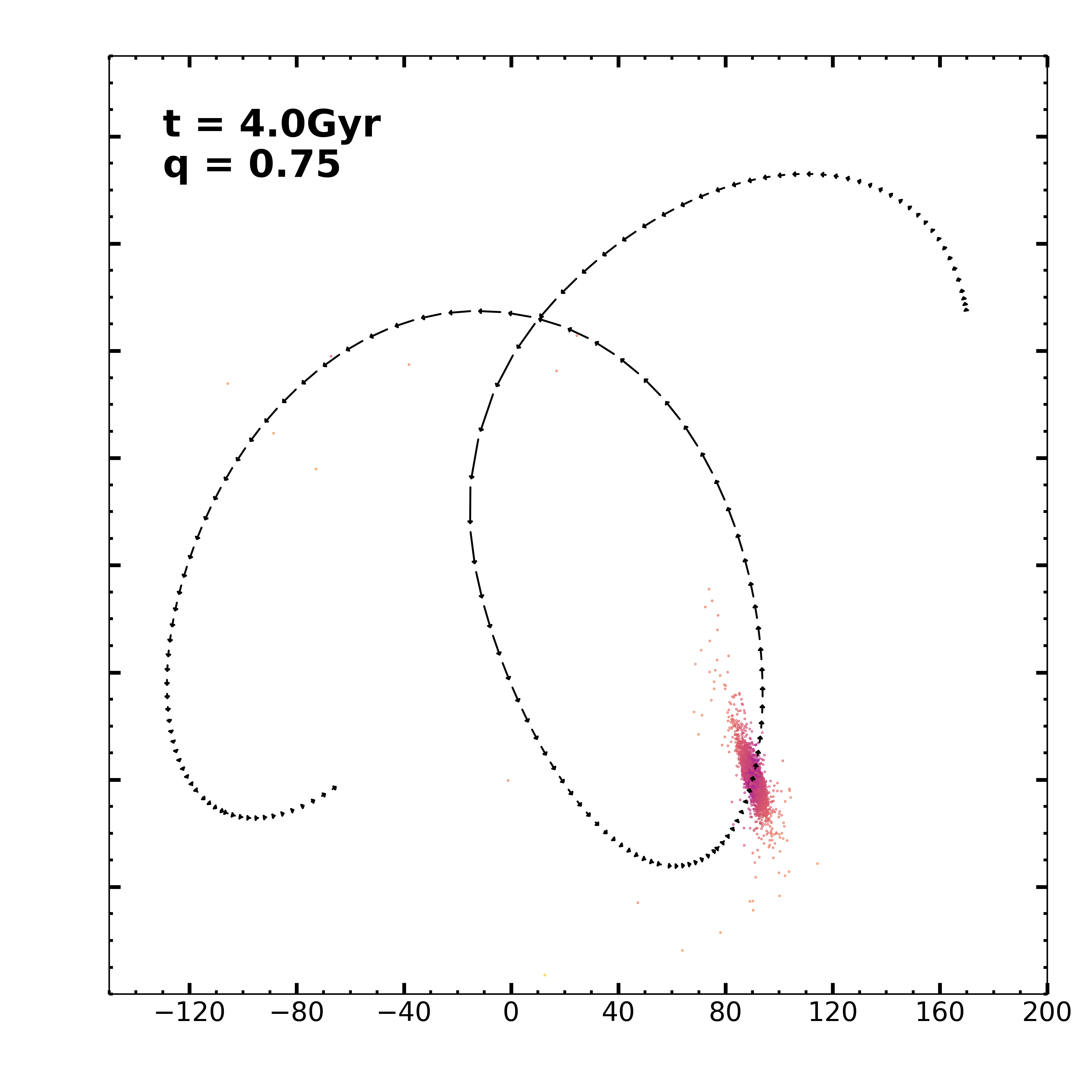} &
    \includegraphics[width=.196\textwidth, height=.196\textwidth]{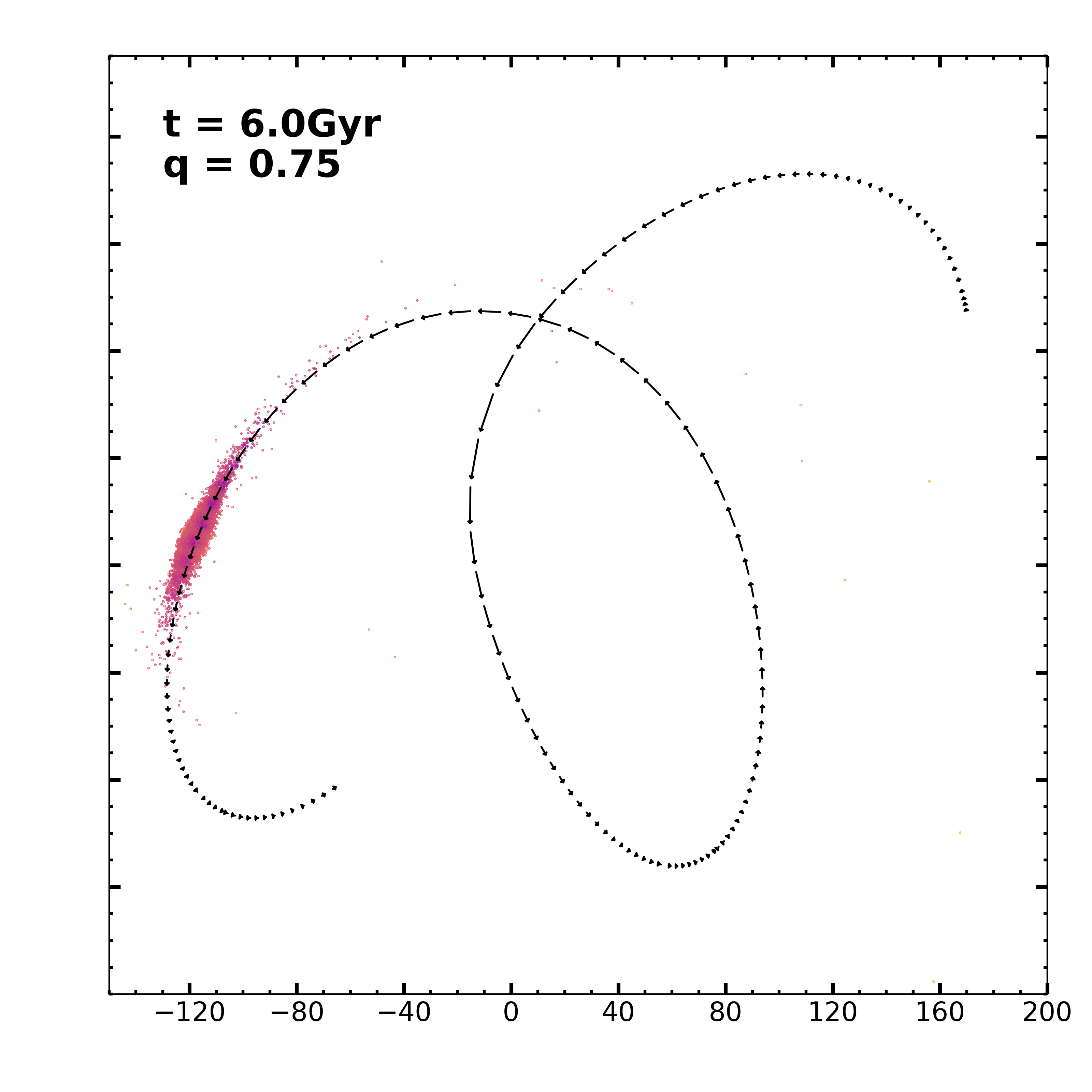} &
    \includegraphics[width=.196\textwidth, height=.196\textwidth]{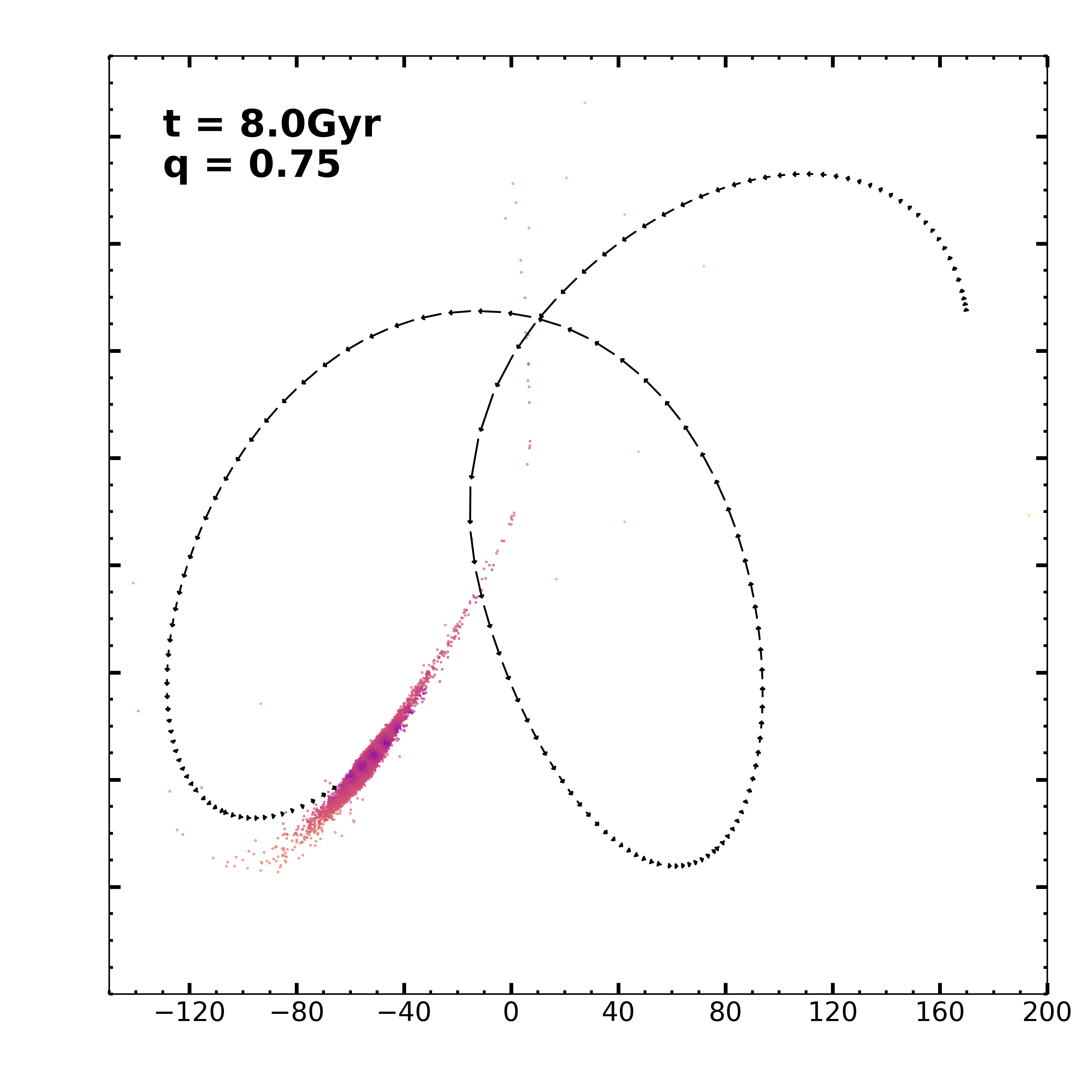} \\[-5pt]
    
    \raisebox{-2pt}[\height][0pt]{\includegraphics[width=.2032\textwidth, height=.2032\textwidth]{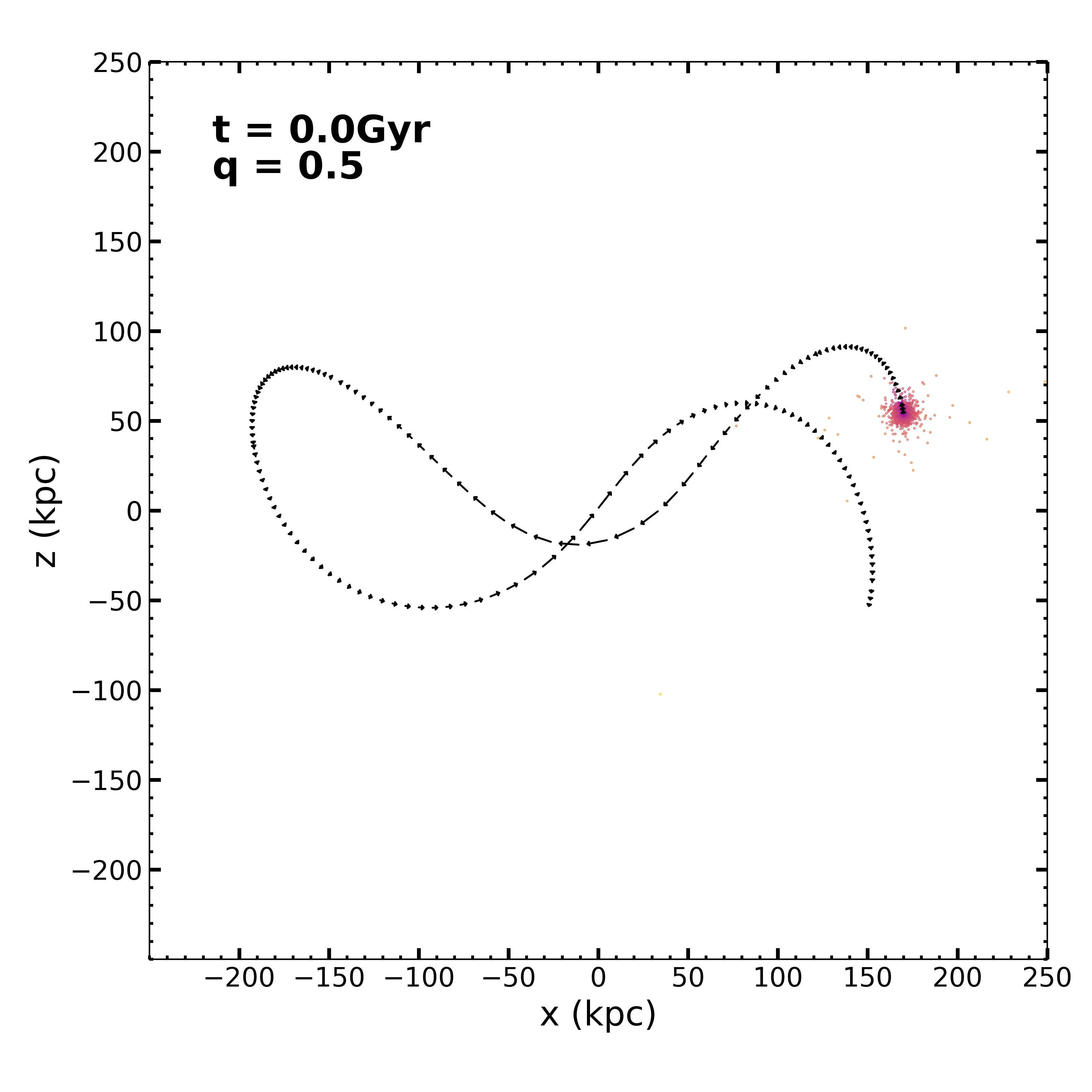}} &
    \includegraphics[width=.196\textwidth, height=.196\textwidth]{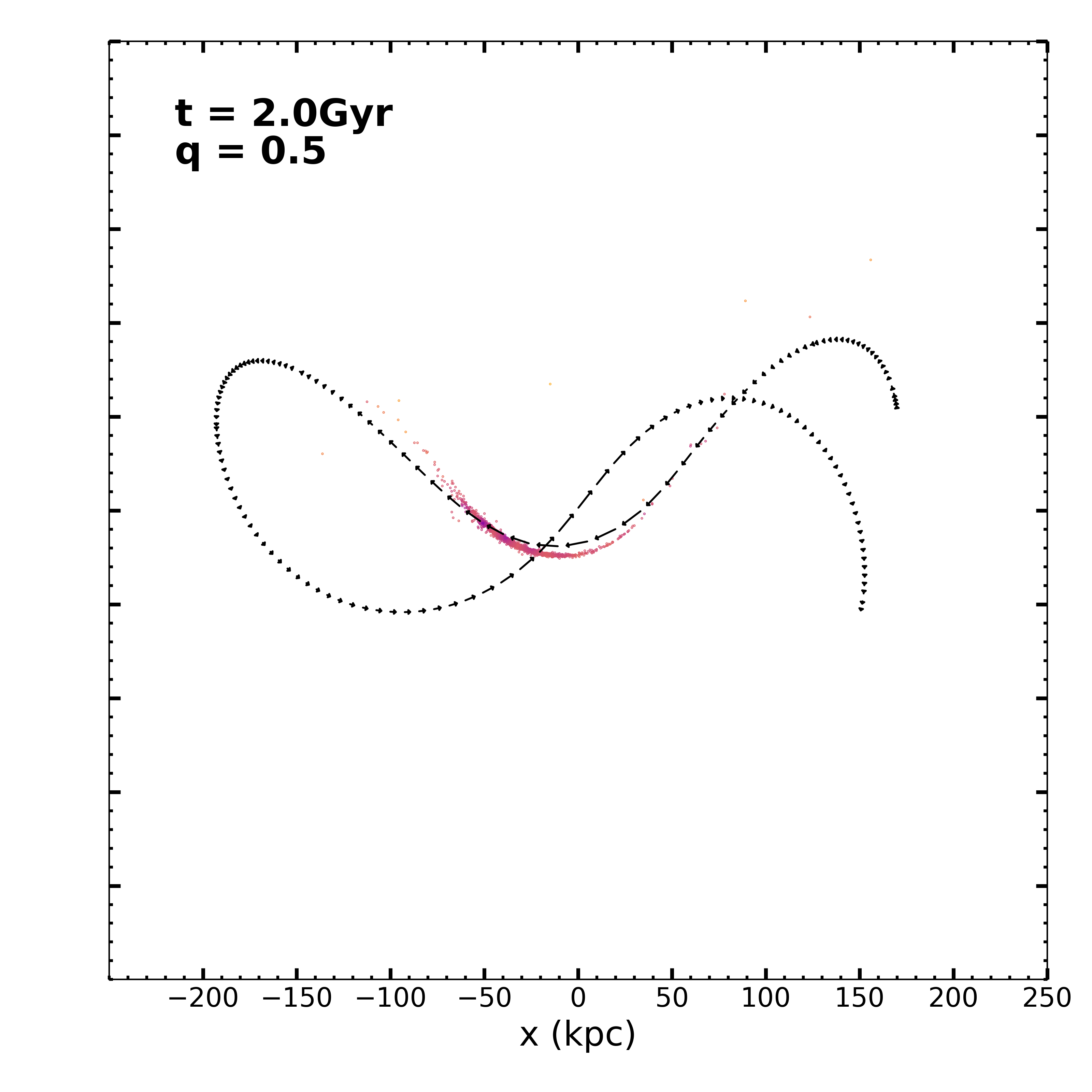} &
    \includegraphics[width=.196\textwidth, height=.196\textwidth]{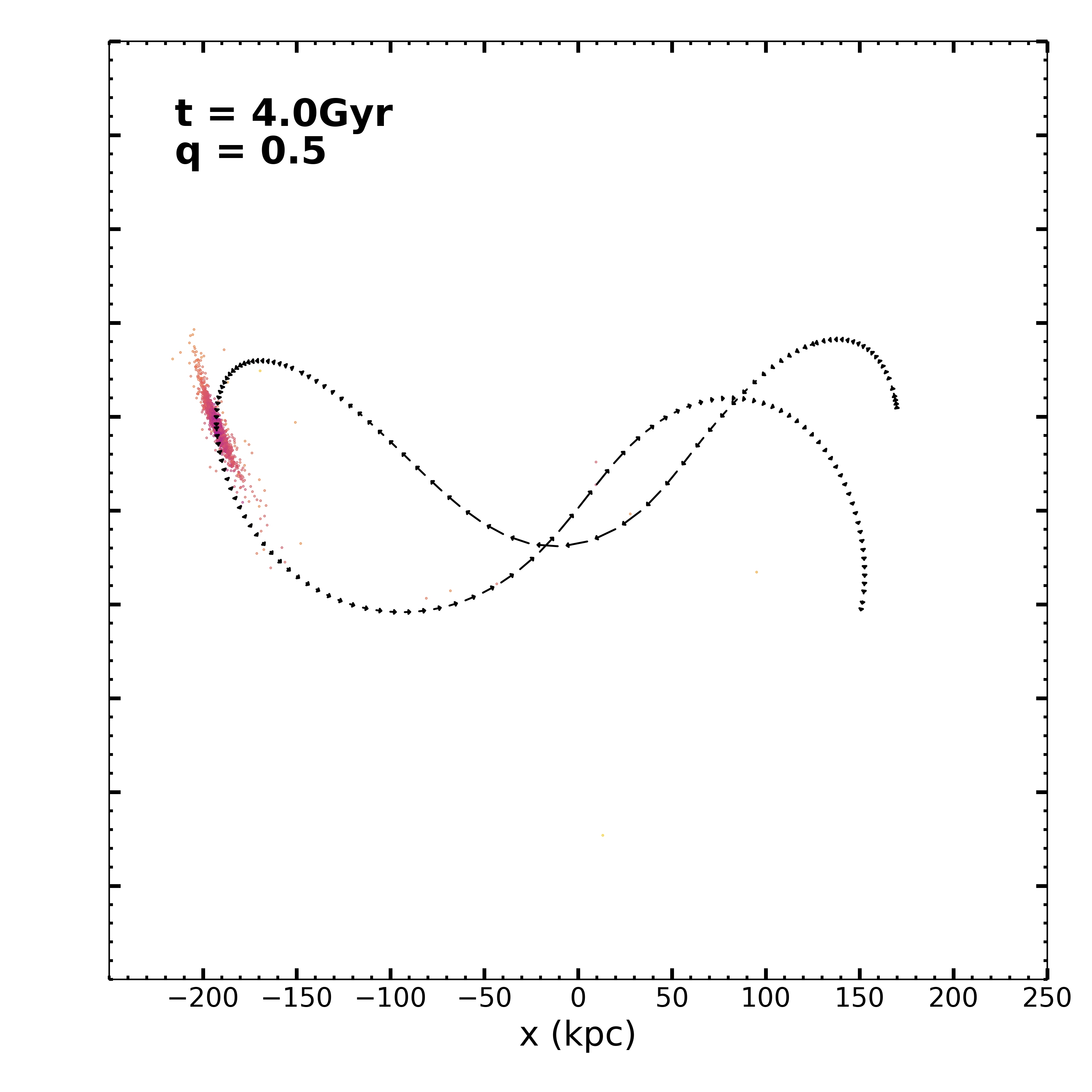} &
    \includegraphics[width=.196\textwidth, height=.196\textwidth]{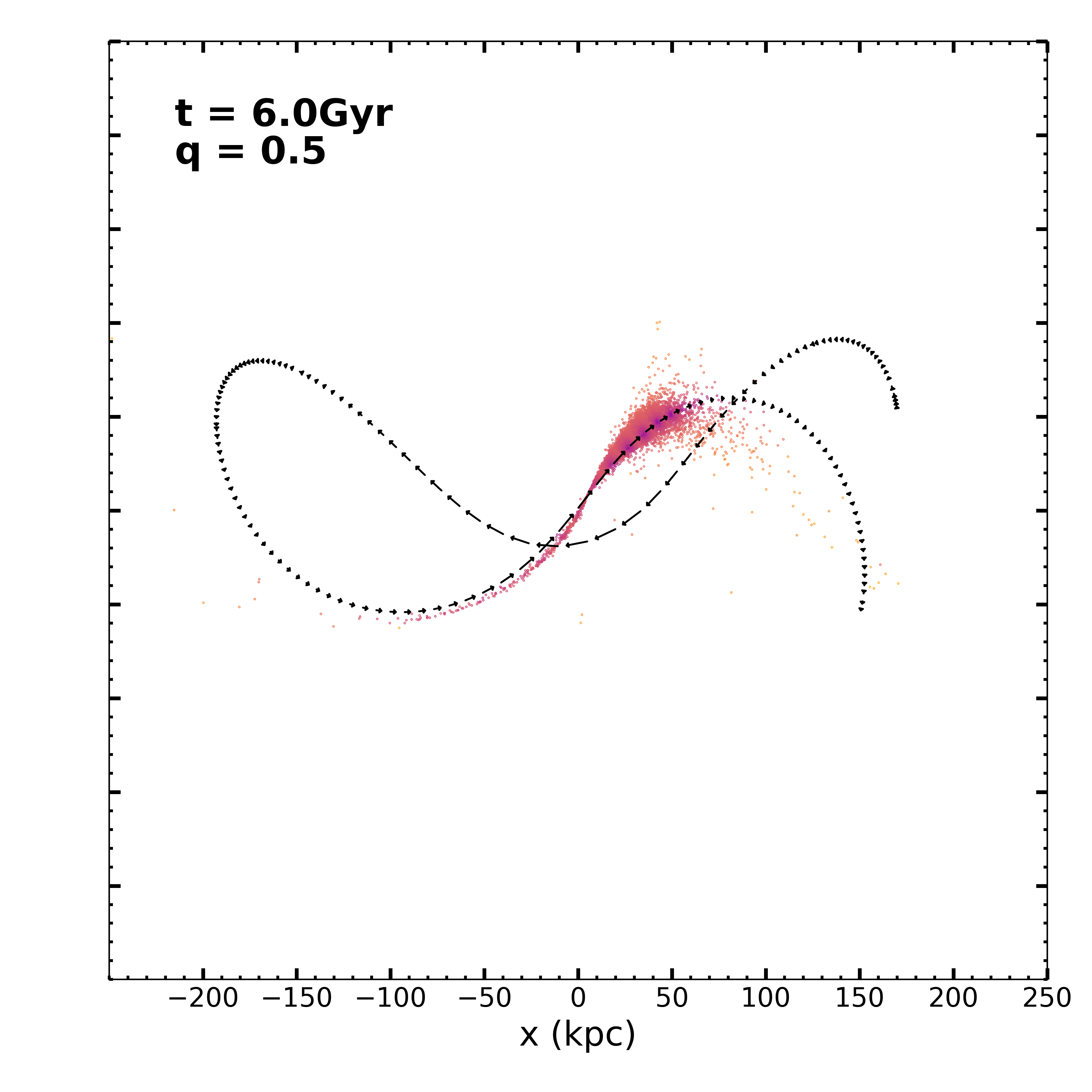} &
    \includegraphics[width=.196\textwidth, height=.196\textwidth]{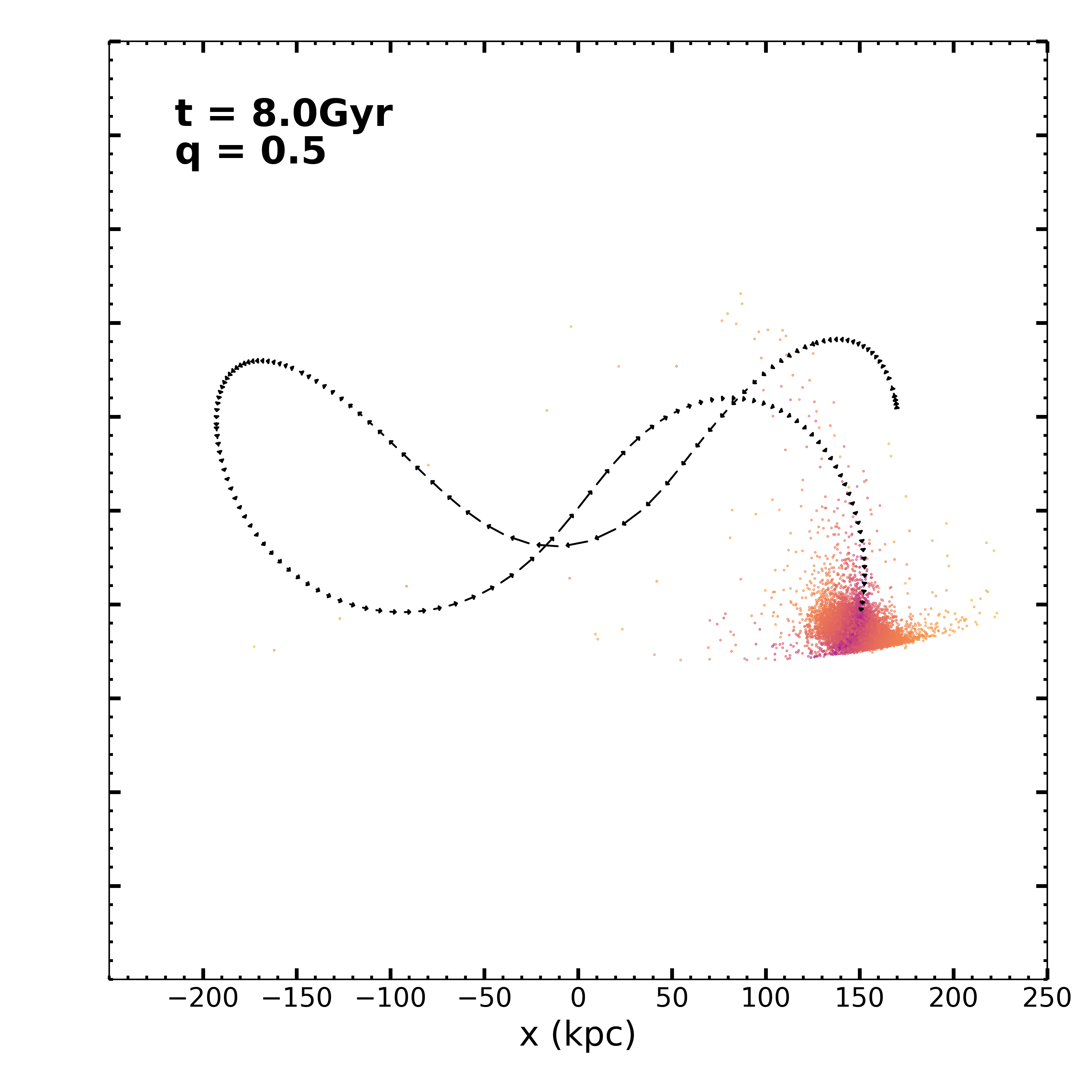}
\end{tabular}
\end{minipage}
\hspace{-3mm}
\begin{minipage}[c]{0.02\textwidth}
\centering
\vspace{-5mm}
\includegraphics[height=0.675\textheight]{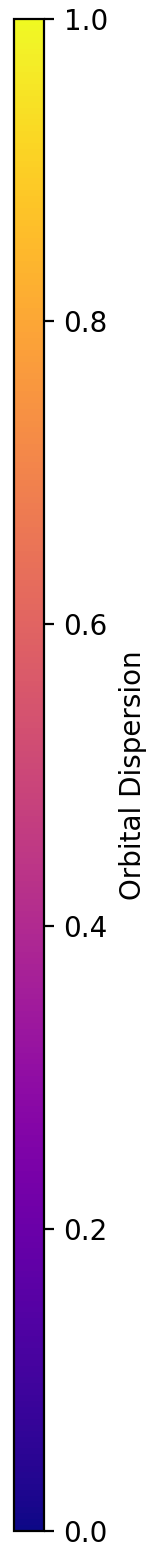}
\end{minipage}

\caption{The orbital paths and spatial dispersions for a a satellite in gradually decreasing $q=1.5$, $1$, $0.9$, $0.75$, $0.5$ (increasing oblate flattening) potentials. The black arrows depict the orbital path of the determined centre of the disrupting satellite. The gradient is the spatial dispersion of each structures' particles from the orbital path, normalised across all time and $q$ snapshots. }
\label{Orbital_paths}
\end{figure*}

As an illustration of this phenomena, we present Figure \ref{Orbital_paths}, depicting the evolution of a satellite in gradually increased flattening $(q=1.5,1,0.9,0.75,0.5)$ potentials. The small black arrows trace the orbital path of the disrupting subhalo’s centre, determined through a convergent mean distance-to-centre method. We also apply a smoothing function to centres to produce a continuous path. The colour gradient shows the normalised orbital spatial dispersion of the disrupted material. Since stellar streams act as tracers of their orbital path, we define the dispersion as the positional deviation of each particle from the orbit. This is computed as the logarithmic distance between each point and its nearest position along the corresponding orbital segment. The segment is centred on the structure’s current centre and extends forward and backward along the orbit up to the 95th percentile of the particles’ radial distances from the centre. Dispersion values exceeding this limit are set to the normalised maximum value of $1$ to minimise the influence of unbound particles. We normalise the dispersion across time and all values of $q$ to better illustrate its distribution for their respective conditions. 

Perhaps most strikingly, we observe the relative inversion in orbital path for the $q=0.5$ case. Here, the subhalo collapses into an orbit on the \textit{inside} of the host potential in contrast to the other iterations of $q$. This leads the subhalo into a highly eccentric infall towards the centre of the host's potential. At $t=6$ Gyrs , we observe the resulting tidal shock as it imparts a fanning effect onto the remnant. The resulting structure possess a shape more reminiscent of a tidal shell. The remaining $q$ values paint a picture of change in orbital path with gradual increase in flattening. We see an increase in eccentricity driven by the asymmetric pull of the flattened potential. This in turn shortens the disruption timescale as evident through the faster orbital progression for the lower $q$ values.

\begin{figure}
\centering
\includegraphics[width=\linewidth]{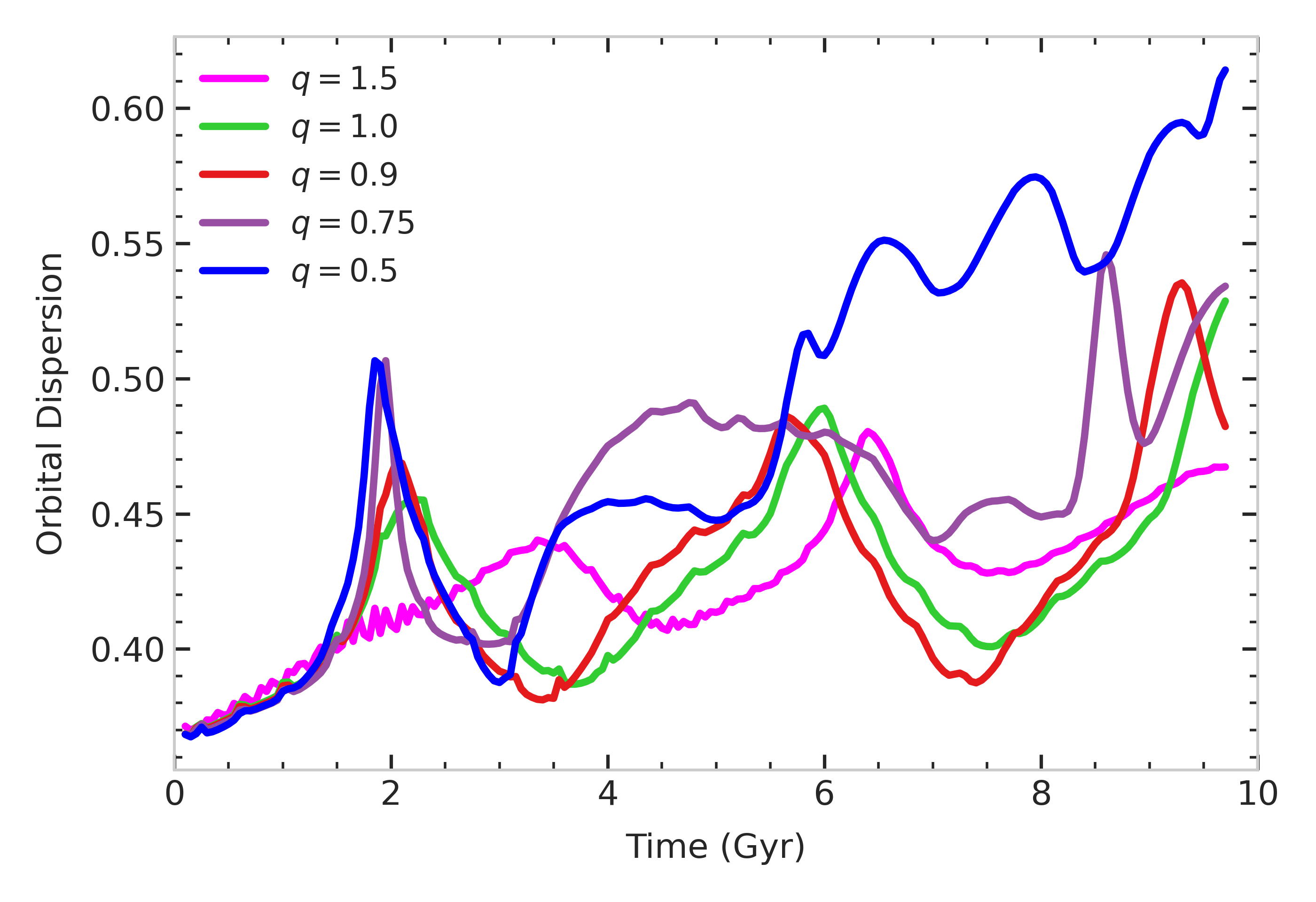}
\caption{The change in average spatial orbital dispersion for each $q$ scenario depicted in figure \ref{Orbital_paths}.}
\label{norm_dispersion}
\end{figure}

Figure \ref{norm_dispersion} displays the change in orbital dispersion with time for the structures presented in figure \ref{Orbital_paths}. We see initial pericentre pass-by for all the structures characterised by a local peak in dispersion at approximately $2$ Gyrs with the exception of the $q=1.5$ which finds a peak near $3.5$ Gyrs. For the former cases, the dispersion evolution forks along different paths for each value of $q$ proceeding the $2$ Gyr pass-by. The $q=1\text{ and }0.9$ scenarios produce an oscillatory pattern with increasing peaks in dispersion. It is however notable that $q=0.9$ leads ahead of the $1$ case indicating a quicker disruption for the flattened halo. The $q=0.75 \text{ and } 0.5$ curves experience significantly greater disruption with $q=0.5$ steadily increasing as a result of the tidal shock. The dispersions are derived from the unsmoothed subhalo centres which become discontinuous upon significant disruption. This produces the highly oscillatory dispersion we observe for the $q=0.5$ after $6$ Gyrs as the tidal shock has destroyed any discernible core. This is also noticeable to a smaller extent in the $q=1.5$ case as the curve oscillates between two very close but separable density peaks within the core. Looking past this however, we continue to observe a consistent disruption pattern, with the prolate case lagging behind in the dispersion timescale compared to the spherical and oblate flattening cases. 

The impact of flattening becomes more apparent when we inspect the orbits further. For the example orbits provided in figure \ref{Orbital_paths}, we observe the shortest pericentre radii to be $63.93$, $57.50$, $42.40$, $14.88$, and $2.18$ kpc for $q=1.5$, $q = 1$, $0.9$, $0.75$, and $0.5$, respectively. For the $q=0.9, 0.75$ cases, these are obtained in their first pericenter passage at approximately $2$ Gyrs. Meanwhile, the prolate $q=1.5$ case experiences it during its second passage at $6.5$ Gyrs. The $q=1, 0.5$ cases also experience it in their second passage, albeit slightly earlier, at approximately $6$ Gyrs. This aligns with the change in dispersion with $q$ we see in figure \ref{norm_dispersion}. Greater (oblate) flattening produces orbits with closer pass-byes to the hosts' centre, exposing the in-falling subhalos to greater tidal forces. Consequently, we see an increase in dispersion as the subhalos break apart quicker. 

As have been previously discussed, stellar streams tend to act as tracers of their orbits. An interesting morphological consequence of this is the more extreme twisting of structures we see for greater values of flattening. This enhanced twisting reflects the stronger tidal interactions that occur during closer pericenter passages, as well as the asymmetric precession that happens at lower values of $q$. In Figure \ref{Orbital_paths} we see a single arc-like stream for $q=1.5, 1, 0.9, 0.75$ morph into an `S'-like shape at $t=6$ Gyrs for $q=0.5$. This suggests that orbits closer to the centre would experience more precessions in the asymmetric potential, leading to both greater dispersion and more convoluted deformations of the orbital path, and consequently, the shape of the stellar stream.

From an observational perspective, streams with globular cluster progenitors are most likely to experience these conditions, given their typically closer initial orbits. However, there are two important caveats to consider and acknowledge. Firstly, flattening conditions as extreme as $0.5$ are unphysical and highly unlikely. Nevertheless, we find that flattening to impart a discernable impact at any value of $q$. Second, the impact of flattening is heavily dependent on the satellite’s initial conditions. As examples, the eccentricity and flattening of the orbit are degenerate for radial orbits, and highly eccentric orbits are expected to form shells through tidal shocks regardless of flattening. Untangling this degeneracy in the initial conditions remains an area for further investigation.

\subsection{Core disruption times}

\begin{figure}
\centering
\includegraphics[width=\linewidth]{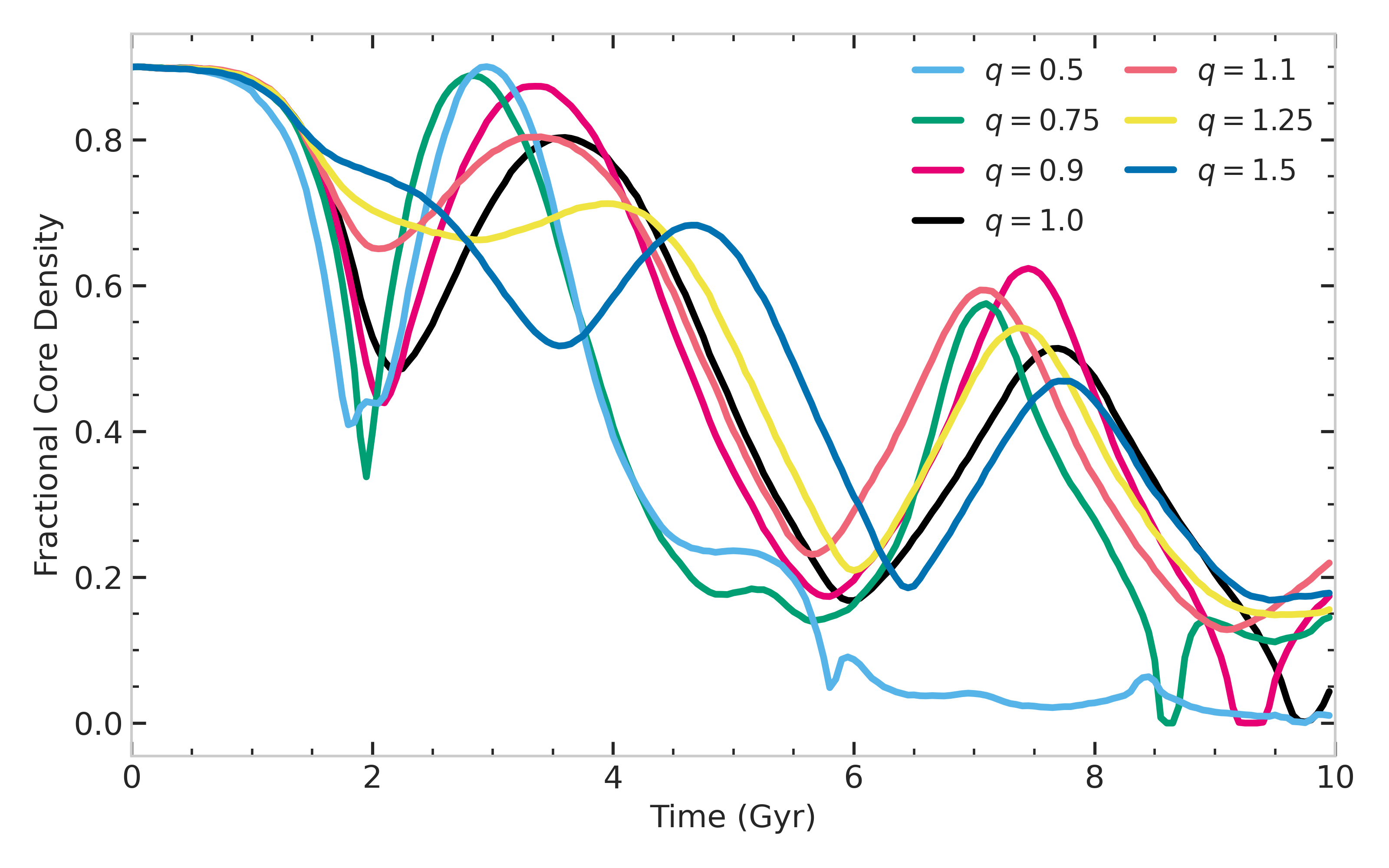}
\caption{The change in fractional core density over time for a satellite with the same initial conditions as Figure \ref{Orbital_paths} for flattening values $q = 1.5$, $1.25$, $1.1$, $1$, $0.9$, $0.75$, and $0.5$.}
\label{fig:core disruption times}
\end{figure}

As a point of comparison, we also inspect the disruption of the core over time as presented in figure \ref{fig:core disruption times}. Here we see the fractional core density of the same subhalos projected in Figures \ref{Orbital_paths}, \ref{norm_dispersion}. However, we extend it to $q=1.1$ and $1.25$ to further study prolate conditions. We define the fraction core density as the mass enclosed within the 90\% mass radius of the initial Plummer sphere. The resulting density evolution traces the rate at which the central bound region is disrupted under different flattening conditions.

Once again, the first pericenter passage stands out, appearing as the initial sharp drop for all the $q$ values at approximately $2$ Gyrs. However, upon closer look we see order in the time at which each curve approaches local minima. The most flattened $q=0.5$ curve reaches the minima first followed by the curves of the other $q$ values in increasing order, reflecting what we see in Figure \ref{norm_dispersion}. Then we see a subsequent increase proceeding this event as the subhalos bounce away from the centre, physically translating to the matter being (temporarily) re-bound to the subhalo core. The prolate scenarios provide an interesting addition as we see a lag in their first pericenter passage reflected as a longer wavelength in the sinusoidal oscillation of the fractional core density.

Between $t \sim 3$--$6~\mathrm{Gyr}$, all systems experience a steady overall decline in density. Models with $q = 0.5$ and $q = 0.75$ follow similar rates of decrease, while $q = 0.9$ and $q = 1.0$ show another grouping with comparable density evolution, despite earlier differences in their rebound amplitudes. After this regime of gradual decline, the flattened case with $q = 0.5$ undergoes a rapid collapse: its density never recovers beyond a brief plateau and subsequently drops sharply. This may be attributed to the tidal shock experienced by the satellite, permanently disrupting the core. The remaining cases ($q = 0.75$, $0.9$, $1.0$, $1.1$, $1.25$ and $1.5$) follow a similar unbinding and re-binding pattern in the process of their second passage. Despite local minima near a fractional core density of $\sim 0$, the $q = 0.5$ $q = 0.75$ and $q = 1$ show a increase indicating a future re-binding event as the remnant continues into the third passage.

\section{Discussion and Conclusion}
\label{Section: Discussion and Conclusion}

We have demonstrated how introducing flattening along the $z$-axis in a static three-component NFW halo significantly alters the morphology and classification of tidal debris. By comparing spherical haloes with extremely oblate and prolate configurations, and by evolving the same set of initial conditions, we assessed how three representative orbital types: a radial orbit generating shells, an eccentric orbit producing a stream, and an intermediate orbit respond to changes in the underlying dark matter distribution. All three initial conditions are not perfectly aligned and each contains a small velocity component along the $z$-axis to probe flattening effects in greater detail.

Our first assessment relied on visual identification in face-on and edge-on density projections after $8.6~Gyr$ of tidal structure evolution, emulating an observable on-sky projection. In the spherical halo, we find a progenitor with a radial orbit producing two plume-like features in the face-on view. 
However, a change of visual to position to that of an edge on observation, reveals these to in fact be well-defined shells. The eccentric orbit produces a canonical stream whose trajectory shifts over time, while the intermediate progenitor forms a stream that oscillates about its own orbital plane. In three dimensions this latter case traces a ring-like path, explaining the absence of shell formation despite its shallow oscillatory imprint in two-dimensional projections.
Unravelling the morphology of a particular structure from visual inspection alone is difficult for this very reason. One approach to solve this problem is kinematically, as done in \citep{Hendel_2015}. They demonstrate a coherent separation in stream and shell-like structures in energy-momentum space. Here, we present a complementary alternative approach to the problem via a data driven density model. As previously discussed, the AstroLink derived ordered-density plot provides a novel perspective to the problem. Combining this with inspections in kinematically derived feature spaces allow us to clarify misinterpretations that arise when relying solely on visual projections. Figures \ref{fig:rad_plot} and \ref{fig:energy angle plot} demonstrate the correspondence between kinematic and density driven approaches to structure classification. However, it must be acknowledged that while both approaches are effective for separating distinct shell and stream structures, phase mixing and intermediary structures continue to present a challenge due to overlap in kinematic and density feature space. \cite{paper_1} investigate this matter further and provides a more comprehensive look at the Astrolink classification method. For our purposes here, we lean on these tools in our investigation into the effect of perspective and flattening on the classification of these tidal objects.
\\
Starting with the spherical case, we are able to recover two clear shells formed through radial orbit disruption along with two distinct streams: one each from the eccentric and intermediate progenitors in position space. 
Radial distance and energy-angle diagrams further reinforce these identifications: shells appear as extended, radially coherent features. A notable feature is that the stream S4 extends to large radial distances, superficially resembling a shell near apocentre. However, its narrow velocity distribution and coherent angle structure show that it remains a stream rather than a true caustic. S3 displays characteristic stream features, including a distinct guitar-pick shape in radial phase-space.

The oblate halo presents a strikingly different picture. In the face-on projection, three pronounced shell-like features appear; however, these same features display the wide orbital width and elongated, diffuse morphology characteristic of thick streams in the edge-on view. 
Ordered-density plots reveal that material which initially forms shell-like structures becomes stretched along the $z$-axis due to the extremely oblate mass distribution, and over time dilutes into debris that resembles stream-like material. S4 is already well into its tidal disruption phase, while S3 occupies a slightly earlier but still comparable stage. 
Energy-angle space reflects this behaviour: both streams have broadened substantially, with the S4 appearing almost fully phase-mixed, consistent with expectations for flattened potentials where differential precession is enhanced. In radial space, the material contributing to the three shell-like features corresponds to two distinct peaks, but thicker streams with wide orbits can also produce similar signatures. The third shell-like feature is in fact one of the stream arms that has extended to a different orientation as a result of twisting induced by precession. 

The prolate halo exhibits yet another regime of behaviour. The radial orbit progenitor produces strong shell-like features, while the streams appear much more diffuse compared to the other cases and even present broadly shell-like characteristics in the edge-on projection. The overall distribution shows a high degree of phase mixing. 
Ordered-density plots, energy-angle diagrams, and radial profiles collectively confirm that these structures are best described as radially dominated debris. Notably, the eccentric and intermediate progenitors acquire shell-like signatures with their elongation in radial distance, despite originating from non-radial orbits.

These results illustrate that although the initial conditions dictate the type of structure a progenitor can in principle form, the halo's mass distribution exerts a significant influence on the final morphology. We find that increasing the flattening to extreme values produces pronounced eccentric or radial stretching of the tidal debris. As a result, shells begin to resemble angularly dispersed streams, while streams acquire features reminiscent of shells. This complicates visual classification, especially as phase mixing progresses and structures diffuse. In both the prolate and oblate haloes, the most diffused shell material blends into the halo background, making structure identification increasingly difficult in regions where multiple debris populations overlap. Phase mixing and faster shell diffusion is observed in moderate oblate halos at a similar time frame, signifying how identification and classification of tidal structures, especially shells can be a difficult affair.

\subsection{Initial Conditions and Halo Shape}
\label{subsection:Initial_Conditions_and_Halo_Shape}
From our analysis, it is evident that structure formation and evolution in these simulations depend sensitively on both the initial conditions of the progenitor subhalos and the flattening parameter, \(q\), of the host dark matter halo. Variations in \(q\) affect not only the timing of tidal feature formation but also the overall morphology of tidal debris. While both shells and streams respond to changes in the halo mass distribution, they do so through different physical pathways: shells are more directly linked to the radialisation, formation plane, and oscillatory history of their progenitors, whereas streams are governed primarily by precession in their orbits, vertical excursions, and differential orbital phase-mixing. Initial conditions with higher velocity components along the $z$-axis react more strongly to the flattening used in this work. Determining which tracer is ultimately more sensitive to subtle variations in a halo distribution that evolves with time remains an open question, and will be particularly important one once combined with observational constraints from the Milky Way and external galaxy's tidal features.

Studying tidal structures aids in understanding more about how they form and evolve and also the properties of the host's dark matter halo. In most cases, the initial condition of the subhalo combined with the halo conditions of the host galaxy affects the type and rate of structure formation from minor mergers with the initial conditions having a greater role in the type of structure that forms. \citep{Hendel_2015} established the foundational framework for this relationship, demonstrating that the degree of orbital circularity governs whether a disrupting satellite produces a stream or a shell: satellites on near-radial orbits preferentially form shells, while those on more circular orbits produce coherent tidal streams. This dependence on orbital circularity implies that the same progenitor, placed on orbits of different eccentricity within the same potential, will produce morphologically distinct debris structures. While the geometry of the host potential imprints changes in the structure, \citep{Dehnen_2018} showed that for satellites orbiting close to the Galactic disc, the strong amplitude-dependence of the vertical orbital frequency \(\Omega_{z}\)causes stripped particles to acquire a range of frequency ratios and the type of structure formed have their own ratio range. They also importantly present tidal ribbons caused by the 'two'-dimensional spread of tidal debris rather than forming more conventional 'one'-dimensional streams, resembling closely to the case of our radial progenitor when the halo is oblate.

Taken together, these studies establish that tidal debris morphology is a sensitive function of both the subhalo's initial orbital configuration and the flattening of the host potential, motivating the systematic variation of halo axis ratio $q$ explored in this work. By evolving three different types of orbit conditions, either extremes of radial and eccentric examples with an intermediate scenario being presented in this paper, there remain several similar structures explored by the authors across 2000 simulations. The subhalo conditions presented in this paper showcases a sub-sample of changing the orbital conditions. In line with previous works, we reinforce the notion of varied halo shape leads to a variation in the distribution and shape of structures. We also find that, although halo shape has a non-negligible effect, initial conditions of the subhalo have the dominating effect in structure morphology. Further quantifying the extent of these effects remains in line of future work. 

The presence of a much heavier subhalo, such as the LMC in the Milky Way, can produce strong substructure in streams, particularly for progenitors on near-circular orbits \citep{2025_Weerasooriya}, while \citet{2011_wetzel} found that disrupted subhalos at high redshift tend to follow increasingly radial orbits. Perturbations can also be induced in existing tidal structures by minor mergers; streams on wider orbits were found to experience stronger perturbation effects that persist for longer \citep{2025_Guillaume}. We also note that dynamical friction plays only a minor role at the mass scales considered in this study \citep{1999_Bosch}. 

The shell-forming progenitor in our simulations is the most massive of the three subhalos (\(10^7\,{\rm M}_{\odot}\)), and is therefore naturally less susceptible to significant perturbations from the two lower mass stream forming progenitors though it can itself induce noticeable perturbations in the streams’ trajectories. In an ideal scenario, where the shell progenitor is evolved in isolation on a purely radial orbit, we would expect a clean Type~I shell system to appear during the first oscillation. However, when the two additional subhalos on eccentric and intermediate orbits are included, the early development of this shell system is temporarily disrupted. The radial progenitor's trajectory is initially perturbed, but it subsequently returns to an almost radial orbit and produces a first, and later a second shell in the expected oscillatory sequence. What is particularly striking is that despite the gravitational influence of the two other satellites and the extremities of halo flattening, almost all three values of \(q\) still produce observable shell-like systems. 
\subsection{Flattened Halos}
 NFW halo is generally preferred for visualising the spherical density profiles, introducing a flattening parameter $q$ in the NFW profile can lead to unrealistic mass distributions. Our study however, focuses on examining structure classification and formation, and we therefore plan to transition to a triaxial model in future work, allowing for variations in axis ratios along all axes. An oblate dark matter halo with a $z$-flattening ratio of $0.92 \pm 0.08$ is preferred, despite the use of a $y$-axis ratio in \cite{Zhu_2025}, where the inner halo was observed to experience a flip attributed to possible minor mergers in that region. Such changes in halo shape can influence stream evolution. For example, studies using a Miyamoto-Nagai potential show that streams forming in different regions of the separatrix-divergence experience different rates of dissolution \citep{Yavetz2020}. To model a more realistic dark matter mass distribution, a Lee-Suto profile \citep{2003_suto} type halo could be adopted, and introducing triaxiality into logarithmic profiles \citep{2000_Evans} is another option. Time-dependent host potentials may further contribute to changes in stream morphology; in particular, long stellar streams show angular misalignment near apocentre in evolving potentials \citep{BuistHelmi2015}. Distorted halos in general are also known to induce torsion in stellar streams, where prolate halos can generate a combination of planar and helical stream orbits \citep{Bariego_2024}.

The initial condition of the progenitor can also heavily impact the core disruption times for different $q$ values. As a result, although a significant portion of core disruption can be attributed to the progenitor's initial conditions, specific values of $q$ may distort the halo mass distribution in a way that favours certain initial configurations. This is particularly relevant when the core disruption trend of streams appears to contradict the expected increase in disruption time with increasing $q$. One such case from our simulations shows that streams with high angular components respond differently to the halo flattening, sometimes resulting in a longer disruption times for a specific q value than expected. In general, it is extremely difficult for a subhalo to undergo core disruption purely through tidal stripping; typically, multiple tidal shocks associated with repeated pericentric passages are the primary drivers of core mass loss \citep{2018_Bosch}. Other physical processes such as resonant shocks and torques can also contribute. For example, in the case of a massive subhalo like the LMC interacting with the Milky Way, \citep{2009M_Choi} found that the LMC's core had not experienced significant stellar mass loss.

\subsection{Observation and future aspects}

With the advent of new observational results from LSST . missions such as Nancy Grace Roman Telescope \citep{nancy_g_roman}, as well as missions dedicated to detecting low surface brightness structures such as ESA's ARRAKHIS \citep{Arrakhis}, we expect unprecedented detail and improved characterisation of tidal debris. These missions will significantly enhance our ability to cluster stellar structures in 6D phase space. We plan to extend our work by applying these methods to tidal structures evolved in cosmological simulations, by identifying and classifying them on different scales. Our line of work also stretches further in exploring the effects of stream shell evolution with perturbations from LMC and could also be inspired by the works of \cite{Brooks2024} on their simulation based inference model on the dynamics of MW-LMC system and another simulation based inference works on extragalactic streams by \cite{Nibauer_2025} with the aim of inferring the underlying dark matter distribution of their host galaxies.As previously mentioned in \ref{subsection:Initial_Conditions_and_Halo_Shape}, we are working on a more quantitative analysis to probe the effect of initial conditions of the subhalo and the host galaxy's conditions to a detailed scale, by varying the dependence on quantities that play a crucial role like the virial radius of the halo, concentration, mass and other parameters. Our goal also aims to look at and study the degrees of effects both of the above said factors play in minor mergers and tidal structure formations.

\section{Acknowledgements}

We are grateful for the feedback and comments provided by the Gravitational Astrophysics group in University of Sydney's School of Physics. We would also like to thank the anonymous reviewers for their valuable feedback on our paper.
This research was undertaken with the assistance of resources from the National Computational Infrastructure (NCI Australia), an NCRIS enabled capability supported by the Australian Government and the Sydney Informatics Hub, a Core Research Facility of the University of Sydney. 
SGB is supported by Australian Academy of Technological Sciences and Engineering's Elevate scholarship. VE is supported by the University of Sydney Postgraduate Award.  WHO's contribution to this project was made possible by funding from the Carl-Zeiss-Stiftung.
\section*{Data Availability}
The simulation data and the initial condition codes utilised in this paper will be shared upon reasonable request to the corresponding author.



\bibliographystyle{mnras}
\bibliography{main} 

@BOOK{Peebles_1981,
       author = {{Peebles}, P.~J.~E.},
        title = "{The Large-Scale Structure of the Universe}",
         year = 1981,
       publisher = {Princeton University Press},
       adsurl = {https://ui.adsabs.harvard.edu/abs/1981lssu.book.....P}
}

@ARTICLE{SPARC,
       author = {{Lelli}, Federico and {McGaugh}, Stacy S. and {Schombert}, James M.},
        title = "{SPARC: Mass Models for 175 Disk Galaxies with Spitzer Photometry and Accurate Rotation Curves}",
      journal = {\aj},
         year = 2016,
        month = dec,
       volume = {152},
       number = {6},
          eid = {157},
        pages = {157},
          doi = {10.3847/0004-6256/152/6/157},
archivePrefix = {arXiv},
       eprint = {1606.09251},
 primaryClass = {astro-ph.GA},
       adsurl = {https://ui.adsabs.harvard.edu/abs/2016AJ....152..157L}
}

@ARTICLE{Bowden_2016,
       author = {{Bowden}, A. and {Evans}, N.~W. and {Williams}, A.~A.},
        title = "{Is the dark halo of the Milky Way prolate?}",
      journal = {\mnras},
         year = 2016,
        month = jul,
       volume = {460},
       number = {1},
        pages = {329-337},
          doi = {10.1093/mnras/stw994},
archivePrefix = {arXiv},
       eprint = {1604.06885},
 primaryClass = {astro-ph.GA},
       adsurl = {https://ui.adsabs.harvard.edu/abs/2016MNRAS.460..329B}
}

@ARTICLE{Vera_2011,
       author = {{Vera-Ciro}, Carlos A. and {Sales}, Laura V. and {Helmi}, Amina and {Frenk}, Carlos S. and {Navarro}, Julio F. and {Springel}, Volker and {Vogelsberger}, Mark and {White}, Simon D.~M.},
        title = "{The shape of dark matter haloes in the Aquarius simulations: evolution and memory}",
      journal = {\mnras},
         year = 2011,
        month = sep,
       volume = {416},
       number = {2},
        pages = {1377-1391},
          doi = {10.1111/j.1365-2966.2011.19134.x},
archivePrefix = {arXiv},
       eprint = {1104.1566},
 primaryClass = {astro-ph.CO},
       adsurl = {https://ui.adsabs.harvard.edu/abs/2011MNRAS.416.1377V}
}

@ARTICLE{Posti_2019,
       author = {{Posti}, Lorenzo and {Helmi}, Amina},
        title = "{Mass and shape of the Milky Way's dark matter halo with globular clusters from Gaia and Hubble}",
      journal = {\aap},
         year = 2019,
        month = jan,
       volume = {621},
          eid = {A56},
        pages = {A56},
          doi = {10.1051/0004-6361/201833355},
archivePrefix = {arXiv},
       eprint = {1805.01408},
 primaryClass = {astro-ph.GA},
       adsurl = {https://ui.adsabs.harvard.edu/abs/2019A&A...621A..56P}
}

@ARTICLE{Fellhauer_2006,
       author = {{Fellhauer}, M. and {Belokurov}, V. and {Evans}, N.~W. and {Wilkinson}, M.~I. and {Zucker}, D.~B. and {Gilmore}, G. and {Irwin}, M.~J. and {Bramich}, D.~M. and {Vidrih}, S. and {Wyse}, R.~F.~G. and {Beers}, T.~C. and {Brinkmann}, J.},
        title = "{The Origin of the Bifurcation in the Sagittarius Stream}",
      journal = {\apj},
         year = 2006,
        month = nov,
       volume = {651},
       number = {1},
        pages = {167-173},
          doi = {10.1086/507128},
archivePrefix = {arXiv},
       eprint = {astro-ph/0605026},
 primaryClass = {astro-ph},
       adsurl = {https://ui.adsabs.harvard.edu/abs/2006ApJ...651..167F}
}

@ARTICLE{Johnston_2005,
       author = {{Bullock}, James S. and {Johnston}, Kathryn V.},
        title = "{Tracing Galaxy Formation with Stellar Halos. I. Methods}",
      journal = {\apj},
         year = 2005,
        month = dec,
       volume = {635},
       number = {2},
        pages = {931-949},
          doi = {10.1086/497422},
archivePrefix = {arXiv},
       eprint = {astro-ph/0506467},
 primaryClass = {astro-ph},
       adsurl = {https://ui.adsabs.harvard.edu/abs/2005ApJ...635..931B}
}

@INPROCEEDINGS{Bariego_2022,
       author = {{Bariego Quintana}, A. and {Llanes-Estrada}, F.~J. and {Manzanilla Carretero}, O.},
        title = "{Dark-matter halo shapes from fits to SPARC galaxy rotation curves}",
    booktitle = {European Physical Society Conference on High Energy Physics},
         year = 2022,
        month = jan,
          eid = {137},
        pages = {137},
          doi = {10.22323/1.398.0137},
archivePrefix = {arXiv},
       eprint = {2109.11153},
 primaryClass = {hep-ph},
       adsurl = {https://ui.adsabs.harvard.edu/abs/2022epsc.confE.137B}
}

@INPROCEEDINGS{Zeldovich_1978,
       author = {{Zeldovich}, Ia. B.},
        title = "{The Theory of the Large Scale Structure of the Universe}",
    booktitle = {Large Scale Structures in the Universe},
         year = 1978,
       editor = {{Longair}, M.~S. and {Einasto}, J.},
       series = {IAU Symposium},
       volume = {79},
        month = jan,
        pages = {409},
       adsurl = {https://ui.adsabs.harvard.edu/abs/1978IAUS...79..409Z}
}

@PROCEEDINGS{Newberg_2016,
        title = "{Tidal Streams in the Local Group and Beyond}",
    booktitle = {Tidal Streams in the Local Group and Beyond},
         year = 2016,
       editor = {{Newberg}, Heidi Jo and {Carlin}, Jeffrey L.},
       series = {Astrophysics and Space Science Library},
       volume = {420},
        month = jan,
          doi = {10.1007/978-3-319-19336-6},
       adsurl = {https://ui.adsabs.harvard.edu/abs/2016ASSL..420.....N}
}

@ARTICLE{Hernquist_Quinn_1988,
       author = {{Hernquist}, Lars and {Quinn}, P.~J.},
        title = "{Formation of Shell Galaxies. I. Spherical Potentials}",
      journal = {\apj},
         year = 1988,
        month = aug,
       volume = {331},
        pages = {682},
          doi = {10.1086/166592},
       adsurl = {https://ui.adsabs.harvard.edu/abs/1988ApJ...331..682H}
}

@ARTICLE{White_Rees_1978,
       author = {{White}, S.~D.~M. and {Rees}, M.~J.},
        title = "{Core condensation in heavy halos: a two-stage theory for galaxy formation and clustering.}",
      journal = {\mnras},
         year = 1978,
        month = may,
       volume = {183},
        pages = {341-358},
          doi = {10.1093/mnras/183.3.341},
       adsurl = {https://ui.adsabs.harvard.edu/abs/1978MNRAS.183..341W}
}

@ARTICLE{Ibata_1994,
       author = {{Ibata}, R.~A. and {Gilmore}, G. and {Irwin}, M.~J.},
        title = "{A dwarf satellite galaxy in Sagittarius}",
      journal = {\nat},
         year = 1994,
        month = jul,
       volume = {370},
       number = {6486},
        pages = {194-196},
          doi = {10.1038/370194a0},
       adsurl = {https://ui.adsabs.harvard.edu/abs/1994Natur.370..194I}
}

@ARTICLE{Ibata_2001,
       author = {{Ibata}, Rodrigo and {Irwin}, Michael and {Lewis}, Geraint F. and {Stolte}, Andrea},
        title = "{Galactic Halo Substructure in the Sloan Digital Sky Survey: The Ancient Tidal Stream from the Sagittarius Dwarf Galaxy}",
      journal = {\apjl},
         year = 2001,
        month = feb,
       volume = {547},
       number = {2},
        pages = {L133-L136},
          doi = {10.1086/318894},
archivePrefix = {arXiv},
       eprint = {astro-ph/0004255},
 primaryClass = {astro-ph},
       adsurl = {https://ui.adsabs.harvard.edu/abs/2001ApJ...547L.133I}
}

@ARTICLE{Odenkirchen_2003,
       author = {{Odenkirchen}, Michael and {Grebel}, Eva K. and {Dehnen}, Walter and {Rix}, Hans-Walter and {Yanny}, Brian and {Newberg}, Heidi Jo and {Rockosi}, Constance M. and {Mart{\'\i}nez-Delgado}, David and {Brinkmann}, Jon and {Pier}, Jeffrey R.},
        title = "{The Extended Tails of Palomar 5: A 10{\textdegree} Arc of Globular Cluster Tidal Debris}",
      journal = {\aj},
         year = 2003,
        month = nov,
       volume = {126},
       number = {5},
        pages = {2385-2407},
          doi = {10.1086/378601},
archivePrefix = {arXiv},
       eprint = {astro-ph/0307446},
 primaryClass = {astro-ph},
       adsurl = {https://ui.adsabs.harvard.edu/abs/2003AJ....126.2385O}
}

@ARTICLE{Ji_s5_streams_2020,
       author = {{Ji}, Alexander P. and {Li}, Ting S. and {Hansen}, Terese T. and {Casey}, Andrew R. and {Koposov}, Sergey E. and {Pace}, Andrew B. and {Mackey}, Dougal and {Lewis}, Geraint F. and {Simpson}, Jeffrey D. and {Bland-Hawthorn}, Joss and {Cullinane}, Lara R. and {Da Costa}, Gary. S. and {Hattori}, Kohei and {Martell}, Sarah L. and {Kuehn}, Kyler and {Erkal}, Denis and {Shipp}, Nora and {Wan}, Zhen and {Zucker}, Daniel B.},
        title = "{The Southern Stellar Stream Spectroscopic Survey (S$^{5}$): Chemical Abundances of Seven Stellar Streams}",
      journal = {\aj},
         year = 2020,
        month = oct,
       volume = {160},
       number = {4},
          eid = {181},
        pages = {181},
          doi = {10.3847/1538-3881/abacb6},
archivePrefix = {arXiv},
       eprint = {2008.07568},
 primaryClass = {astro-ph.SR},
       adsurl = {https://ui.adsabs.harvard.edu/abs/2020AJ....160..181J}
}

@ARTICLE{SDSS_2000,
       author = {{York}, Donald G. and {Adelman}, J. and {Anderson}, Jr., John E. and {Anderson}, Scott F. and {Annis}, James and {Bahcall}, Neta A. and {Bakken}, J.~A. and {Barkhouser}, Robert and {Bastian}, Steven and {Berman}, Eileen and {Boroski}, William N. and {Bracker}, Steve and {Briegel}, Charlie and {Briggs}, John W. and {Brinkmann}, J. and {Brunner}, Robert and {Burles}, Scott and {Carey}, Larry and {Carr}, Michael A. and {Castander}, Francisco J. and {Chen}, Bing and {Colestock}, Patrick L. and {Connolly}, A.~J. and {Crocker}, J.~H. and {Csabai}, Istv{\'a}n and {Czarapata}, Paul C. and {Davis}, John Eric and {Doi}, Mamoru and {Dombeck}, Tom and {Eisenstein}, Daniel and {Ellman}, Nancy and {Elms}, Brian R. and {Evans}, Michael L. and {Fan}, Xiaohui and {Federwitz}, Glenn R. and {Fiscelli}, Larry and {Friedman}, Scott and {Frieman}, Joshua A. and {Fukugita}, Masataka and {Gillespie}, Bruce and {Gunn}, James E. and {Gurbani}, Vijay K. and {de Haas}, Ernst and {Haldeman}, Merle and {Harris}, Frederick H. and {Hayes}, J. and {Heckman}, Timothy M. and {Hennessy}, G.~S. and {Hindsley}, Robert B. and {Holm}, Scott and {Holmgren}, Donald J. and {Huang}, Chi-hao and {Hull}, Charles and {Husby}, Don and {Ichikawa}, Shin-Ichi and {Ichikawa}, Takashi and {Ivezi{\'c}}, {\v{Z}}eljko and {Kent}, Stephen and {Kim}, Rita S.~J. and {Kinney}, E. and {Klaene}, Mark and {Kleinman}, A.~N. and {Kleinman}, S. and {Knapp}, G.~R. and {Korienek}, John and {Kron}, Richard G. and {Kunszt}, Peter Z. and {Lamb}, D.~Q. and {Lee}, B. and {Leger}, R. French and {Limmongkol}, Siriluk and {Lindenmeyer}, Carl and {Long}, Daniel C. and {Loomis}, Craig and {Loveday}, Jon and {Lucinio}, Rich and {Lupton}, Robert H. and {MacKinnon}, Bryan and {Mannery}, Edward J. and {Mantsch}, P.~M. and {Margon}, Bruce and {McGehee}, Peregrine and {McKay}, Timothy A. and {Meiksin}, Avery and {Merelli}, Aronne and {Monet}, David G. and {Munn}, Jeffrey A. and {Narayanan}, Vijay K. and {Nash}, Thomas and {Neilsen}, Eric and {Neswold}, Rich and {Newberg}, Heidi Jo and {Nichol}, R.~C. and {Nicinski}, Tom and {Nonino}, Mario and {Okada}, Norio and {Okamura}, Sadanori and {Ostriker}, Jeremiah P. and {Owen}, Russell and {Pauls}, A. George and {Peoples}, John and {Peterson}, R.~L. and {Petravick}, Donald and {Pier}, Jeffrey R. and {Pope}, Adrian and {Pordes}, Ruth and {Prosapio}, Angela and {Rechenmacher}, Ron and {Quinn}, Thomas R. and {Richards}, Gordon T. and {Richmond}, Michael W. and {Rivetta}, Claudio H. and {Rockosi}, Constance M. and {Ruthmansdorfer}, Kurt and {Sandford}, Dale and {Schlegel}, David J. and {Schneider}, Donald P. and {Sekiguchi}, Maki and {Sergey}, Gary and {Shimasaku}, Kazuhiro and {Siegmund}, Walter A. and {Smee}, Stephen and {Smith}, J. Allyn and {Snedden}, S. and {Stone}, R. and {Stoughton}, Chris and {Strauss}, Michael A. and {Stubbs}, Christopher and {SubbaRao}, Mark and {Szalay}, Alexander S. and {Szapudi}, Istvan and {Szokoly}, Gyula P. and {Thakar}, Anirudda R. and {Tremonti}, Christy and {Tucker}, Douglas L. and {Uomoto}, Alan and {Vanden Berk}, Dan and {Vogeley}, Michael S. and {Waddell}, Patrick and {Wang}, Shu-i. and {Watanabe}, Masaru and {Weinberg}, David H. and {Yanny}, Brian and {Yasuda}, Naoki and {SDSS Collaboration}},
        title = "{The Sloan Digital Sky Survey: Technical Summary}",
      journal = {\aj},
         year = 2000,
        month = sep,
       volume = {120},
       number = {3},
        pages = {1579-1587},
          doi = {10.1086/301513},
archivePrefix = {arXiv},
       eprint = {astro-ph/0006396},
 primaryClass = {astro-ph},
       adsurl = {https://ui.adsabs.harvard.edu/abs/2000AJ....120.1579Y}
}

@ARTICLE{Gaia_2016,
       author = {{Gaia Collaboration} and {Prusti}, T. and {de Bruijne}, J.~H.~J. and {Brown}, A.~G.~A. and {Vallenari}, A. and {Babusiaux}, C. and {Bailer-Jones}, C.~A.~L. and {Bastian}, U. and {Biermann}, M. and {Evans}, D.~W. and {Eyer}, L. and {Jansen}, F. and {Jordi}, C. and {Klioner}, S.~A. and {Lammers}, U. and {Lindegren}, L. and {Luri}, X. and {Mignard}, F. and {Milligan}, D.~J. and {Panem}, C. and {Poinsignon}, V. and {Pourbaix}, D. and {Randich}, S. and {Sarri}, G. and {Sartoretti}, P. and {Siddiqui}, H.~I. and {Soubiran}, C. and {Valette}, V. and {van Leeuwen}, F. and {Walton}, N.~A. and {Aerts}, C. and {Arenou}, F. and {Cropper}, M. and {Drimmel}, R. and {H{\o}g}, E. and {Katz}, D. and {Lattanzi}, M.~G. and {O'Mullane}, W. and {Grebel}, E.~K. and {Holland}, A.~D. and {Huc}, C. and {Passot}, X. and {Bramante}, L. and {Cacciari}, C. and {Casta{\~n}eda}, J. and {Chaoul}, L. and {Cheek}, N. and {De Angeli}, F. and {Fabricius}, C. and {Guerra}, R. and {Hern{\'a}ndez}, J. and {Jean-Antoine-Piccolo}, A. and {Masana}, E. and {Messineo}, R. and {Mowlavi}, N. and {Nienartowicz}, K. and {Ord{\'o}{\~n}ez-Blanco}, D. and {Panuzzo}, P. and {Portell}, J. and {Richards}, P.~J. and {Riello}, M. and {Seabroke}, G.~M. and {Tanga}, P. and {Th{\'e}venin}, F. and {Torra}, J. and {Els}, S.~G. and {Gracia-Abril}, G. and {Comoretto}, G. and {Garcia-Reinaldos}, M. and {Lock}, T. and {Mercier}, E. and {Altmann}, M. and {Andrae}, R. and {Astraatmadja}, T.~L. and {Bellas-Velidis}, I. and {Benson}, K. and {Berthier}, J. and {Blomme}, R. and {Busso}, G. and {Carry}, B. and {Cellino}, A. and {Clementini}, G. and {Cowell}, S. and {Creevey}, O. and {Cuypers}, J. and {Davidson}, M. and {De Ridder}, J. and {de Torres}, A. and {Delchambre}, L. and {Dell'Oro}, A. and {Ducourant}, C. and {Fr{\'e}mat}, Y. and {Garc{\'\i}a-Torres}, M. and {Gosset}, E. and {Halbwachs}, J. -L. and {Hambly}, N.~C. and {Harrison}, D.~L. and {Hauser}, M. and {Hestroffer}, D. and {Hodgkin}, S.~T. and {Huckle}, H.~E. and {Hutton}, A. and {Jasniewicz}, G. and {Jordan}, S. and {Kontizas}, M. and {Korn}, A.~J. and {Lanzafame}, A.~C. and {Manteiga}, M. and {Moitinho}, A. and {Muinonen}, K. and {Osinde}, J. and {Pancino}, E. and {Pauwels}, T. and {Petit}, J. -M. and {Recio-Blanco}, A. and {Robin}, A.~C. and {Sarro}, L.~M. and {Siopis}, C. and {Smith}, M. and {Smith}, K.~W. and {Sozzetti}, A. and {Thuillot}, W. and {van Reeven}, W. and {Viala}, Y. and {Abbas}, U. and {Abreu Aramburu}, A. and {Accart}, S. and {Aguado}, J.~J. and {Allan}, P.~M. and {Allasia}, W. and {Altavilla}, G. and {{\'A}lvarez}, M.~A. and {Alves}, J. and {Anderson}, R.~I. and {Andrei}, A.~H. and {Anglada Varela}, E. and {Antiche}, E. and {Antoja}, T. and {Ant{\'o}n}, S. and {Arcay}, B. and {Atzei}, A. and {Ayache}, L. and {Bach}, N. and {Baker}, S.~G. and {Balaguer-N{\'u}{\~n}ez}, L. and {Barache}, C. and {Barata}, C. and {Barbier}, A. and {Barblan}, F. and {Baroni}, M. and {Barrado y Navascu{\'e}s}, D. and {Barros}, M. and {Barstow}, M.~A. and {Becciani}, U. and {Bellazzini}, M. and {Bellei}, G. and {Bello Garc{\'\i}a}, A. and {Belokurov}, V. and {Bendjoya}, P. and {Berihuete}, A. and {Bianchi}, L. and {Bienaym{\'e}}, O. and {Billebaud}, F. and {Blagorodnova}, N. and {Blanco-Cuaresma}, S. and {Boch}, T. and {Bombrun}, A. and {Borrachero}, R. and {Bouquillon}, S. and {Bourda}, G. and {Bouy}, H. and {Bragaglia}, A. and {Breddels}, M.~A. and {Brouillet}, N. and {Br{\"u}semeister}, T. and {Bucciarelli}, B. and {Budnik}, F. and {Burgess}, P. and {Burgon}, R. and {Burlacu}, A. and {Busonero}, D. and {Buzzi}, R. and {Caffau}, E. and {Cambras}, J. and {Campbell}, H. and {Cancelliere}, R. and {Cantat-Gaudin}, T. and {Carlucci}, T. and {Carrasco}, J.~M. and {Castellani}, M. and {Charlot}, P. and {Charnas}, J. and {Charvet}, P. and {Chassat}, F. and {Chiavassa}, A. and {Clotet}, M. and {Cocozza}, G. and {Collins}, R.~S. and {Collins}, P. and {Costigan}, G.},
        title = "{The Gaia mission}",
      journal = {\aap},
         year = 2016,
        month = nov,
       volume = {595},
          eid = {A1},
        pages = {A1},
          doi = {10.1051/0004-6361/201629272},
archivePrefix = {arXiv},
       eprint = {1609.04153},
 primaryClass = {astro-ph.IM},
       adsurl = {https://ui.adsabs.harvard.edu/abs/2016A&A...595A...1G}
}

@ARTICLE{Atlas_3d_2001,
       author = {{Cappellari}, Michele and {Emsellem}, Eric and {Krajnovi{\'c}}, Davor and {McDermid}, Richard M. and {Scott}, Nicholas and {Verdoes Kleijn}, G.~A. and {Young}, Lisa M. and {Alatalo}, Katherine and {Bacon}, R. and {Blitz}, Leo and {Bois}, Maxime and {Bournaud}, Fr{\'e}d{\'e}ric and {Bureau}, M. and {Davies}, Roger L. and {Davis}, Timothy A. and {de Zeeuw}, P.~T. and {Duc}, Pierre-Alain and {Khochfar}, Sadegh and {Kuntschner}, Harald and {Lablanche}, Pierre-Yves and {Morganti}, Raffaella and {Naab}, Thorsten and {Oosterloo}, Tom and {Sarzi}, Marc and {Serra}, Paolo and {Weijmans}, Anne-Marie},
        title = "{The ATLAS$^{3D}$ project - I. A volume-limited sample of 260 nearby early-type galaxies: science goals and selection criteria}",
      journal = {\mnras},
         year = 2011,
        month = may,
       volume = {413},
       number = {2},
        pages = {813-836},
          doi = {10.1111/j.1365-2966.2010.18174.x},
archivePrefix = {arXiv},
       eprint = {1012.1551},
 primaryClass = {astro-ph.CO},
       adsurl = {https://ui.adsabs.harvard.edu/abs/2011MNRAS.413..813C}
}

@ARTICLE{DES_2005,
       author = {{The Dark Energy Survey Collaboration}},
        title = "{The Dark Energy Survey}",
      journal = {arXiv e-prints},
         year = 2005,
        month = oct,
          eid = {astro-ph/0510346},
        pages = {astro-ph/0510346},
          doi = {10.48550/arXiv.astro-ph/0510346},
archivePrefix = {arXiv},
       eprint = {astro-ph/0510346},
 primaryClass = {astro-ph},
       adsurl = {https://ui.adsabs.harvard.edu/abs/2005astro.ph.10346T}
}

@ARTICLE{s5_2019,
       author = {{Li}, T.~S. and {Koposov}, S.~E. and {Zucker}, D.~B. and {Lewis}, G.~F. and {Kuehn}, K. and {Simpson}, J.~D. and {Ji}, A.~P. and {Shipp}, N. and {Mao}, Y. -Y. and {Geha}, M. and {Pace}, A.~B. and {Mackey}, A.~D. and {Allam}, S. and {Tucker}, D.~L. and {Da Costa}, G.~S. and {Erkal}, D. and {Simon}, J.~D. and {Mould}, J.~R. and {Martell}, S.~L. and {Wan}, Z. and {De Silva}, G.~M. and {Bechtol}, K. and {Balbinot}, E. and {Belokurov}, V. and {Bland-Hawthorn}, J. and {Casey}, A.~R. and {Cullinane}, L. and {Drlica-Wagner}, A. and {Sharma}, S. and {Vivas}, A.~K. and {Wechsler}, R.~H. and {Yanny}, B. and {S5 Collaboration}},
        title = "{The southern stellar stream spectroscopic survey (S$^{5}$): Overview, target selection, data reduction, validation, and early science}",
      journal = {\mnras},
         year = 2019,
        month = dec,
       volume = {490},
       number = {3},
        pages = {3508-3531},
          doi = {10.1093/mnras/stz2731},
archivePrefix = {arXiv},
       eprint = {1907.09481},
 primaryClass = {astro-ph.GA},
       adsurl = {https://ui.adsabs.harvard.edu/abs/2019MNRAS.490.3508L}
}

@ARTICLE{APOGEE_2017,
       author = {{Majewski}, Steven R. and {Schiavon}, Ricardo P. and {Frinchaboy}, Peter M. and {Allende Prieto}, Carlos and {Barkhouser}, Robert and {Bizyaev}, Dmitry and {Blank}, Basil and {Brunner}, Sophia and {Burton}, Adam and {Carrera}, Ricardo and {Chojnowski}, S. Drew and {Cunha}, K{\'a}tia and {Epstein}, Courtney and {Fitzgerald}, Greg and {Garc{\'\i}a P{\'e}rez}, Ana E. and {Hearty}, Fred R. and {Henderson}, Chuck and {Holtzman}, Jon A. and {Johnson}, Jennifer A. and {Lam}, Charles R. and {Lawler}, James E. and {Maseman}, Paul and {M{\'e}sz{\'a}ros}, Szabolcs and {Nelson}, Matthew and {Nguyen}, Duy Coung and {Nidever}, David L. and {Pinsonneault}, Marc and {Shetrone}, Matthew and {Smee}, Stephen and {Smith}, Verne V. and {Stolberg}, Todd and {Skrutskie}, Michael F. and {Walker}, Eric and {Wilson}, John C. and {Zasowski}, Gail and {Anders}, Friedrich and {Basu}, Sarbani and {Beland}, Stephane and {Blanton}, Michael R. and {Bovy}, Jo and {Brownstein}, Joel R. and {Carlberg}, Joleen and {Chaplin}, William and {Chiappini}, Cristina and {Eisenstein}, Daniel J. and {Elsworth}, Yvonne and {Feuillet}, Diane and {Fleming}, Scott W. and {Galbraith-Frew}, Jessica and {Garc{\'\i}a}, Rafael A. and {Garc{\'\i}a-Hern{\'a}ndez}, D. An{\'\i}bal and {Gillespie}, Bruce A. and {Girardi}, L{\'e}o and {Gunn}, James E. and {Hasselquist}, Sten and {Hayden}, Michael R. and {Hekker}, Saskia and {Ivans}, Inese and {Kinemuchi}, Karen and {Klaene}, Mark and {Mahadevan}, Suvrath and {Mathur}, Savita and {Mosser}, Beno{\^\i}t and {Muna}, Demitri and {Munn}, Jeffrey A. and {Nichol}, Robert C. and {O'Connell}, Robert W. and {Parejko}, John K. and {Robin}, A.~C. and {Rocha-Pinto}, Helio and {Schultheis}, Matthias and {Serenelli}, Aldo M. and {Shane}, Neville and {Silva Aguirre}, Victor and {Sobeck}, Jennifer S. and {Thompson}, Benjamin and {Troup}, Nicholas W. and {Weinberg}, David H. and {Zamora}, Olga},
        title = "{The Apache Point Observatory Galactic Evolution Experiment (APOGEE)}",
      journal = {\aj},
         year = 2017,
        month = sep,
       volume = {154},
       number = {3},
          eid = {94},
        pages = {94},
          doi = {10.3847/1538-3881/aa784d},
archivePrefix = {arXiv},
       eprint = {1509.05420},
 primaryClass = {astro-ph.IM},
       adsurl = {https://ui.adsabs.harvard.edu/abs/2017AJ....154...94M}
}

@ARTICLE{Blumenthal_1984,
       author = {{Blumenthal}, G.~R. and {Faber}, S.~M. and {Primack}, J.~R. and {Rees}, M.~J.},
        title = "{Formation of galaxies and large-scale structure with cold dark matter.}",
      journal = {\nat},
         year = 1984,
        month = oct,
       volume = {311},
        pages = {517-525},
          doi = {10.1038/311517a0},
       adsurl = {https://ui.adsabs.harvard.edu/abs/1984Natur.311..517B}
}

@ARTICLE{Frenk_1988,
       author = {{Frenk}, Carlos S. and {White}, Simon D.~M. and {Davis}, Marc and {Efstathiou}, George},
        title = "{The Formation of Dark Halos in a Universe Dominated by Cold Dark Matter}",
      journal = {\apj},
         year = 1988,
        month = apr,
       volume = {327},
        pages = {507},
          doi = {10.1086/166213},
       adsurl = {https://ui.adsabs.harvard.edu/abs/1988ApJ...327..507F}
}

@ARTICLE{Planck_2020,
       author = {{Planck Collaboration} and {Aghanim}, N. and {Akrami}, Y. and {Ashdown}, M. and {Aumont}, J. and {Baccigalupi}, C. and {Ballardini}, M. and {Banday}, A.~J. and {Barreiro}, R.~B. and {Bartolo}, N. and {Basak}, S. and {Battye}, R. and {Benabed}, K. and {Bernard}, J. -P. and {Bersanelli}, M. and {Bielewicz}, P. and {Bock}, J.~J. and {Bond}, J.~R. and {Borrill}, J. and {Bouchet}, F.~R. and {Boulanger}, F. and {Bucher}, M. and {Burigana}, C. and {Butler}, R.~C. and {Calabrese}, E. and {Cardoso}, J. -F. and {Carron}, J. and {Challinor}, A. and {Chiang}, H.~C. and {Chluba}, J. and {Colombo}, L.~P.~L. and {Combet}, C. and {Contreras}, D. and {Crill}, B.~P. and {Cuttaia}, F. and {de Bernardis}, P. and {de Zotti}, G. and {Delabrouille}, J. and {Delouis}, J. -M. and {Di Valentino}, E. and {Diego}, J.~M. and {Dor{\'e}}, O. and {Douspis}, M. and {Ducout}, A. and {Dupac}, X. and {Dusini}, S. and {Efstathiou}, G. and {Elsner}, F. and {En{\ss}lin}, T.~A. and {Eriksen}, H.~K. and {Fantaye}, Y. and {Farhang}, M. and {Fergusson}, J. and {Fernandez-Cobos}, R. and {Finelli}, F. and {Forastieri}, F. and {Frailis}, M. and {Fraisse}, A.~A. and {Franceschi}, E. and {Frolov}, A. and {Galeotta}, S. and {Galli}, S. and {Ganga}, K. and {G{\'e}nova-Santos}, R.~T. and {Gerbino}, M. and {Ghosh}, T. and {Gonz{\'a}lez-Nuevo}, J. and {G{\'o}rski}, K.~M. and {Gratton}, S. and {Gruppuso}, A. and {Gudmundsson}, J.~E. and {Hamann}, J. and {Handley}, W. and {Hansen}, F.~K. and {Herranz}, D. and {Hildebrandt}, S.~R. and {Hivon}, E. and {Huang}, Z. and {Jaffe}, A.~H. and {Jones}, W.~C. and {Karakci}, A. and {Keih{\"a}nen}, E. and {Keskitalo}, R. and {Kiiveri}, K. and {Kim}, J. and {Kisner}, T.~S. and {Knox}, L. and {Krachmalnicoff}, N. and {Kunz}, M. and {Kurki-Suonio}, H. and {Lagache}, G. and {Lamarre}, J. -M. and {Lasenby}, A. and {Lattanzi}, M. and {Lawrence}, C.~R. and {Le Jeune}, M. and {Lemos}, P. and {Lesgourgues}, J. and {Levrier}, F. and {Lewis}, A. and {Liguori}, M. and {Lilje}, P.~B. and {Lilley}, M. and {Lindholm}, V. and {L{\'o}pez-Caniego}, M. and {Lubin}, P.~M. and {Ma}, Y. -Z. and {Mac{\'\i}as-P{\'e}rez}, J.~F. and {Maggio}, G. and {Maino}, D. and {Mandolesi}, N. and {Mangilli}, A. and {Marcos-Caballero}, A. and {Maris}, M. and {Martin}, P.~G. and {Martinelli}, M. and {Mart{\'\i}nez-Gonz{\'a}lez}, E. and {Matarrese}, S. and {Mauri}, N. and {McEwen}, J.~D. and {Meinhold}, P.~R. and {Melchiorri}, A. and {Mennella}, A. and {Migliaccio}, M. and {Millea}, M. and {Mitra}, S. and {Miville-Desch{\^e}nes}, M. -A. and {Molinari}, D. and {Montier}, L. and {Morgante}, G. and {Moss}, A. and {Natoli}, P. and {N{\o}rgaard-Nielsen}, H.~U. and {Pagano}, L. and {Paoletti}, D. and {Partridge}, B. and {Patanchon}, G. and {Peiris}, H.~V. and {Perrotta}, F. and {Pettorino}, V. and {Piacentini}, F. and {Polastri}, L. and {Polenta}, G. and {Puget}, J. -L. and {Rachen}, J.~P. and {Reinecke}, M. and {Remazeilles}, M. and {Renzi}, A. and {Rocha}, G. and {Rosset}, C. and {Roudier}, G. and {Rubi{\~n}o-Mart{\'\i}n}, J.~A. and {Ruiz-Granados}, B. and {Salvati}, L. and {Sandri}, M. and {Savelainen}, M. and {Scott}, D. and {Shellard}, E.~P.~S. and {Sirignano}, C. and {Sirri}, G. and {Spencer}, L.~D. and {Sunyaev}, R. and {Suur-Uski}, A. -S. and {Tauber}, J.~A. and {Tavagnacco}, D. and {Tenti}, M. and {Toffolatti}, L. and {Tomasi}, M. and {Trombetti}, T. and {Valenziano}, L. and {Valiviita}, J. and {Van Tent}, B. and {Vibert}, L. and {Vielva}, P. and {Villa}, F. and {Vittorio}, N. and {Wandelt}, B.~D. and {Wehus}, I.~K. and {White}, M. and {White}, S.~D.~M. and {Zacchei}, A. and {Zonca}, A.},
        title = "{Planck 2018 results. VI. Cosmological parameters}",
      journal = {\aap},
         year = 2020,
        month = sep,
       volume = {641},
          eid = {A6},
        pages = {A6},
          doi = {10.1051/0004-6361/201833910},
archivePrefix = {arXiv},
       eprint = {1807.06209},
 primaryClass = {astro-ph.CO},
       adsurl = {https://ui.adsabs.harvard.edu/abs/2020A&A...641A...6P}
}

@ARTICLE{Simon_geha_2007,
       author = {{Simon}, Joshua D. and {Geha}, Marla},
        title = "{The Kinematics of the Ultra-faint Milky Way Satellites: Solving the Missing Satellite Problem}",
      journal = {\apj},
         year = 2007,
        month = nov,
       volume = {670},
       number = {1},
        pages = {313-331},
          doi = {10.1086/521816},
archivePrefix = {arXiv},
       eprint = {0706.0516},
 primaryClass = {astro-ph},
       adsurl = {https://ui.adsabs.harvard.edu/abs/2007ApJ...670..313S}
}

@ARTICLE{Kaufman_white_1999,
       author = {{Kauffmann}, G. and {White}, S.~D.~M. and {Guiderdoni}, B.},
        title = "{The formation and evolution of galaxies within merging dark matter haloes.}",
      journal = {\mnras},
         year = 1993,
        month = sep,
       volume = {264},
        pages = {201-218},
          doi = {10.1093/mnras/264.1.201},
       adsurl = {https://ui.adsabs.harvard.edu/abs/1993MNRAS.264..201K}
}

@ARTICLE{Missing_satellites,
       author = {{Klypin}, Anatoly and {Kravtsov}, Andrey V. and {Valenzuela}, Octavio and {Prada}, Francisco},
        title = "{Where Are the Missing Galactic Satellites?}",
      journal = {\apj},
         year = 1999,
        month = sep,
       volume = {522},
       number = {1},
        pages = {82-92},
          doi = {10.1086/307643},
archivePrefix = {arXiv},
       eprint = {astro-ph/9901240},
 primaryClass = {astro-ph},
       adsurl = {https://ui.adsabs.harvard.edu/abs/1999ApJ...522...82K}
}

@ARTICLE{Kravtsov_2004,
       author = {{Kravtsov}, Andrey V. and {Gnedin}, Oleg Y. and {Klypin}, Anatoly A.},
        title = "{The Tumultuous Lives of Galactic Dwarfs and the Missing Satellites Problem}",
      journal = {\apj},
         year = 2004,
        month = jul,
       volume = {609},
       number = {2},
        pages = {482-497},
          doi = {10.1086/421322},
archivePrefix = {arXiv},
       eprint = {astro-ph/0401088},
 primaryClass = {astro-ph},
       adsurl = {https://ui.adsabs.harvard.edu/abs/2004ApJ...609..482K}
}

@ARTICLE{Law_Majewski_2010,
       author = {{Law}, David R. and {Majewski}, Steven R.},
        title = "{The Sagittarius Dwarf Galaxy: A Model for Evolution in a Triaxial Milky Way Halo}",
      journal = {\apj},
         year = 2010,
        month = may,
       volume = {714},
       number = {1},
        pages = {229-254},
          doi = {10.1088/0004-637X/714/1/229},
archivePrefix = {arXiv},
       eprint = {1003.1132},
 primaryClass = {astro-ph.GA},
       adsurl = {https://ui.adsabs.harvard.edu/abs/2010ApJ...714..229L}
}

@ARTICLE{paper_1,
       author = {{Ekanayaka}, Viraj and {Gireesh Babu}, Smrithi and {Oliver}, William H. and {Lewis}, Geraint F.},
        title = "{Streams and Shells Decoded: A Density-Driven Approach to Stellar Clustering in Galactic Halos with AstroLink}",
      journal = {arXiv e-prints},
         year = 2025,
        month = jul,
          eid = {arXiv:2507.22333},
        pages = {arXiv:2507.22333},
          doi = {10.48550/arXiv.2507.22333},
archivePrefix = {arXiv},
       eprint = {2507.22333},
 primaryClass = {astro-ph.GA},
       adsurl = {https://ui.adsabs.harvard.edu/abs/2025arXiv250722333E}
}

@ARTICLE{2024_Oliver,
       author = {{Oliver}, William H. and {Elahi}, Pascal J. and {Lewis}, Geraint F. and {Buck}, Tobias},
        title = "{The hierarchical structure of galactic haloes: differentiating clusters from stochastic clumping with ASTROLINK}",
      journal = {\mnras},
         year = 2024,
        month = may,
       volume = {530},
       number = {3},
        pages = {2637-2647},
          doi = {10.1093/mnras/stae1029},
archivePrefix = {arXiv},
       eprint = {2312.14632},
 primaryClass = {astro-ph.GA},
       adsurl = {https://ui.adsabs.harvard.edu/abs/2024MNRAS.530.2637O}
}

@ARTICLE{2021_springel,
       author = {{Springel}, Volker and {Pakmor}, R{\"u}diger and {Zier}, Oliver and {Reinecke}, Martin},
        title = "{Simulating cosmic structure formation with the GADGET-4 code}",
      journal = {\mnras},
         year = 2021,
        month = sep,
       volume = {506},
       number = {2},
        pages = {2871-2949},
          doi = {10.1093/mnras/stab1855},
archivePrefix = {arXiv},
       eprint = {2010.03567},
 primaryClass = {astro-ph.IM},
       adsurl = {https://ui.adsabs.harvard.edu/abs/2021MNRAS.506.2871S}
}

@ARTICLE{Wang_2011,
       author = {{Wang}, J. and {Navarro}, J.~F. and {Frenk}, C.~S. and {White}, S.~D.~M. and {Springel}, V. and {Jenkins}, A. and {Helmi}, A. and {Ludlow}, A. and {Vogelsberger}, M.},
        title = "{Assembly history and structure of galactic cold dark matter haloes}",
      journal = {\mnras},
         year = 2011,
        month = may,
       volume = {413},
       number = {2},
        pages = {1373-1382},
          doi = {10.1111/j.1365-2966.2011.18220.x},
archivePrefix = {arXiv},
       eprint = {1008.5114},
 primaryClass = {astro-ph.CO},
       adsurl = {https://ui.adsabs.harvard.edu/abs/2011MNRAS.413.1373W}
}

@ARTICLE{springel_2005,
       author = {{Springel}, Volker and {White}, Simon D.~M. and {Jenkins}, Adrian and {Frenk}, Carlos S. and {Yoshida}, Naoki and {Gao}, Liang and {Navarro}, Julio and {Thacker}, Robert and {Croton}, Darren and {Helly}, John and {Peacock}, John A. and {Cole}, Shaun and {Thomas}, Peter and {Couchman}, Hugh and {Evrard}, August and {Colberg}, J{\"o}rg and {Pearce}, Frazer},
        title = "{Simulations of the formation, evolution and clustering of galaxies and quasars}",
      journal = {\nat},
         year = 2005,
        month = jun,
       volume = {435},
       number = {7042},
        pages = {629-636},
          doi = {10.1038/nature03597},
archivePrefix = {arXiv},
       eprint = {astro-ph/0504097},
 primaryClass = {astro-ph},
       adsurl = {https://ui.adsabs.harvard.edu/abs/2005Natur.435..629S}
}

@ARTICLE{Frenk_white_2012,
       author = {{Frenk}, C.~S. and {White}, S.~D.~M.},
        title = "{Dark matter and cosmic structure}",
      journal = {Annalen der Physik},
         year = 2012,
        month = oct,
       volume = {524},
       number = {9-10},
        pages = {507-534},
          doi = {10.1002/andp.201200212},
archivePrefix = {arXiv},
       eprint = {1210.0544},
 primaryClass = {astro-ph.CO},
       adsurl = {https://ui.adsabs.harvard.edu/abs/2012AnP...524..507F}
}

@ARTICLE{Springel_2008,
       author = {{Springel}, V. and {Wang}, J. and {Vogelsberger}, M. and {Ludlow}, A. and {Jenkins}, A. and {Helmi}, A. and {Navarro}, J.~F. and {Frenk}, C.~S. and {White}, S.~D.~M.},
        title = "{The Aquarius Project: the subhaloes of galactic haloes}",
      journal = {\mnras},
         year = 2008,
        month = dec,
       volume = {391},
       number = {4},
        pages = {1685-1711},
          doi = {10.1111/j.1365-2966.2008.14066.x},
archivePrefix = {arXiv},
       eprint = {0809.0898},
 primaryClass = {astro-ph},
       adsurl = {https://ui.adsabs.harvard.edu/abs/2008MNRAS.391.1685S}
}

@ARTICLE{Navarro_NFW_1997,
       author = {{Navarro}, Julio F. and {Frenk}, Carlos S. and {White}, Simon D.~M.},
        title = "{A Universal Density Profile from Hierarchical Clustering}",
      journal = {\apj},
         year = 1997,
        month = dec,
       volume = {490},
       number = {2},
        pages = {493-508},
          doi = {10.1086/304888},
archivePrefix = {arXiv},
       eprint = {astro-ph/9611107},
 primaryClass = {astro-ph},
       adsurl = {https://ui.adsabs.harvard.edu/abs/1997ApJ...490..493N}
}

@ARTICLE{Power_2003,
       author = {{Power}, C. and {Navarro}, J.~F. and {Jenkins}, A. and {Frenk}, C.~S. and {White}, S.~D.~M. and {Springel}, V. and {Stadel}, J. and {Quinn}, T.},
        title = "{The inner structure of {\ensuremath{\Lambda}}CDM haloes - I. A numerical convergence study}",
      journal = {\mnras},
         year = 2003,
        month = jan,
       volume = {338},
       number = {1},
        pages = {14-34},
          doi = {10.1046/j.1365-8711.2003.05925.x},
archivePrefix = {arXiv},
       eprint = {astro-ph/0201544},
 primaryClass = {astro-ph},
       adsurl = {https://ui.adsabs.harvard.edu/abs/2003MNRAS.338...14P}
}

@ARTICLE{Frenk_2004,
       author = {{Navarro}, J.~F. and {Hayashi}, E. and {Power}, C. and {Jenkins}, A.~R. and {Frenk}, C.~S. and {White}, S.~D.~M. and {Springel}, V. and {Stadel}, J. and {Quinn}, T.~R.},
        title = "{The inner structure of {\ensuremath{\Lambda}}CDM haloes - III. Universality and asymptotic slopes}",
      journal = {\mnras},
         year = 2004,
        month = apr,
       volume = {349},
       number = {3},
        pages = {1039-1051},
          doi = {10.1111/j.1365-2966.2004.07586.x},
archivePrefix = {arXiv},
       eprint = {astro-ph/0311231},
 primaryClass = {astro-ph},
       adsurl = {https://ui.adsabs.harvard.edu/abs/2004MNRAS.349.1039N}
}

@ARTICLE{Reino_2021,
       author = {{Reino}, Stella and {Rossi}, Elena M. and {Sanderson}, Robyn E. and {Sellentin}, Elena and {Helmi}, Amina and {Koppelman}, Helmer H. and {Sharma}, Sanjib},
        title = "{Galactic potential constraints from clustering in action space of combined stellar stream data}",
      journal = {\mnras},
         year = 2021,
        month = apr,
       volume = {502},
       number = {3},
        pages = {4170-4193},
          doi = {10.1093/mnras/stab304},
archivePrefix = {arXiv},
       eprint = {2007.00356},
 primaryClass = {astro-ph.GA},
       adsurl = {https://ui.adsabs.harvard.edu/abs/2021MNRAS.502.4170R}
}

@ARTICLE{Bullock_2001,
       author = {{Bullock}, J.~S. and {Kolatt}, T.~S. and {Sigad}, Y. and {Somerville}, R.~S. and {Kravtsov}, A.~V. and {Klypin}, A.~A. and {Primack}, J.~R. and {Dekel}, A.},
        title = "{Profiles of dark haloes: evolution, scatter and environment}",
      journal = {\mnras},
         year = 2001,
        month = mar,
       volume = {321},
       number = {3},
        pages = {559-575},
          doi = {10.1046/j.1365-8711.2001.04068.x},
archivePrefix = {arXiv},
       eprint = {astro-ph/9908159},
 primaryClass = {astro-ph},
       adsurl = {https://ui.adsabs.harvard.edu/abs/2001MNRAS.321..559B}
}

@article{Hernquist_1990,
    author = {{Hernquist}, L.},
    title = "{An analytical model for spherical galaxies and bulges}",
    journal = {The Astrophysical Journal},
    volume = {356},
    pages = {359},
    year = {1990},
    doi = {10.1086/168846},
    adsurl = {https://ui.adsabs.harvard.edu/abs/1990ApJ...356..359H},
}

@article{Miyamoto_Nagai_1975,
    author = {Miyamoto, M. and Nagai, R.},
    title = "{Three-dimensional models for the distribution of mass in galaxies}",
    journal = {Publications of the Astronomical Society of Japan},
    volume = {27},
    pages = {533},
    year = {1975},
    doi = {10.1093/pasj/27.4.533},
    url = {https://ui.adsabs.harvard.edu/abs/1975PASJ...27..533M},
}

@ARTICLE{Bland-Hawthorn_2016,
       author = {{Bland-Hawthorn}, Joss and {Gerhard}, Ortwin},
        title = "{The Galaxy in Context: Structural, Kinematic, and Integrated Properties}",
      journal = {\araa},
         year = 2016,
        month = sep,
       volume = {54},
        pages = {529-596},
          doi = {10.1146/annurev-astro-081915-023441},
archivePrefix = {arXiv},
       eprint = {1602.07702},
 primaryClass = {astro-ph.GA},
       adsurl = {https://ui.adsabs.harvard.edu/abs/2016ARA&A..54..529B}
}

@ARTICLE{NFW_1995,
       author = {{Navarro}, Julio F. and {Frenk}, Carlos S. and {White}, Simon D.~M.},
        title = "{The assembly of galaxies in a hierarchically clustering universe}",
      journal = {\mnras},
         year = 1995,
        month = jul,
       volume = {275},
       number = {1},
        pages = {56-66},
          doi = {10.1093/mnras/275.1.56},
archivePrefix = {arXiv},
       eprint = {astro-ph/9408067},
 primaryClass = {astro-ph},
       adsurl = {https://ui.adsabs.harvard.edu/abs/1995MNRAS.275...56N}
}

@ARTICLE{Ibata_2001b,
       author = {{Ibata}, Rodrigo and {Lewis}, Geraint F. and {Irwin}, Michael and {Totten}, Edward and {Quinn}, Thomas},
        title = "{Great Circle Tidal Streams: Evidence for a Nearly Spherical Massive Dark Halo around the Milky Way}",
      journal = {\apj},
         year = 2001,
        month = apr,
       volume = {551},
       number = {1},
        pages = {294-311},
          doi = {10.1086/320060},
archivePrefix = {arXiv},
       eprint = {astro-ph/0004011},
 primaryClass = {astro-ph},
       adsurl = {https://ui.adsabs.harvard.edu/abs/2001ApJ...551..294I}
}

@ARTICLE{Helmi_2004,
       author = {{Helmi}, Amina},
        title = "{Velocity Trends in the Debris of Sagittarius and the Shape of the Dark Matter Halo of Our Galaxy}",
      journal = {\apjl},
         year = 2004,
        month = aug,
       volume = {610},
       number = {2},
        pages = {L97-L100},
          doi = {10.1086/423340},
archivePrefix = {arXiv},
       eprint = {astro-ph/0406396},
 primaryClass = {astro-ph},
       adsurl = {https://ui.adsabs.harvard.edu/abs/2004ApJ...610L..97H}
}

@ARTICLE{kupper_2015,
       author = {{K{\"u}pper}, Andreas H.~W. and {Balbinot}, Eduardo and {Bonaca}, Ana and {Johnston}, Kathryn V. and {Hogg}, David W. and {Kroupa}, Pavel and {Santiago}, Basilio X.},
        title = "{Globular Cluster Streams as Galactic High-Precision Scales{\textemdash}the Poster Child Palomar 5}",
      journal = {\apj},
         year = 2015,
        month = apr,
       volume = {803},
       number = {2},
          eid = {80},
        pages = {80},
          doi = {10.1088/0004-637X/803/2/80},
archivePrefix = {arXiv},
       eprint = {1502.02658},
 primaryClass = {astro-ph.GA},
       adsurl = {https://ui.adsabs.harvard.edu/abs/2015ApJ...803...80K}
}

@ARTICLE{Nibauer_2025,
       author = {{Nibauer}, Jacob and {Bonaca}, Ana},
        title = "{Galactic Accelerations from the GD-1 Stream Suggest a Tilted Dark Matter Halo}",
      journal = {\apjl},
         year = 2025,
        month = may,
       volume = {985},
       number = {1},
          eid = {L22},
        pages = {L22},
          doi = {10.3847/2041-8213/add0a9},
archivePrefix = {arXiv},
       eprint = {2504.07187},
 primaryClass = {astro-ph.GA},
       adsurl = {https://ui.adsabs.harvard.edu/abs/2025ApJ...985L..22N}
}

@ARTICLE{Bovy_2016,
       author = {{Bovy}, Jo and {Bahmanyar}, Anita and {Fritz}, Tobias K. and {Kallivayalil}, Nitya},
        title = "{The Shape of the Inner Milky Way Halo from Observations of the Pal 5 and GD--1 Stellar Streams}",
      journal = {\apj},
         year = 2016,
        month = dec,
       volume = {833},
       number = {1},
          eid = {31},
        pages = {31},
          doi = {10.3847/1538-4357/833/1/31},
archivePrefix = {arXiv},
       eprint = {1609.01298},
 primaryClass = {astro-ph.GA},
       adsurl = {https://ui.adsabs.harvard.edu/abs/2016ApJ...833...31B}
}

@ARTICLE{Koposov_2023,
       author = {{Koposov}, Sergey E. and {Erkal}, Denis and {Li}, Ting S. and {Da Costa}, Gary S. and {Cullinane}, Lara R. and {Ji}, Alexander P. and {Kuehn}, Kyler and {Lewis}, Geraint F. and {Pace}, Andrew B. and {Shipp}, Nora and {Zucker}, Daniel B. and {Bland-Hawthorn}, Joss and {Lilleengen}, Sophia and {Martell}, Sarah L. and {S5 Collaboration}},
        title = "{S $^{5}$: Probing the Milky Way and Magellanic Clouds potentials with the 6D map of the Orphan-Chenab stream}",
      journal = {\mnras},
         year = 2023,
        month = jun,
       volume = {521},
       number = {4},
        pages = {4936-4962},
          doi = {10.1093/mnras/stad551},
archivePrefix = {arXiv},
       eprint = {2211.04495},
 primaryClass = {astro-ph.GA},
       adsurl = {https://ui.adsabs.harvard.edu/abs/2023MNRAS.521.4936K}
}

@ARTICLE{Bonaca_2018,
       author = {{Bonaca}, Ana and {Hogg}, David W.},
        title = "{The Information Content in Cold Stellar Streams}",
      journal = {\apj},
         year = 2018,
        month = nov,
       volume = {867},
       number = {2},
          eid = {101},
        pages = {101},
          doi = {10.3847/1538-4357/aae4da},
archivePrefix = {arXiv},
       eprint = {1804.06854},
 primaryClass = {astro-ph.GA},
       adsurl = {https://ui.adsabs.harvard.edu/abs/2018ApJ...867..101B}
}

@ARTICLE{Koposov_2010,
       author = {{Koposov}, Sergey E. and {Rix}, Hans-Walter and {Hogg}, David W.},
        title = "{Constraining the Milky Way Potential with a Six-Dimensional Phase-Space Map of the GD-1 Stellar Stream}",
      journal = {\apj},
         year = 2010,
        month = mar,
       volume = {712},
       number = {1},
        pages = {260-273},
          doi = {10.1088/0004-637X/712/1/260},
archivePrefix = {arXiv},
       eprint = {0907.1085},
 primaryClass = {astro-ph.GA},
       adsurl = {https://ui.adsabs.harvard.edu/abs/2010ApJ...712..260K}
}

@ARTICLE{Merrifield_1998,
       author = {{Merrifield}, Michael R. and {Kuijken}, Konrad},
        title = "{Measuring galaxy potentials using shell kinematics}",
      journal = {\mnras},
         year = 1998,
        month = jul,
       volume = {297},
       number = {4},
        pages = {1292-1296},
          doi = {10.1046/j.1365-8711.1998.01625.x},
archivePrefix = {arXiv},
       eprint = {astro-ph/9803053},
 primaryClass = {astro-ph},
       adsurl = {https://ui.adsabs.harvard.edu/abs/1998MNRAS.297.1292M}
}

@ARTICLE{Fardal_2007,
       author = {{Fardal}, M.~A. and {Guhathakurta}, P. and {Babul}, A. and {McConnachie}, A.~W.},
        title = "{Investigating the Andromeda stream - III. A young shell system in M31}",
      journal = {\mnras},
         year = 2007,
        month = sep,
       volume = {380},
       number = {1},
        pages = {15-32},
          doi = {10.1111/j.1365-2966.2007.11929.x},
archivePrefix = {arXiv},
       eprint = {astro-ph/0609050},
 primaryClass = {astro-ph},
       adsurl = {https://ui.adsabs.harvard.edu/abs/2007MNRAS.380...15F}
}

@ARTICLE{Bilek_2022,
       author = {{B{\'\i}lek}, Michal and {Fensch}, J{\'e}r{\'e}my and {Ebrov{\'a}}, Ivana and {Nagesh}, Srikanth T. and {Famaey}, Benoit and {Duc}, Pierre-Alain and {Kroupa}, Pavel},
        title = "{Origin of the spectacular tidal shells of galaxy NGC 474}",
      journal = {\aap},
         year = 2022,
        month = apr,
       volume = {660},
          eid = {A28},
        pages = {A28},
          doi = {10.1051/0004-6361/202141709},
archivePrefix = {arXiv},
       eprint = {2111.14886},
 primaryClass = {astro-ph.GA},
       adsurl = {https://ui.adsabs.harvard.edu/abs/2022A&A...660A..28B}
}

@ARTICLE{Fensch_2020,
       author = {{Fensch}, J{\'e}r{\'e}my and {Duc}, Pierre-Alain and {Lim}, Sungsoon and {Emsellem}, {\'E}ric and {B{\'\i}lek}, Michal and {Durrell}, Patrick and {Liu}, Chengze and {Peng}, {\'E}ric and {Smith}, Rory},
        title = "{Shedding light on the formation mechanism of shell galaxy NGC 474 with MUSE}",
      journal = {\aap},
         year = 2020,
        month = dec,
       volume = {644},
          eid = {A164},
        pages = {A164},
          doi = {10.1051/0004-6361/202038550},
archivePrefix = {arXiv},
       eprint = {2007.03318},
 primaryClass = {astro-ph.GA},
       adsurl = {https://ui.adsabs.harvard.edu/abs/2020A&A...644A.164F}
}

@ARTICLE{Belokurov_2018,
       author = {{Belokurov}, V. and {Erkal}, D. and {Evans}, N.~W. and {Koposov}, S.~E. and {Deason}, A.~J.},
        title = "{Co-formation of the disc and the stellar halo}",
      journal = {\mnras},
         year = 2018,
        month = jul,
       volume = {478},
       number = {1},
        pages = {611-619},
          doi = {10.1093/mnras/sty982},
archivePrefix = {arXiv},
       eprint = {1802.03414},
 primaryClass = {astro-ph.GA},
       adsurl = {https://ui.adsabs.harvard.edu/abs/2018MNRAS.478..611B}
}

@ARTICLE{Helmi_2018,
       author = {{Helmi}, Amina and {Babusiaux}, Carine and {Koppelman}, Helmer H. and {Massari}, Davide and {Veljanoski}, Jovan and {Brown}, Anthony G.~A.},
        title = "{The merger that led to the formation of the Milky Way's inner stellar halo and thick disk}",
      journal = {\nat},
         year = 2018,
        month = oct,
       volume = {563},
       number = {7729},
        pages = {85-88},
          doi = {10.1038/s41586-018-0625-x},
archivePrefix = {arXiv},
       eprint = {1806.06038},
 primaryClass = {astro-ph.GA},
       adsurl = {https://ui.adsabs.harvard.edu/abs/2018Natur.563...85H}
}

@ARTICLE{Donlon_2020,
       author = {{Donlon}, II, Thomas and {Newberg}, Heidi Jo and {Sanderson}, Robyn and {Widrow}, Lawrence M.},
        title = "{The Milky Way's Shell Structure Reveals the Time of a Radial Collision}",
      journal = {\apj},
         year = 2020,
        month = oct,
       volume = {902},
       number = {2},
          eid = {119},
        pages = {119},
          doi = {10.3847/1538-4357/abb5f6},
archivePrefix = {arXiv},
       eprint = {2006.08764},
 primaryClass = {astro-ph.GA},
       adsurl = {https://ui.adsabs.harvard.edu/abs/2020ApJ...902..119D}
}

@ARTICLE{Donlon_2022,
       author = {{Donlon}, II, Thomas and {Newberg}, Heidi Jo and {Kim}, Bokyoung and {L{\'e}pine}, Sebastien},
        title = "{The Local Stellar Halo is Not Dominated by a Single Radial Merger Event}",
      journal = {\apjl},
         year = 2022,
        month = jun,
       volume = {932},
       number = {2},
          eid = {L16},
        pages = {L16},
          doi = {10.3847/2041-8213/ac7531},
archivePrefix = {arXiv},
       eprint = {2110.11465},
 primaryClass = {astro-ph.GA},
       adsurl = {https://ui.adsabs.harvard.edu/abs/2022ApJ...932L..16D}
}

@article{Pearson2015,
  author = {Pearson, Sarah and Fardal, Mark A. and Bovy, Jo},
  title = {Streams evolving on chaotic orbits in nonspherical potentials can evolve into large, diffuse “fans”},
  journal = {ResearchGate or ApJ reference},
  year = {2015}
}

@article{PriceWhelan2016,
  author = {Price-Whelan, Adrian M. and others},
  title = {Morphology of streams as signatures of resonances in nonspherical potentials},
  journal = {MNRAS or arXiv reference},
  year = {2016}
}

@article{Yavetz2020,
  author = {Yavetz, Tomer D. and Johnston, Kathryn V. and Pearson, Sarah and Price-Whelan, Adrian M. and Weinberg, Martin D.},
  title = {Separatrix Divergence of Stellar Streams in Galactic Potentials},
  journal = {arXiv},
  year = {2020}
}

@article{Brooks2024,
  author = {Brooks, Richard A. N. and Sanders, Jason L. and Lilleengen, Sophia and Petersen, Michael S. and Pontzen, Andrew},
  title = {Action and energy clustering of stellar streams in deforming Milky Way dark matter haloes},
  journal = {MNRAS},
  year = {2024}
}

@article{BuistHelmi2015,
  author = {Buist, Hans J. T. and Helmi, Amina},
  title = {The evolution of streams in a time-dependent potential},
  journal = {A\&A},
  volume = {584},
  pages = {A120},
  year = {2015}
}

@ARTICLE{Zhu_2025,
       author = {{Zhu}, Ling and {Cai}, Runsheng and {Kang}, Xi and {Xue}, Xiang-Xiang and {Yang}, Chengqun and {Zhang}, Lan and {Mao}, Shude and {Liu}, Chao},
        title = "{A Vertically Orientated Dark Matter Halo Marks a Flip of the Galactic Disk}",
      journal = {arXiv e-prints},
         year = 2025,
        month = oct,
          eid = {arXiv:2510.08684},
        pages = {arXiv:2510.08684},
          doi = {10.48550/arXiv.2510.08684},
archivePrefix = {arXiv},
       eprint = {2510.08684},
 primaryClass = {astro-ph.GA},
       adsurl = {https://ui.adsabs.harvard.edu/abs/2025arXiv251008684Z}
}

@ARTICLE{1991_Thomson,
       author = {{Thomson}, R.~C.},
        title = "{Shell formation in elliptical galaxies.}",
      journal = {\mnras},
         year = 1991,
        month = nov,
       volume = {253},
        pages = {256},
          doi = {10.1093/mnras/253.2.256},
       adsurl = {https://ui.adsabs.harvard.edu/abs/1991MNRAS.253..256T}
}

@ARTICLE{2009M_Choi,
       author = {{Choi}, Jun-Hwan and {Weinberg}, Martin D. and {Katz}, Neal},
        title = "{The dynamics of satellite disruption in cold dark matter haloes}",
      journal = {\mnras},
         year = 2009,
        month = dec,
       volume = {400},
       number = {3},
        pages = {1247-1263},
          doi = {10.1111/j.1365-2966.2009.15556.x},
archivePrefix = {arXiv},
       eprint = {0812.0009},
 primaryClass = {astro-ph},
       adsurl = {https://ui.adsabs.harvard.edu/abs/2009MNRAS.400.1247C}
}

@ARTICLE{2025_Guillaume,
       author = {{Guillaume}, Claire and {Renaud}, Florent and {Martin}, Nicolas F. and {Famaey}, Benoit and {Di Matteo}, Paola and {Thomas}, Guillaume F. and {Ferrone}, Salvatore and {Ibata}, Rodrigo and {Pagnini}, Giulia},
        title = "{Asymmetries in stellar streams induced by a galactic merger}",
      journal = {arXiv e-prints},
         year = 2025,
        month = oct,
          eid = {arXiv:2510.06329},
        pages = {arXiv:2510.06329},
          doi = {10.48550/arXiv.2510.06329},
archivePrefix = {arXiv},
       eprint = {2510.06329},
 primaryClass = {astro-ph.GA},
       adsurl = {https://ui.adsabs.harvard.edu/abs/2025arXiv251006329G}
}

@ARTICLE{Bariego_2024,
       author = {{Bariego-Quintana}, Adriana and {Llanes-Estrada}, Felipe J.},
        title = "{The torsion of stellar streams and the overall shape of galactic gravity's source}",
      journal = {\aap},
         year = 2024,
        month = jul,
       volume = {687},
          eid = {A46},
        pages = {A46},
          doi = {10.1051/0004-6361/202347502},
archivePrefix = {arXiv},
       eprint = {2307.07402},
 primaryClass = {astro-ph.GA},
       adsurl = {https://ui.adsabs.harvard.edu/abs/2024A&A...687A..46B}
}

@ARTICLE{2025_Weerasooriya,
       author = {{Weerasooriya}, Sachi and {Starkenburg}, Tjitske and {Cunningham}, Emily C. and {Johnston}, Kathryn V},
        title = "{Dancing Streams In Merging Halos: Stellar Streams in a MW--LMC-like merger}",
      journal = {arXiv e-prints},
         year = 2025,
        month = may,
          eid = {arXiv:2505.14792},
        pages = {arXiv:2505.14792},
          doi = {10.48550/arXiv.2505.14792},
archivePrefix = {arXiv},
       eprint = {2505.14792},
 primaryClass = {astro-ph.GA},
       adsurl = {https://ui.adsabs.harvard.edu/abs/2025arXiv250514792W}
}

@ARTICLE{2011_wetzel,
       author = {{Wetzel}, Andrew R.},
        title = "{On the orbits of infalling satellite haloes}",
      journal = {\mnras},
         year = 2011,
        month = mar,
       volume = {412},
       number = {1},
        pages = {49-58},
          doi = {10.1111/j.1365-2966.2010.17877.x},
archivePrefix = {arXiv},
       eprint = {1001.4792},
 primaryClass = {astro-ph.CO},
       adsurl = {https://ui.adsabs.harvard.edu/abs/2011MNRAS.412...49W}
}

@ARTICLE{2018_Bosch,
       author = {{van den Bosch}, Frank C. and {Ogiya}, Go and {Hahn}, Oliver and {Burkert}, Andreas},
        title = "{Disruption of dark matter substructure: fact or fiction?}",
      journal = {\mnras},
         year = 2018,
        month = mar,
       volume = {474},
       number = {3},
        pages = {3043-3066},
          doi = {10.1093/mnras/stx2956},
archivePrefix = {arXiv},
       eprint = {1711.05276},
 primaryClass = {astro-ph.GA},
       adsurl = {https://ui.adsabs.harvard.edu/abs/2018MNRAS.474.3043V}
}

@ARTICLE{Aarseth_1974,
       author = {{Aarseth}, S.~J. and {Henon}, M. and {Wielen}, R.},
        title = "{A Comparison of Numerical Methods for the Study of Star Cluster Dynamics}",
      journal = {\aap},
         year = 1974,
        month = dec,
       volume = {37},
       number = {1},
        pages = {183-187},
       adsurl = {https://ui.adsabs.harvard.edu/abs/1974A&A....37..183A}
}

@INPROCEEDINGS{Arrakhis,
       author = {{Guzman}, Rafael and {G{\'o}mez-Flechoso}, Mari{\'a}ngeles and {Mart{\'\i}nez-Delgado}, David and {Roca-F{\'a}brega}, Santiago and {Serrano}, Santiago and {Diego}, Jose-Maria and {Torrey}, Paul and {Camaz{\'o}n-Pinilla}, Alejandro and {Serrano-Borlaff}, Alejandro and {Moore}, Ben and {Agertz}, Oscar and {Guedel}, Manuel},
        title = "{The DUNES and ARRAKHIS space missions to study Dark Matter}",
    booktitle = {EAS2022, European Astronomical Society Annual Meeting},
         year = 2022,
        month = jul,
          eid = {1507},
        pages = {1507},
       adsurl = {https://ui.adsabs.harvard.edu/abs/2022eas..conf.1507G}
}

@ARTICLE{nancy_g_roman,
       author = {{Spergel}, D. and {Gehrels}, N. and {Baltay}, C. and {Bennett}, D. and {Breckinridge}, J. and {Donahue}, M. and {Dressler}, A. and {Gaudi}, B.~S. and {Greene}, T. and {Guyon}, O. and {Hirata}, C. and {Kalirai}, J. and {Kasdin}, N.~J. and {Macintosh}, B. and {Moos}, W. and {Perlmutter}, S. and {Postman}, M. and {Rauscher}, B. and {Rhodes}, J. and {Wang}, Y. and {Weinberg}, D. and {Benford}, D. and {Hudson}, M. and {Jeong}, W. -S. and {Mellier}, Y. and {Traub}, W. and {Yamada}, T. and {Capak}, P. and {Colbert}, J. and {Masters}, D. and {Penny}, M. and {Savransky}, D. and {Stern}, D. and {Zimmerman}, N. and {Barry}, R. and {Bartusek}, L. and {Carpenter}, K. and {Cheng}, E. and {Content}, D. and {Dekens}, F. and {Demers}, R. and {Grady}, K. and {Jackson}, C. and {Kuan}, G. and {Kruk}, J. and {Melton}, M. and {Nemati}, B. and {Parvin}, B. and {Poberezhskiy}, I. and {Peddie}, C. and {Ruffa}, J. and {Wallace}, J.~K. and {Whipple}, A. and {Wollack}, E. and {Zhao}, F.},
        title = "{Wide-Field InfrarRed Survey Telescope-Astrophysics Focused Telescope Assets WFIRST-AFTA 2015 Report}",
      journal = {arXiv e-prints},
         year = 2015,
        month = mar,
          eid = {arXiv:1503.03757},
        pages = {arXiv:1503.03757},
          doi = {10.48550/arXiv.1503.03757},
archivePrefix = {arXiv},
       eprint = {1503.03757},
 primaryClass = {astro-ph.IM},
       adsurl = {https://ui.adsabs.harvard.edu/abs/2015arXiv150303757S}
}

@ARTICLE{1999_Bosch,
       author = {{van den Bosch}, Frank C. and {Lewis}, Geraint F. and {Lake}, George and {Stadel}, Joachim},
        title = "{Substructure in Dark Halos: Orbital Eccentricities and Dynamical Friction}",
      journal = {\apj},
         year = 1999,
        month = apr,
       volume = {515},
       number = {1},
        pages = {50-68},
          doi = {10.1086/307023},
archivePrefix = {arXiv},
       eprint = {astro-ph/9811229},
 primaryClass = {astro-ph},
       adsurl = {https://ui.adsabs.harvard.edu/abs/1999ApJ...515...50V}
}

@INPROCEEDINGS{2003_suto,
       author = {{Suto}, Y.},
        title = "{Density Profiles and Clustering of Dark Halos and Clusters of Galaxies}",
    booktitle = {Matter and Energy in Clusters of Galaxies},
         year = 2003,
       editor = {{Bowyer}, Stuart and {Hwang}, Chorng-Yuan},
       series = {Astronomical Society of the Pacific Conference Series},
       volume = {301},
        month = jan,
        pages = {379},
          doi = {10.48550/arXiv.astro-ph/0207202},
archivePrefix = {arXiv},
       eprint = {astro-ph/0207202},
 primaryClass = {astro-ph},
       adsurl = {https://ui.adsabs.harvard.edu/abs/2003ASPC..301..379S}
}

@ARTICLE{2000_Evans,
       author = {{Evans}, N.~W. and {Carollo}, C.~M. and {de Zeeuw}, P.~T.},
        title = "{Triaxial haloes and particle dark matter detection}",
      journal = {\mnras},
         year = 2000,
        month = nov,
       volume = {318},
       number = {4},
        pages = {1131-1143},
          doi = {10.1046/j.1365-8711.2000.03787.x},
archivePrefix = {arXiv},
       eprint = {astro-ph/0008156},
 primaryClass = {astro-ph},
       adsurl = {https://ui.adsabs.harvard.edu/abs/2000MNRAS.318.1131E}
}

@ARTICLE{Hendel_2015,
       author = {{Hendel}, David and {Johnston}, Kathryn V.},
        title = "{Tidal debris morphology and the orbits of satellite galaxies}",
      journal = {\mnras},
         year = 2015,
        month = dec,
       volume = {454},
       number = {3},
        pages = {2472-2485},
          doi = {10.1093/mnras/stv2035},
archivePrefix = {arXiv},
       eprint = {1509.06369},
 primaryClass = {astro-ph.GA},
       adsurl = {https://ui.adsabs.harvard.edu/abs/2015MNRAS.454.2472H}
}

@ARTICLE{Dehnen_2018,
       author = {{Dehnen}, Walter and {Hasanuddin}},
        title = "{Tidal ribbons}",
      journal = {\mnras},
         year = 2018,
        month = oct,
       volume = {479},
       number = {4},
        pages = {4720-4726},
          doi = {10.1093/mnras/sty1726},
archivePrefix = {arXiv},
       eprint = {1805.08481},
 primaryClass = {astro-ph.GA},
       adsurl = {https://ui.adsabs.harvard.edu/abs/2018MNRAS.479.4720D}
}

@ARTICLE{Nibauer_2023,
       author = {{Nibauer}, Jacob and {Bonaca}, Ana and {Johnston}, Kathryn V.},
        title = "{Constraining the Gravitational Potential from the Projected Morphology of Extragalactic Tidal Streams}",
      journal = {\apj},
         year = 2023,
        month = sep,
       volume = {954},
       number = {2},
          eid = {195},
        pages = {195},
          doi = {10.3847/1538-4357/ace9bc},
archivePrefix = {arXiv},
       eprint = {2303.17406},
 primaryClass = {astro-ph.GA},
       adsurl = {https://ui.adsabs.harvard.edu/abs/2023ApJ...954..195N}
}

@ARTICLE{Chemaly_2026,
       author = {{Chemaly}, David and {Sola}, Elisabeth and {Belokurov}, Vasily and {Koposov}, Sergey and {Meyong}, GyuChul and {Zhang}, HanYuan and {Erkal}, Denis},
        title = "{Hierarchical bayesian inference: constraining population distribution of dark matter halo shapes via stellar streams}",
      journal = {arXiv e-prints},
         year = 2026,
        month = jan,
          eid = {arXiv:2601.15373},
        pages = {arXiv:2601.15373},
          doi = {10.48550/arXiv.2601.15373},
archivePrefix = {arXiv},
       eprint = {2601.15373},
 primaryClass = {astro-ph.GA},
       adsurl = {https://ui.adsabs.harvard.edu/abs/2026arXiv260115373C}
}

@ARTICLE{Walder_2025,
       author = {{Walder}, Madison and {Erkal}, Denis and {Collins}, Michelle and {Mart{\'\i}nez-Delgado}, David},
        title = "{Probing the Dark Matter Halos of External Galaxies with Stellar Streams}",
      journal = {\apj},
         year = 2025,
        month = nov,
       volume = {994},
       number = {1},
          eid = {36},
        pages = {36},
          doi = {10.3847/1538-4357/adfa1f},
archivePrefix = {arXiv},
       eprint = {2402.13314},
 primaryClass = {astro-ph.GA},
       adsurl = {https://ui.adsabs.harvard.edu/abs/2025ApJ...994...36W}
}



\appendix
\section{Density Projections of Subhalos in Multiple Projections}
\label{Density Projections of Subhalos in multiple projections}

\begin{figure*}
\centering
    \setlength{\tabcolsep}{1pt} 
    \renewcommand{\arraystretch}{0} 
\includegraphics[width=0.9\textwidth, height=.5\textwidth]{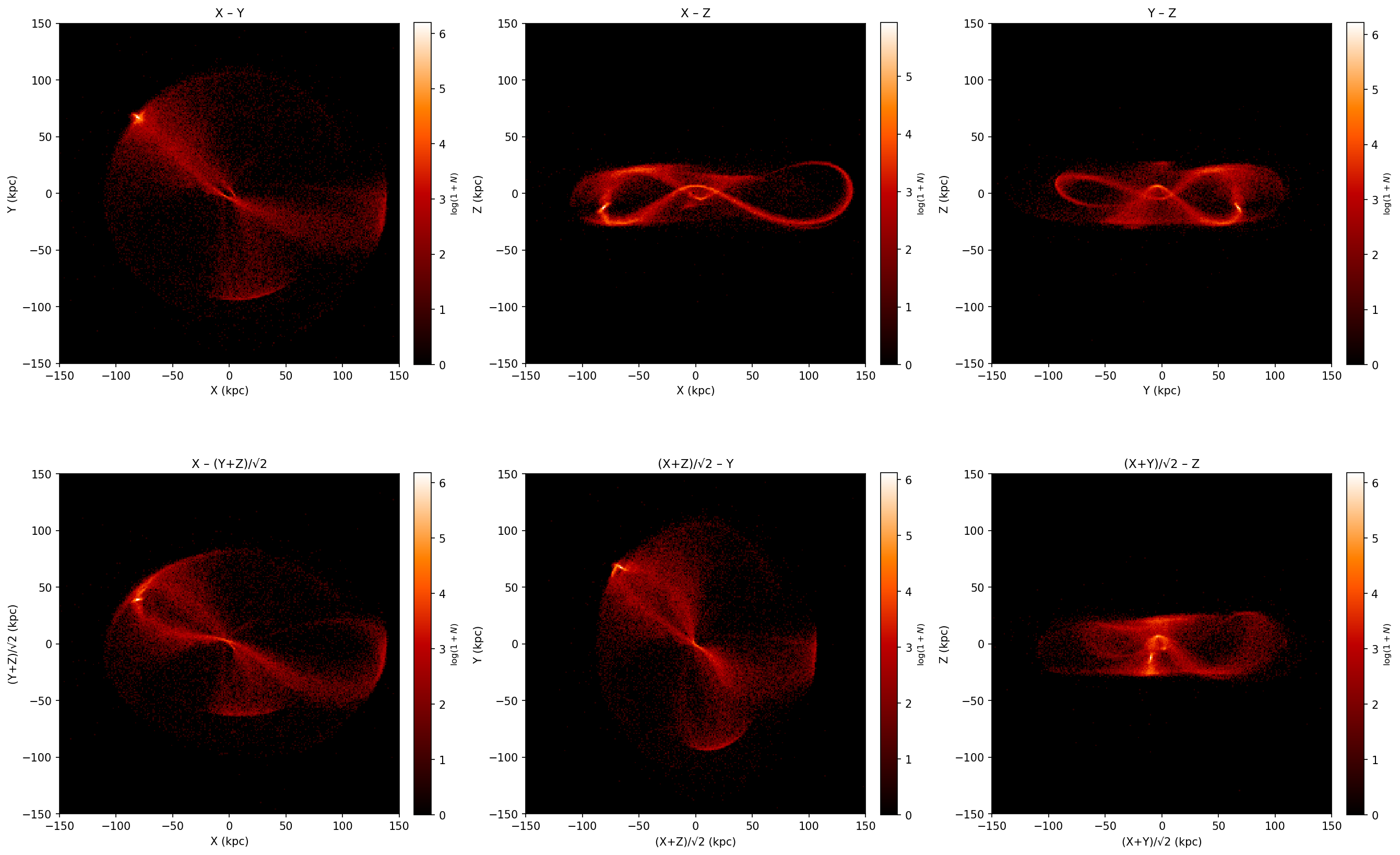} 

\caption{Six projected position distributions of the radial orbit progenitor 
in an extremely oblate host halo ($q = 0.5$). Row~1 shows the standard orthogonal projections: $XY$ (left), $XZ$ (centre), and $YZ$ (right). Row~2 shows intermediate projections rotated 45° between adjacent planes: $X$ 
vs $(Y{+}Z)/\sqrt{2}$ rotated about $\hat{X}$ (left), $Y$ vs $(X{+}Z)/\sqrt{2}$ rotated about $\hat{Y}$ (centre), and $(X{+}Y)/\sqrt{2}$ vs $Z$ rotated about $\hat{Z}$ (right).} 
\label{appendix:figure_1_0.5_density_projections}
\end{figure*}
Here we present the remaining projections and ordered density plots for the presented sub-sample.
\begin{figure*}
\centering
    \setlength{\tabcolsep}{1pt} 
    \renewcommand{\arraystretch}{0} 
\includegraphics[width=0.9\textwidth, height=.35\textwidth]{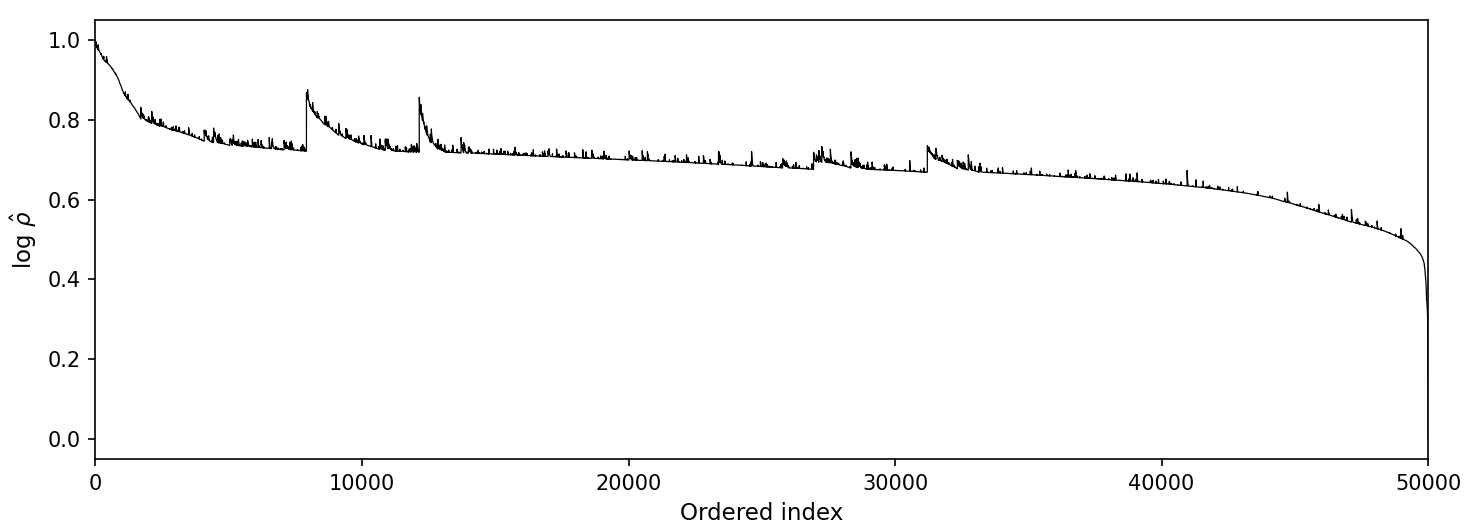} 
\caption{Ordered density distribution of the radial orbit progenitor in an extremely oblate host halo ($q = 0.5$). The overall morphology is stream-like, with localised shell-like signatures in the tail sub-cluster attributable to debris spreading driven by halo flattening.}
\end{figure*}

\begin{figure*}
\centering
    \setlength{\tabcolsep}{1pt} 
    \renewcommand{\arraystretch}{0} 
\includegraphics[width=0.9\textwidth, height=.5\textwidth]{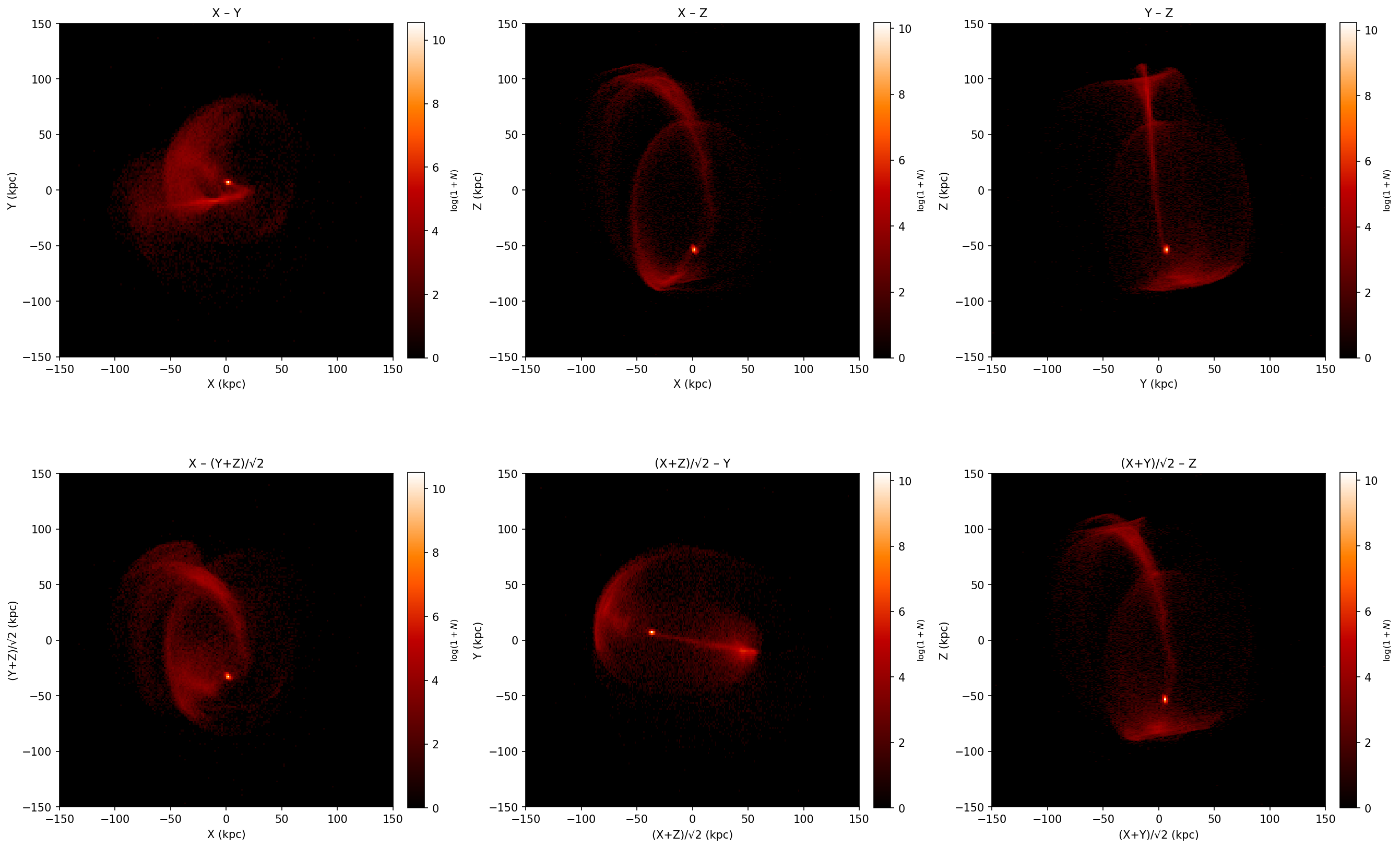} 
\caption{Six projected position distributions of the intermediate orbit progenitor in an extremely oblate host halo ($q = 0.5$), following the same projection scheme as Figure \ref{appendix:figure_1_0.5_density_projections} Consistent with the discussion in Section~\ref{subsection: Subhalo B example}, the structure retains shell-like features in certain projections,a signature of the strong halo flattening at $q = 0.5$ imprinting on the debris morphology.}
\end{figure*}

\begin{figure*}
\centering
    \setlength{\tabcolsep}{1pt} 
    \renewcommand{\arraystretch}{0} 
\includegraphics[width=0.9\textwidth, height=.5\textwidth]{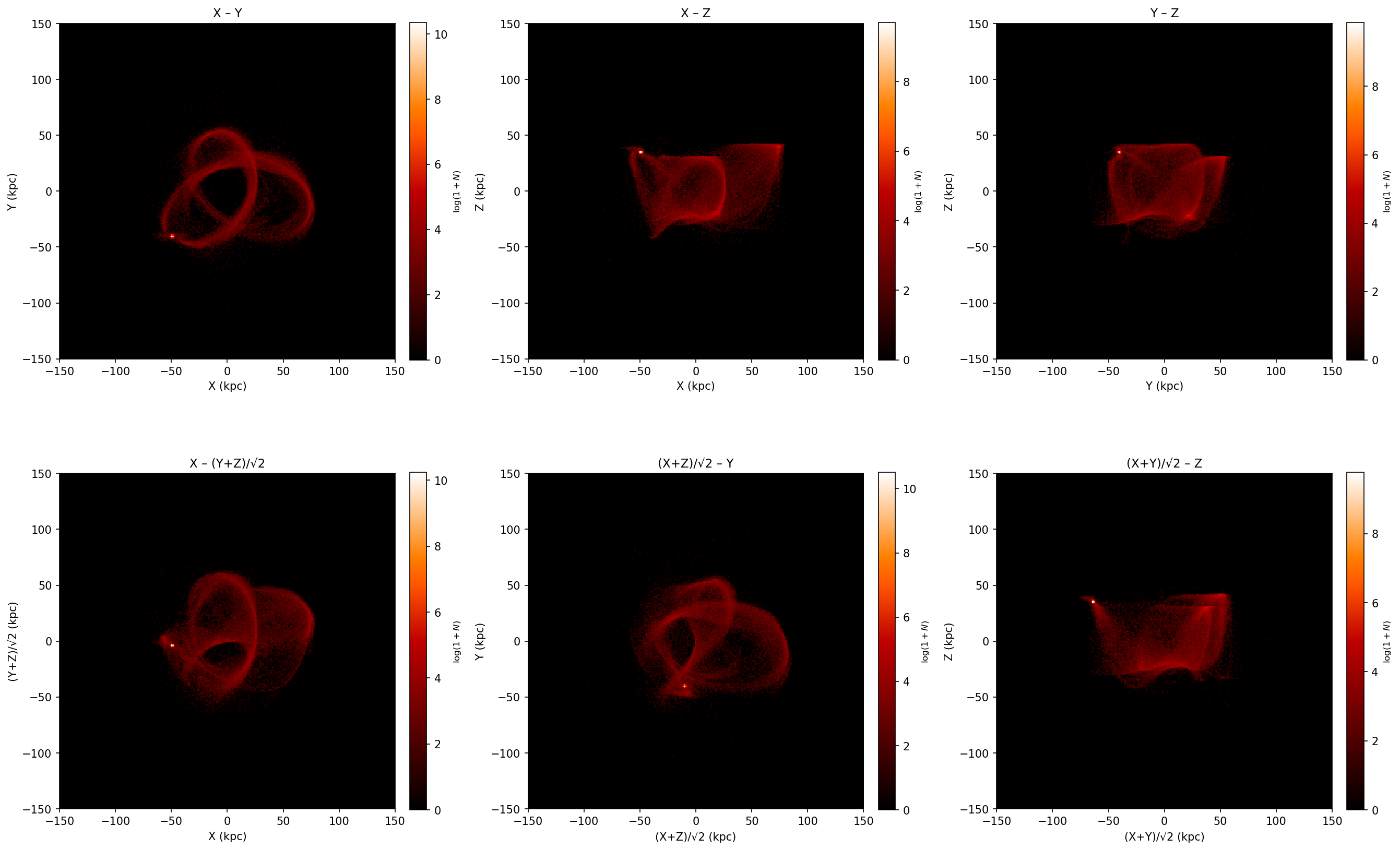} 
\caption{Projected position distributions of the eccentric orbit progenitor in an extremely oblate host halo ($q = 0.5$), following the same projection scheme as Figure \ref{appendix:figure_1_0.5_density_projections} The structure displays coherent stream-like morphology across all projections.}
\label{appendix fig: Subahlo C 0.5 density projections}
\end{figure*}

\begin{figure*}
\centering
    \setlength{\tabcolsep}{1pt} 
    \renewcommand{\arraystretch}{0} 
\includegraphics[width=0.9\textwidth, height=0.35\textwidth]{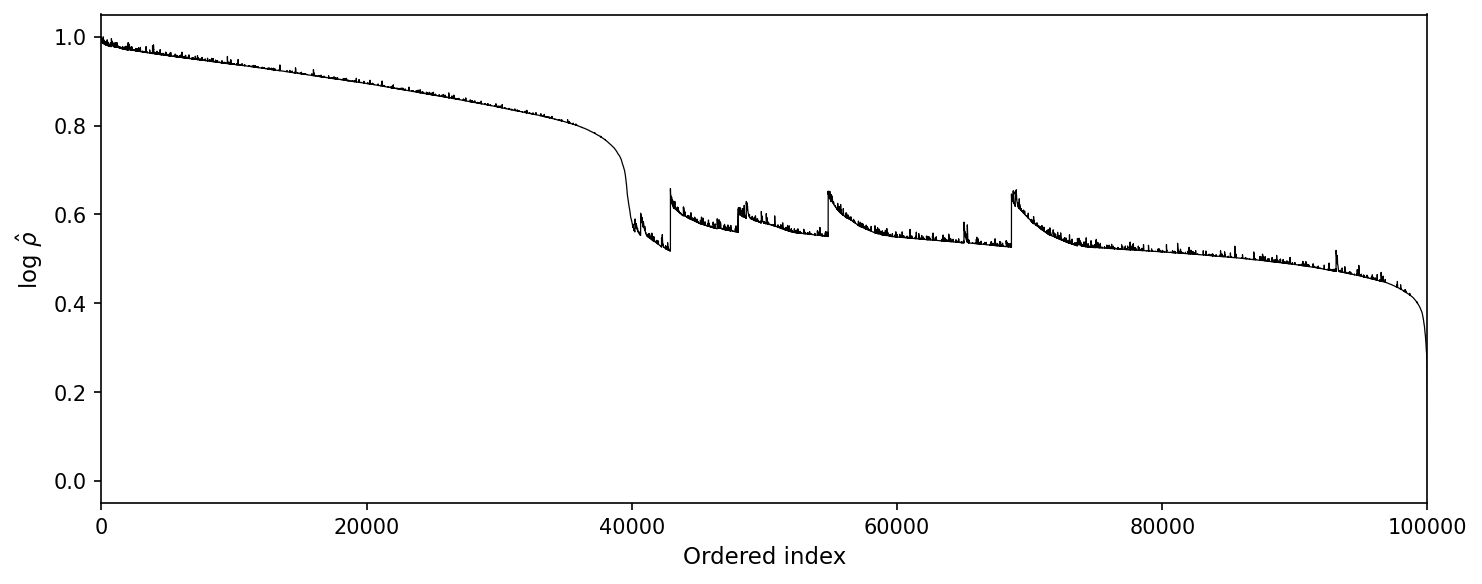} 
\caption{Ordered density distribution of the eccentric orbit progenitor in an extremely oblate host halo ($q = 0.5$). All subclusters identified by AstroLink exhibit stream-like density profiles, consistent with the visual morphology in Figure \ref{appendix fig: Subahlo C 0.5 density projections}.}
\end{figure*}

\begin{figure*}
\centering
    \setlength{\tabcolsep}{1pt} 
    \renewcommand{\arraystretch}{0} 
\includegraphics[width=0.9\textwidth, height=.5\textwidth]{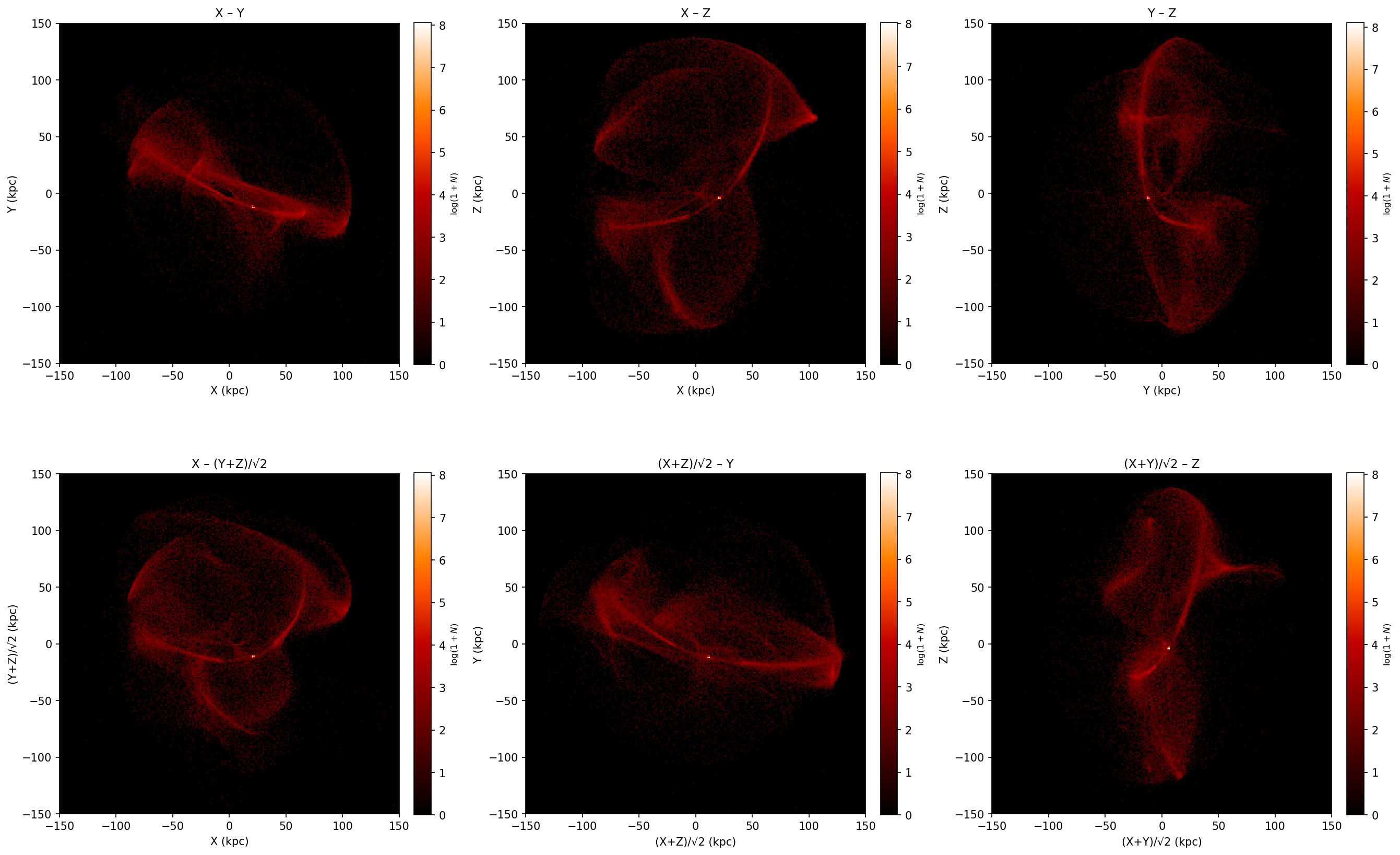} 
\caption{Projected position distributions of the radial orbit progenitor in an extremely prolate host halo ($q = 1.5$), following the same projection scheme as Figure \ref{appendix:figure_1_0.5_density_projections}. The debris is significantly more phase-mixed relative to the oblate case, exhibiting a superposition of stream and shell-like features across all projections.}
\label{appendix_fig_6:q=1.5_radial}
\end{figure*}
\begin{figure*}
\centering
    \setlength{\tabcolsep}{1pt} 
    \renewcommand{\arraystretch}{0} 
\includegraphics[width=0.9\textwidth, height=.35\textwidth]{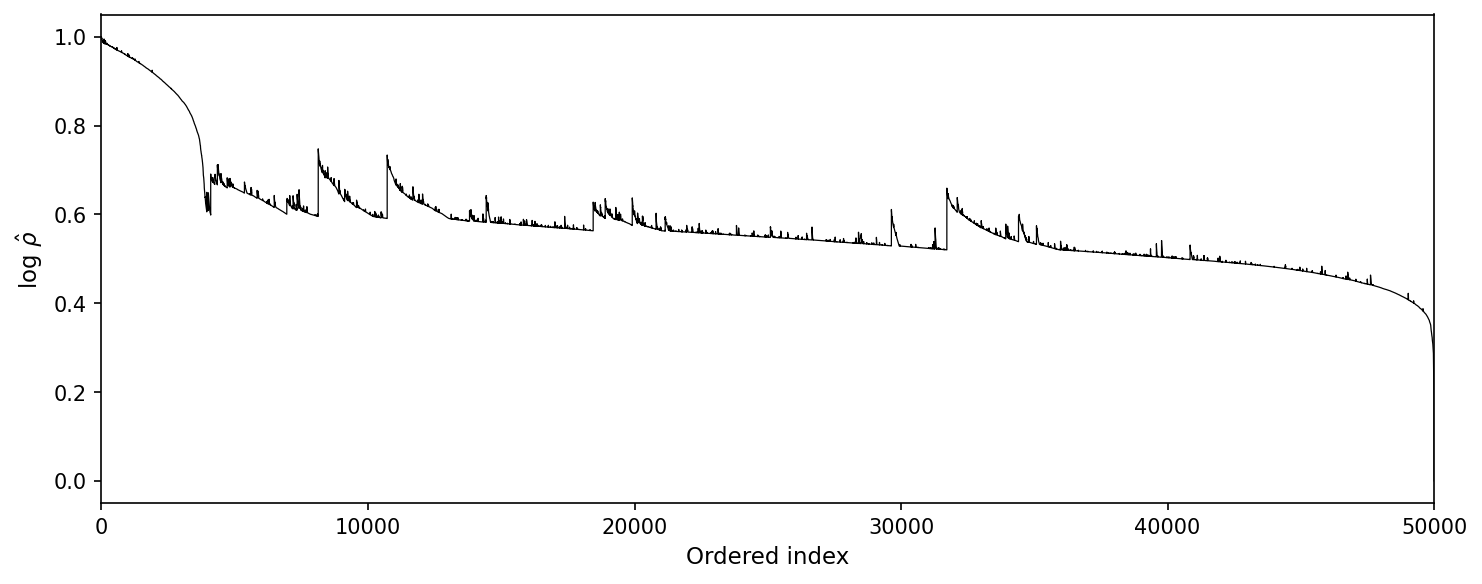} 
\caption{Ordered density distribution of the radial orbit progenitor in an extremely prolate host halo ($q = 1.5$). The leading sub-cluster exhibits a stream-like density gradient, while subsequent subclusters display shell-like drop-offs, indicating a composite morphology in which stream-like features connect distinct shell formations visible in Figure~ \ref{appendix_fig_6:q=1.5_radial}. }
\end{figure*}

\begin{figure*}
\centering
    \setlength{\tabcolsep}{1pt} 
    \renewcommand{\arraystretch}{0} 
\includegraphics[width=0.9\textwidth, height=.5\textwidth]{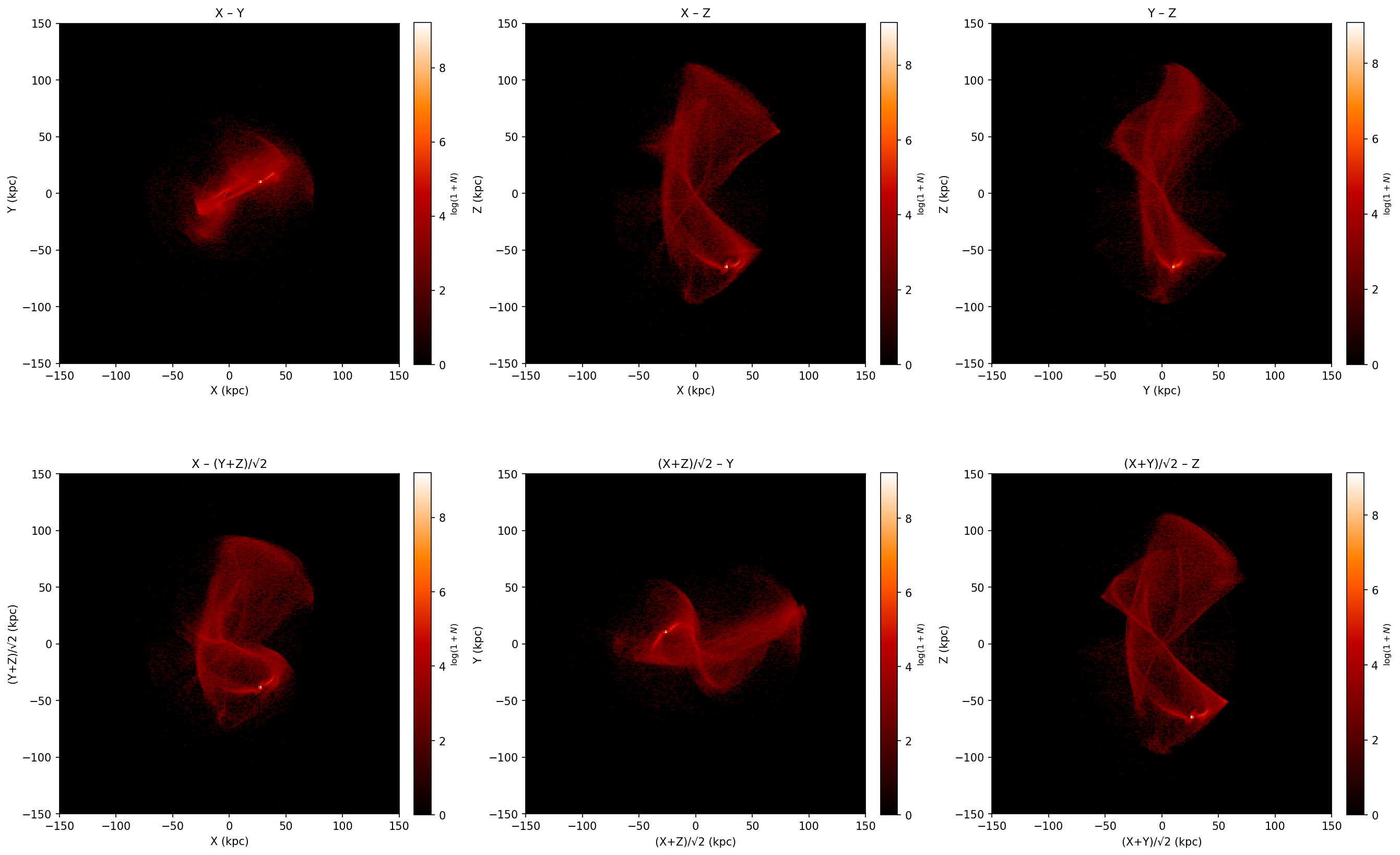} 
\caption{Projected position distributions of the intermediate orbit progenitor in an extremely prolate host halo ($q = 1.5$), following the same projection scheme as Figure \ref{appendix:figure_1_0.5_density_projections}. The structure is more phase-mixed and spatially diffuse relative to lower-$q$ counterparts. }
\end{figure*}

\begin{figure*}
\centering
    \setlength{\tabcolsep}{1pt} 
    \renewcommand{\arraystretch}{0} 
\includegraphics[width=0.9\textwidth, height=.35\textwidth]{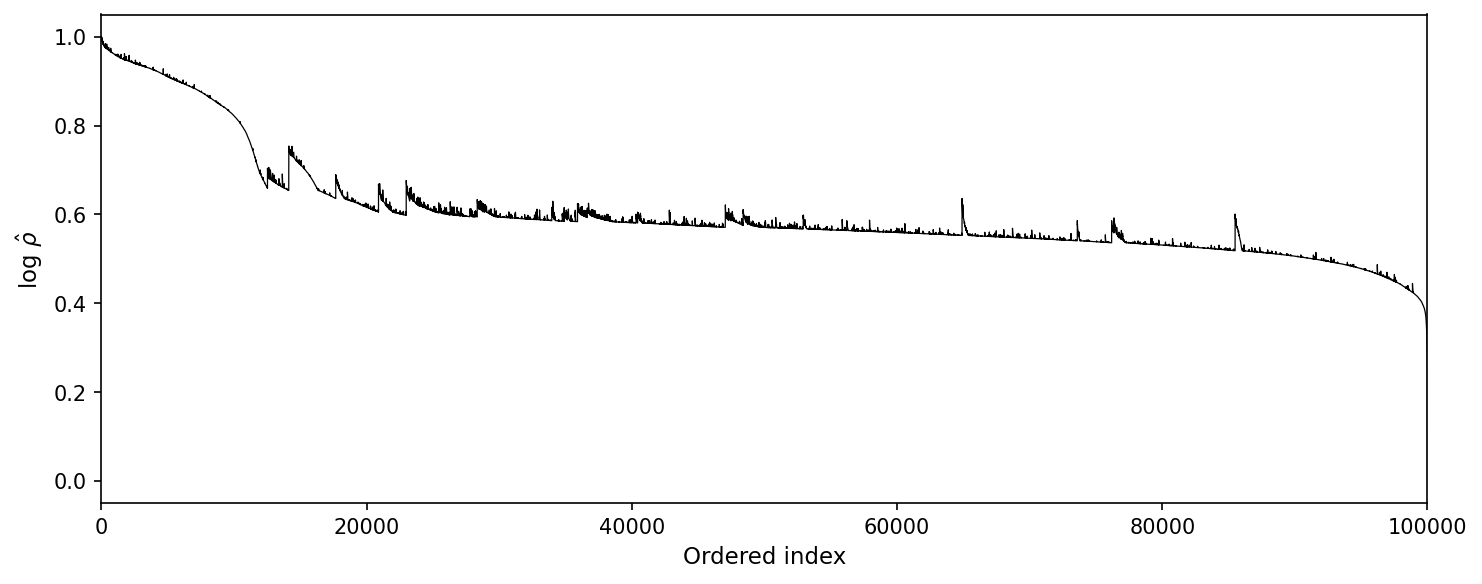} 
\caption{Ordered density distribution of the intermediate orbit progenitor in an extremely prolate host halo ($q = 1.5$). A dominant stream-like subcluster is identified, with minor shell-like features too small to constitute independent substructures. The overall morphology is classified as a diffuse stream.}
\end{figure*}
\begin{figure*}
\centering
    \setlength{\tabcolsep}{1pt} 
    \renewcommand{\arraystretch}{0} 
\includegraphics[width=0.9\textwidth, height=.5\textwidth]{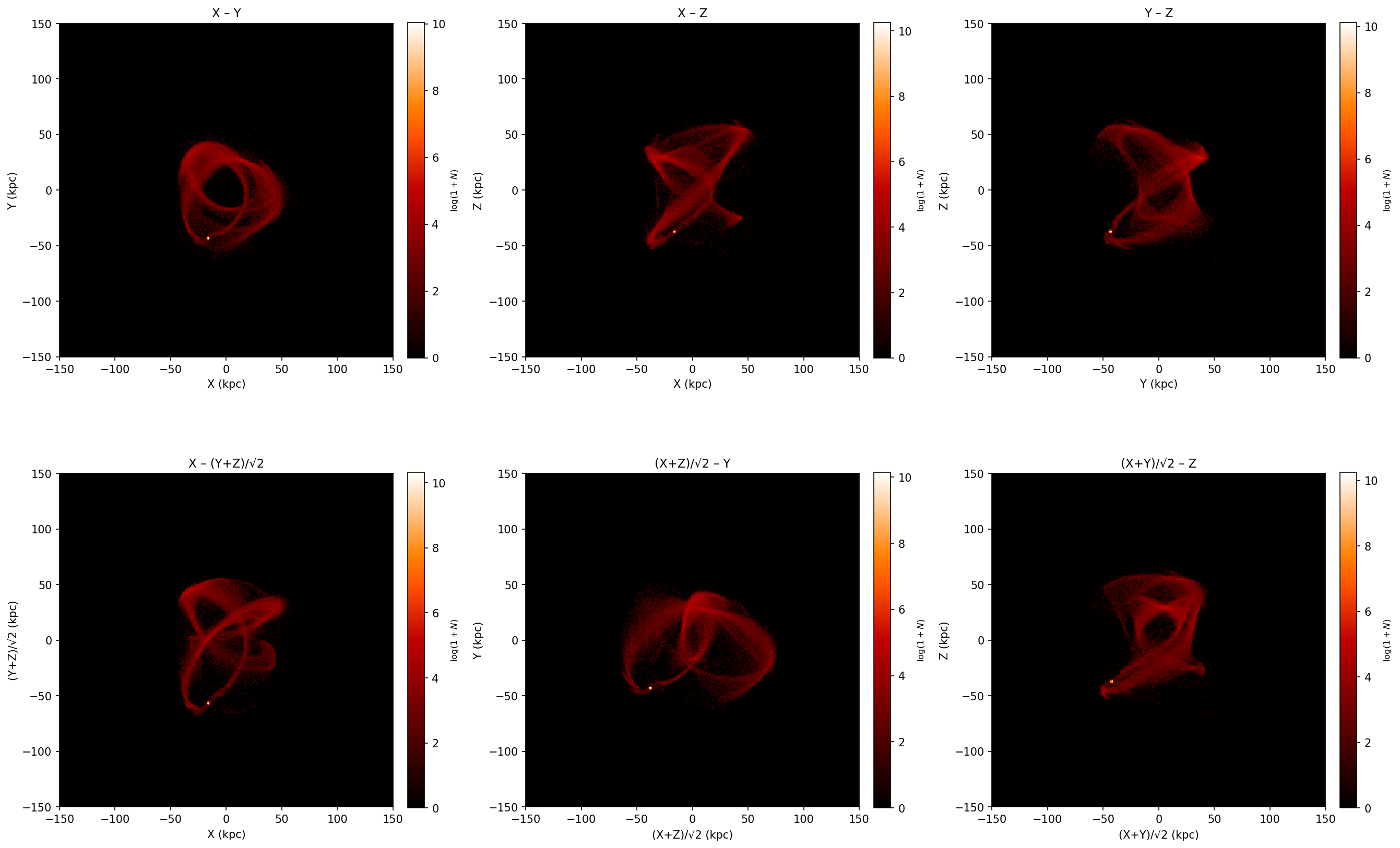} 
\caption{Projected position distributions of the eccentric orbit progenitor in an extremely prolate host halo ($q = 1.5$), following the same projection scheme as Figure \ref{appendix:figure_1_0.5_density_projections}. The stream-like morphology is more coherent and better preserved across projections relative to the radial and intermediate cases at the same $q$.}
\end{figure*}

\begin{figure*}
\centering
    \setlength{\tabcolsep}{1pt} 
    \renewcommand{\arraystretch}{0} 
\includegraphics[width=0.9\textwidth, height=.35\textwidth]{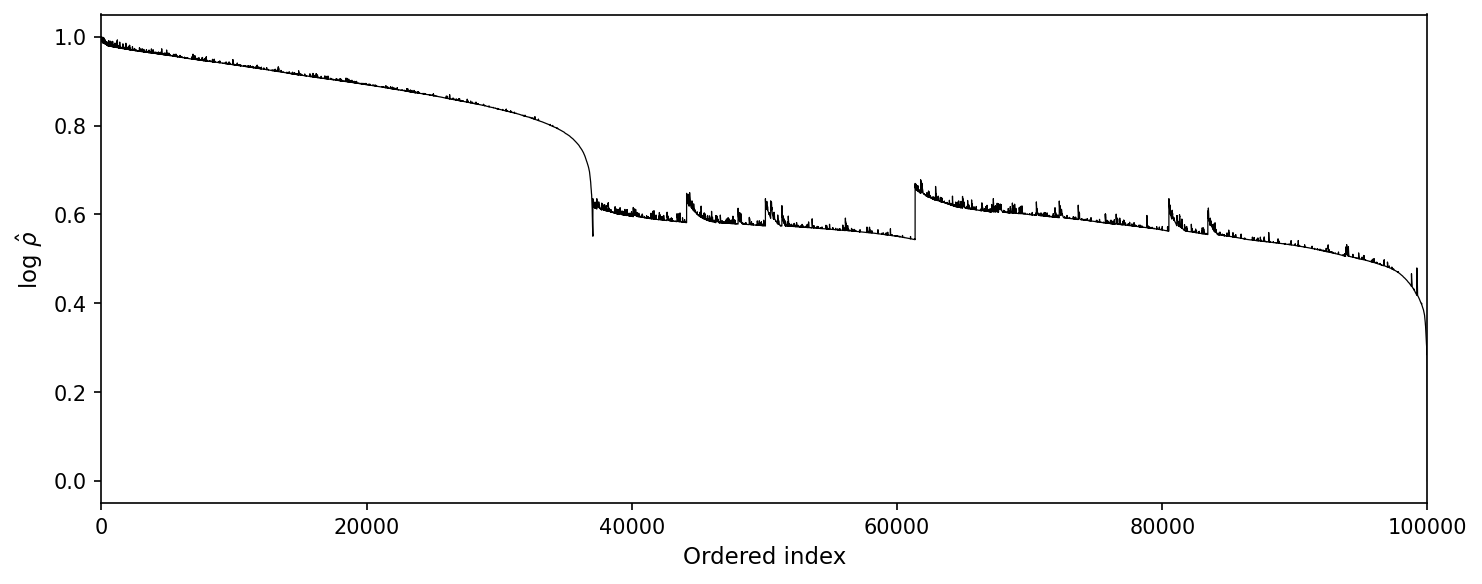} 
\caption{Ordered density distribution of the eccentric orbit progenitor in an extremely prolate host halo ($q = 1.5$). All subclusters identified by AstroLink exhibit stream-like profiles, and the overall ordered density distribution is consistent with a coherent tidal stream classification.}
\end{figure*}

\begin{figure*}
\centering
    \setlength{\tabcolsep}{1pt} 
    \renewcommand{\arraystretch}{0} 
\includegraphics[width=0.9\textwidth, height=.5\textwidth]{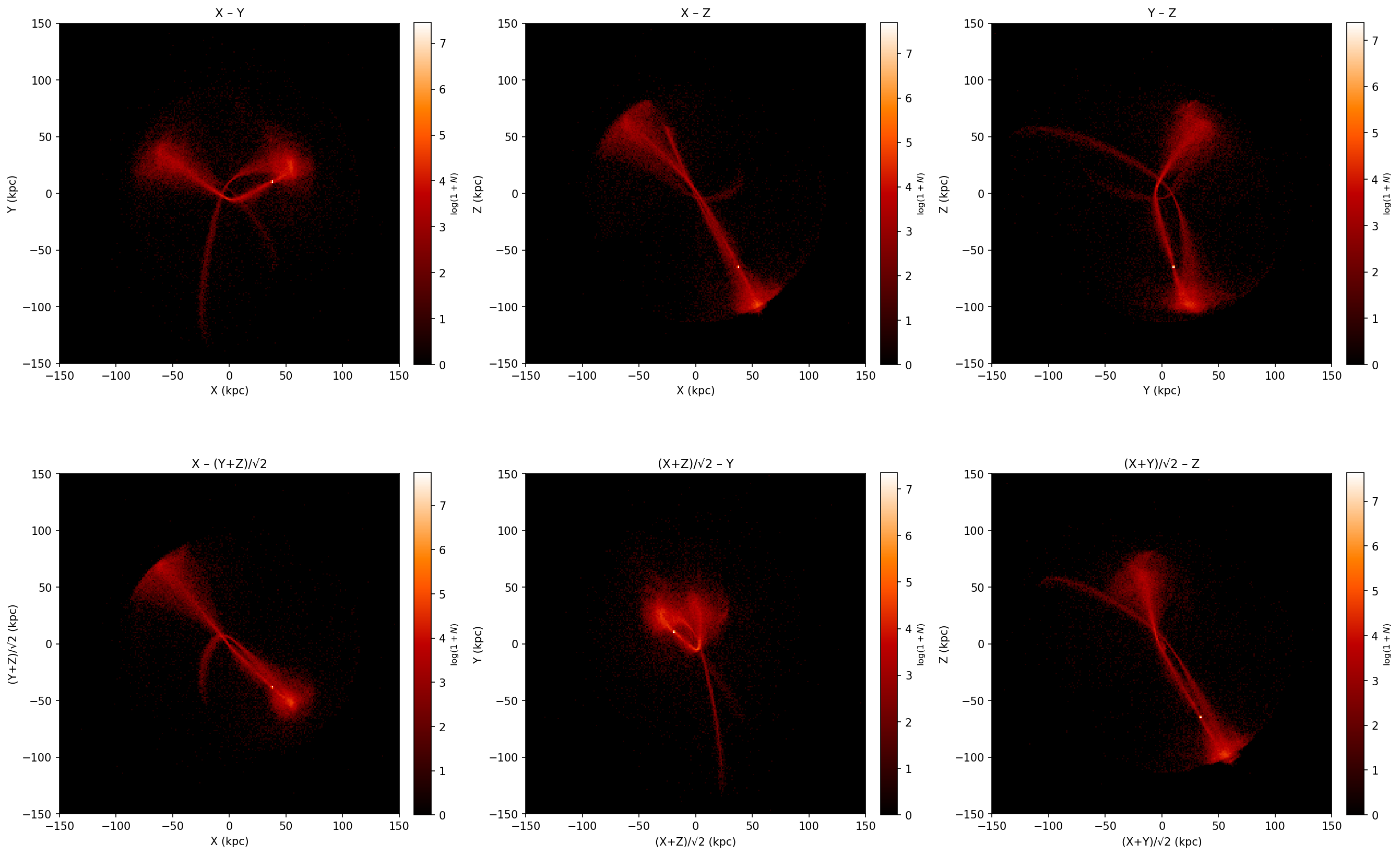} 
\caption{Projected position distributions of the radial orbit progenitor evolved in a spherical NFW host halo ($q = 1.0$), following the same projection scheme as Figure \ref{appendix:figure_1_0.5_density_projections}. The debris presents a mixed morphology  the core displays plume-like shell features while the extended tails can be misidentified as streams depending on the viewing orientation, illustrating the projection dependent ambiguity inherent to this 
structure.}
\end{figure*}

\begin{figure*}
\centering
    \setlength{\tabcolsep}{1pt} 
    \renewcommand{\arraystretch}{0} 
\includegraphics[width=0.9\textwidth, height=.35\textwidth]{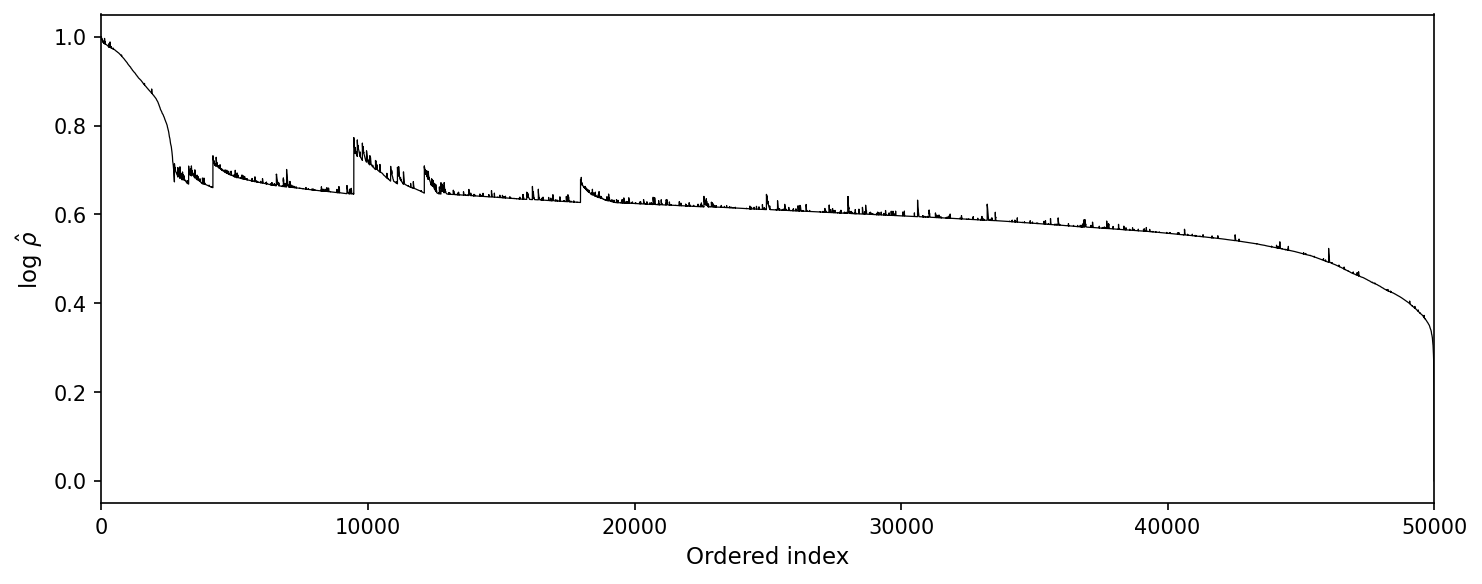} 
\caption{Ordered density distribution of the radial orbit progenitor in a spherical NFW host halo ($q = 1.0$). While individual subclusters carry stream and tail-like signatures, the overall density drop-off of the complete structure is characteristic of a shell, confirming the shell classification for this progenitor in the absence of halo flattening. }
\end{figure*}

\begin{figure*}
\centering
    \setlength{\tabcolsep}{1pt} 
    \renewcommand{\arraystretch}{0} 
\includegraphics[width=.9\textwidth, height=.5\textwidth]{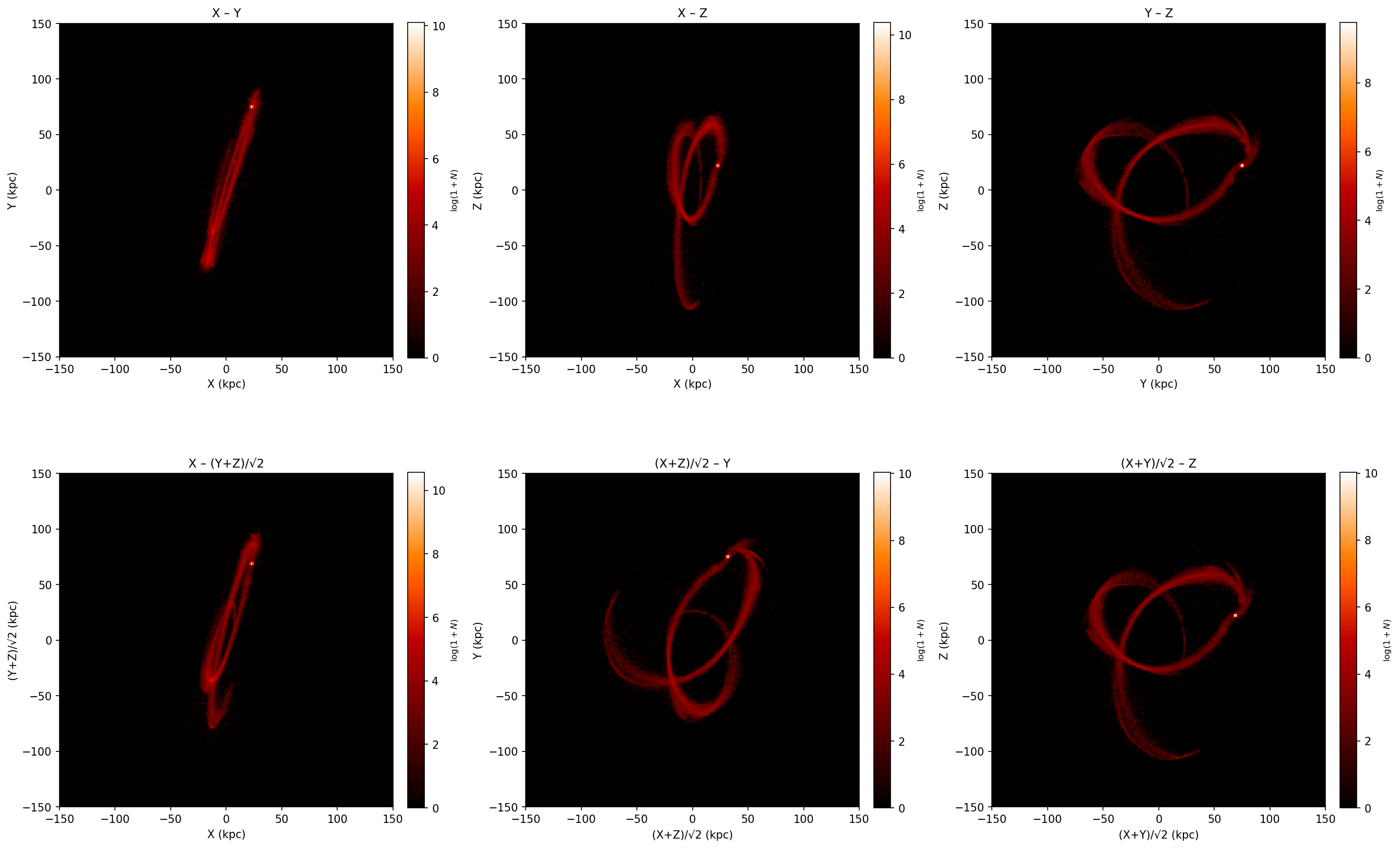} 
\caption{Projected position distributions of the intermediate orbit progenitor evolved in a spherical NFW host halo ($q = 1.0$), following the same projection scheme as Figure \ref{appendix:figure_1_0.5_density_projections}. The structure presents a clear and coherent stream-like morphology across all projections, with none of the shell-like contamination seen in the oblate and prolate cases, reflecting the absence of potential-driven debris spreading in the spherical limit.}
\end{figure*}
\begin{figure*}
\centering
    \setlength{\tabcolsep}{1pt} 
    \renewcommand{\arraystretch}{0} 
\includegraphics[width=.9\textwidth, height=.35\textwidth]{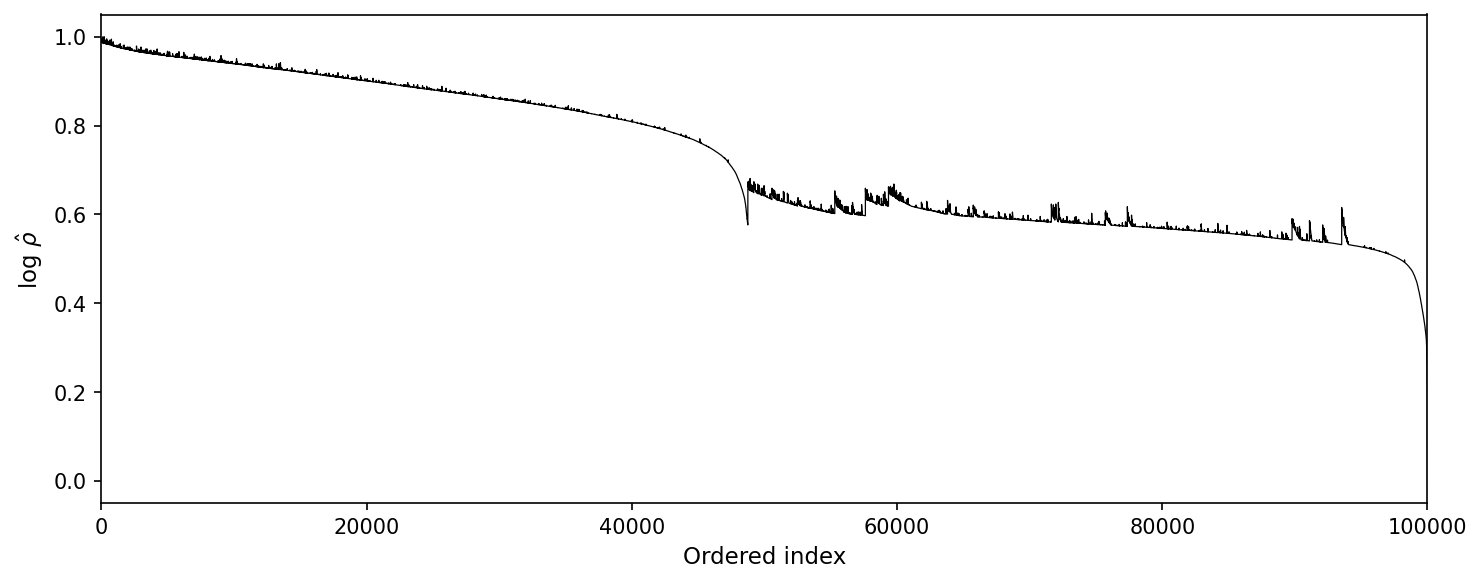} 
\caption{Ordered density distribution of the intermediate orbit progenitor in a spherical NFW host halo ($q = 1.0$). All subclusters identified by AstroLink trace stream-like profiles, and the global ordered density distribution is fully consistent with a tidal stream classification.}
\end{figure*}

\begin{figure*}
\centering
    \setlength{\tabcolsep}{1pt} 
    \renewcommand{\arraystretch}{0} 
\includegraphics[width=.9\textwidth, height=.5\textwidth]{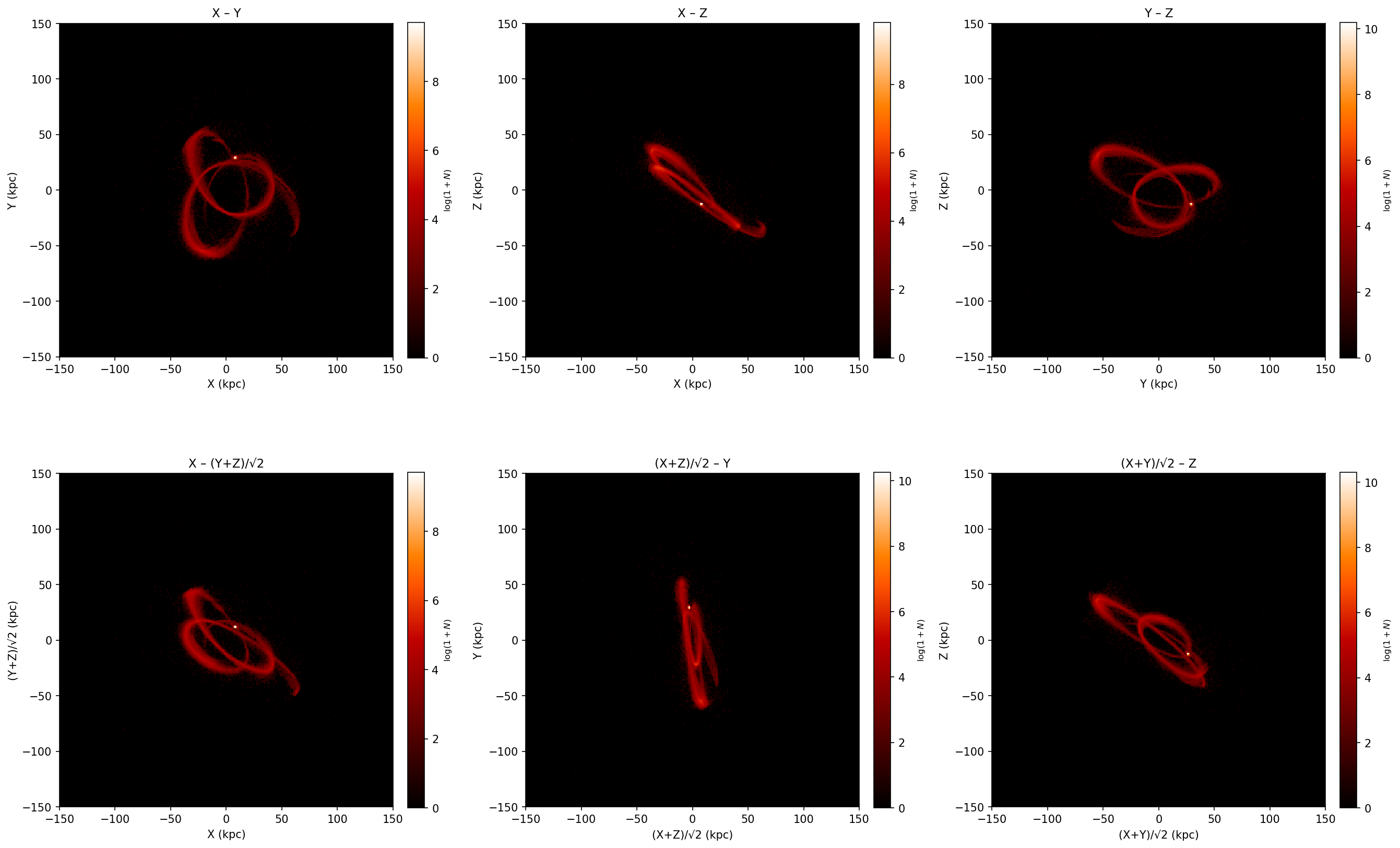} 
\caption{Projected position distributions of the eccentric orbit progenitor evolved in a spherical NFW host halo ($q = 1.0$), following the same projection scheme as Figure \ref{appendix:figure_1_0.5_density_projections}. The structure has developed well-defined and coherent stream-like features across all projections, with no morphological ambiguity.}
\end{figure*}
\begin{figure*}
\centering
    \setlength{\tabcolsep}{1pt} 
    \renewcommand{\arraystretch}{0} 
\includegraphics[width=0.9\textwidth, height=.35\textwidth]{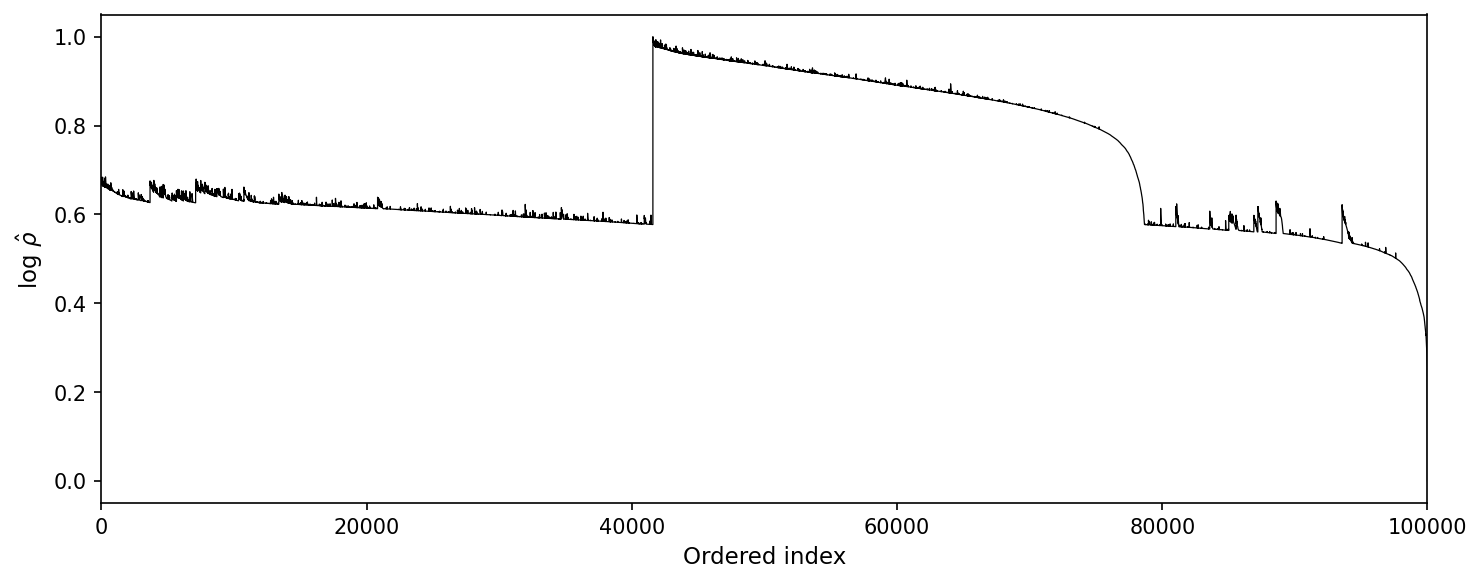} 
\caption{Ordered density distribution of the eccentric orbit progenitor in a spherical NFW host halo ($q = 1.0$). The ordered density profile unambiguously traces a strong tidal stream, with all subclusters identified by AstroLink displaying consistent stream-like density gradients throughout.}
\end{figure*}

\bsp	
\label{lastpage}
\end{document}